\documentclass[preprint,12pt]{elsarticle}

\usepackage{amssymb}
\usepackage{amsmath}
\usepackage{hyperref}
\hypersetup{hypertex=true,
	colorlinks=true,
	linkcolor=blue,
	anchorcolor=blue,
	citecolor=blue}
\usepackage{booktabs}    
\usepackage{multirow}
\usepackage{array}       
\usepackage{caption}     
\usepackage{subcaption}  
\usepackage{siunitx}     

\journal{Journal of Computational Physics}

\begin{document}
	
	\begin{frontmatter}
		
		
		\title{An improved lattice Boltzmann method with a novel conservative boundary scheme for viscoelastic fluid flows}
		
		\author[aff0,aff1,aff2]{Yuan Yu\corref{cor1}}
		\ead{yuyuan@xtu.edu.cn}
		\author[aff0]{Siwei Chen}
		\author[aff3]{Lei Wang}
		\author[aff0]{Hai-zhuan Yuan\corref{cor1}}
		\ead{yhz@xtu.edu.cn}
		\author[aff0]{Shi Shu}

		\cortext[cor1]{Corresponding author.}
		\affiliation[aff0]{organization={School of Mathematics and Computational Science,
				Xiangtan University},
			city={Xiangtan}, 
			postcode={411105},
			country={China}}
		\affiliation[aff1]{organization={National Center for Applied Mathematics in Hunan},
			city={Xiangtan}, 
			postcode={411105}, 
			country={China}}
		\affiliation[aff2]{organization={Hunan Key Laboratory for Computation and Simulation in Science and Engineering, Xiangtan University}, 
			city={Xiangtan}, 
			postcode={411105},
			country={China}}
		\affiliation[aff3]{organization={School of Mathematics and Physics, China University of Geosciences}, 
			city={Wuhan}, 
			postcode={430074},
			country={China}}
		
		\begin{abstract}
			The high Weissenberg number problem has been a persistent challenge in the numerical simulation of viscoelastic fluid flows. This paper presents an improved lattice Boltzmann method for solving viscoelastic flow problems at high Weissenberg numbers. The proposed approach employs two independent two-relaxation-time regularized lattice Boltzmann models to solve the hydrodynamic field and conformation tensor field of viscoelastic fluid flows, respectively. The viscoelastic stress computed from the conformation tensor is directly embedded into the hydrodynamic field using a newly proposed local velocity discretization scheme, thereby avoiding spatial gradient calculations. The constitutive equations are treated as convection-diffusion equations and solved using an improved convection-diffusion model specifically designed for this purpose, incorporating a novel auxiliary source term that eliminates the need for spatial and temporal derivative computations. Additionally, a conservative non-equilibrium bounce-back (CNEBB) scheme is proposed for implementing solid wall boundary conditions in the constitutive equations. The robustness of the present algorithm is validated through a series of benchmark problems. The simplified four-roll mill problem demonstrates that the method effectively improves numerical accuracy and stability in bulk regions containing stress singularities. The Poiseuille flow problem validates the accuracy of the current algorithm with the CNEBB boundary scheme at extremely high Weissenberg numbers (tested up to Wi = 10,000). The flow past a circular cylinder problem confirms the superior stability and applicability of the algorithm for complex curved boundary problems compared to other existing common schemes.
		\end{abstract}
		
		
		
		
	\end{frontmatter}
	
	
	\section{Introduction}\label{}

	Viscoelastic fluids are ubiquitous in industrial applications and natural phenomena, ranging from polymer processing~\cite{leonov2012nonlinear} and food manufacturing~\cite{siacor2021additive} to biological systems~\cite{pajic2024friction} and enhanced oil recovery~\cite{xie2020nonwetting,xie2022oscillative}. These fluids exhibit both viscous and elastic properties, displaying complex rheological behaviors such as normal stress differences, memory effects, and shear-thinning or shear-thickening characteristics~\cite{cheneler2016viscoelasticity}. The accurate numerical simulation of viscoelastic fluid flows is crucial for understanding and optimizing various industrial processes, as well as advancing our fundamental knowledge in computational rheology. However, despite over four decades of intensive research, the numerical simulation of viscoelastic fluids remains one of the most challenging problems in computational fluid dynamics~\cite{alves2021numerical}.
	
	The primary obstacle in simulating viscoelastic fluid flows is the high Weissenberg number problem (HWNP)~\cite{keunings1986high,alves2021numerical}, which manifests as severe numerical instabilities when the Weissenberg number (Wi) – representing the ratio of elastic relaxation time to the characteristic flow time scale – exceeds a critical threshold. The HWNP is characterized by several detrimental phenomena: exponential growth of stress components, loss of solution convergence, catastrophic numerical breakdown, and mesh-dependent solutions that violate grid convergence principles~\cite{alves2021numerical}. The underlying mechanism stems from the hyperbolic nature of the constitutive equations in regions of strong extensional flow, where elastic stresses undergo rapid spatial and temporal variations that challenge conventional discretization schemes. This fundamental limitation significantly constrains the practical utility of computational methods for viscoelastic flows, particularly in industrial applications where moderate to high Weissenberg numbers are prevalent and elastic effects dominate flow behavior.
	
	To address the HWNP, researchers have developed various strategies that can be broadly categorized into three approaches. The first approach involves modifying the constitutive equations to prevent unbounded stress growth. Notable examples include the Giesekus model~\cite{giesekus1982simple}, which introduces a nonlinear mobility coefficient to account for intermolecular friction between polymer chains, and finite extensibility models such as FENE-P and FENE-CR~\cite{bird1980polymer,chilcott1988creeping}, which impose maximum stretch ratios on molecular chains to prevent excessive stress growth under large deformations. These modifications improve numerical stability while capturing important physical phenomena such as shear-thinning behavior and finite chain extensibility. In contrast, the Oldroyd-B model~\cite{oldroyd1950formulation}, which incorporates memory effects through the upper-convected derivative but lacks stress saturation mechanisms, allows unlimited stress growth in extensional flows. This characteristic makes the Oldroyd-B model particularly challenging for numerical simulations at high Wi, yet it remains a fundamental benchmark for testing numerical methods because it exposes core difficulties without artificial stabilization~\cite{alves2021numerical}.
	
	The second approach focuses on improving numerical discretization schemes. The Elastic-Viscous Stress Splitting (EVSS) method~\cite{rajagopalan1990finite,sun1996adaptive,fortin2000discrete} separates total stress into elastic and viscous components to reduce coupling oscillations. The Streamline Upwind Petrov-Galerkin (SUPG) method~\cite{marchal1987new,hughes1986new} applies upstream weighting to convection terms to minimize directional errors. Additionally, specialized formulations such as the log-conformation approach~\cite{fattal2005time} ensure positive-definiteness of the conformation tensor to avoid numerical breakdown from exponential stress growth. These numerical improvements have extended the range of computable Weissenberg numbers while preserving original constitutive equations, though stable simulations at extremely high Wi remain challenging for complex geometries.
	
	The third approach involves alternative numerical frameworks, among which the lattice Boltzmann method (LBM) has emerged as a promising candidate~\cite{aidun2010lattice}. LBM offers several advantages: its kinetic foundation provides a natural framework for complex physics, the streaming step corresponds to the exact solution of the d'Alembert advection equation and is therefore free from numerical diffusion, the method is inherently parallel, and boundary conditions can be implemented through local operations. This exact d'Alembert solution property is particularly valuable for transport-dominated problems where traditional methods suffer from numerical diffusion~\cite{marie2009comparison}. Recent theoretical advances by Bellotti~\cite{bellotti2023numerical,bellotti2024initialisation} have provided rigorous mathematical frameworks for analyzing lattice Boltzmann schemes, including their consistency, stability, and convergence properties through modified equation analysis and connections to finite difference methods. These theoretical foundations strengthen the mathematical basis for LBM applications to complex fluid problems. These features make LBM attractive for viscoelastic fluid simulations, where accurate coupling between flow fields and stress evolution is essential.
	
	Research on LBM for viscoelastic fluids has followed several directions. For macroscopic constitutive equations, Malaspinas et al.\cite{malaspinas2010lattice} established the foundation by discretizing both conservation and constitutive equations within the LBM framework, successfully simulating benchmark flows though with reduced accuracy at higher Wi. Su et al.\cite{su2018lattice} developed a tensor distribution function approach that eliminated redundant terms in stress tensor evolution, demonstrating improved stability for high-Wi cavity flows. Recent work by Zhang et al.\cite{zhang2025latticea} introduced a viscoelastic lattice Boltzmann flux solver incorporating stress-splitting techniques while maintaining second-order spatial accuracy. Alternative approaches have targeted the Fokker-Planck equation for polymer conformation distribution\cite{phillips2011lattice,singh2013lattice}, though computational requirements scale unfavorably with system complexity. Hybrid methods combining LBM with traditional schemes have also been explored~\cite{zou2015benchmark,gupta2015hybrid}, achieving improved efficiency while maintaining accuracy.
	
	Despite these advances, existing LBM approaches for viscoelastic fluids face significant limitations in addressing the HWNP. Maximum achievable Weissenberg numbers remain modest for benchmark problems, and key phenomena such as flow instabilities in complex geometries are difficult to capture reliably. While LBM's exact advection property largely eliminates issues related to convective term treatment and stress tensor discretization that plague traditional finite difference and finite element methods, boundary condition implementation emerges as a critical limiting factor. Furthermore, theoretical analysis of LBM initialization and boundary treatments remains an active area of research~\cite{bellotti2023numerical,bellotti2024initialisation}, with proper initialization schemes being crucial for maintaining consistency and stability, particularly for extended state space problems like viscoelastic flows where multiple field variables must be simultaneously evolved. The handling of boundary conditions represents a particularly under-explored aspect in LBM for viscoelastic flows, despite being extensively studied for velocity fields. While boundary conditions for velocity fields have been well-established in LBM, the treatment of the conformation tensor at solid boundaries has received limited attention. The standard bounce-back scheme in LBM, though effective for velocity fields, does not account for the specific physical requirements of the conformation tensor field at solid boundaries. The trace of the conformation tensor, representing the mean-square end-to-end distance of polymer chains, should satisfy conservation properties consistent with polymer physics to maintain physical fidelity. Specifically, in the absence of chemical reactions or chain scission, the trace of the conformation tensor should remain constant along material lines, a principle that extends to boundary regions where polymer chains interact with solid surfaces. When these conservation principles are violated at boundaries, artificial polymer stretching or compression occurs, generating spurious stresses that can propagate into the bulk flow and potentially trigger numerical instabilities, particularly at elevated Weissenberg numbers where small perturbations can lead to catastrophic breakdown.
	
	In this work, we develop an improved lattice Boltzmann method for Oldroyd-B viscoelastic fluid flows that addresses several key limitations of existing approaches. Our method incorporates three main innovations. First, we employ separate two-relaxation-time (TRT) regularized lattice Boltzmann models~\cite{yu2025tworelaxationtime,yu2025tworelaxationtime2} for the hydrodynamic and conformation tensor fields, with the viscoelastic stress tensor directly incorporated in velocity-discrete form to eliminate gradient calculations. Second, we develop a new convection-diffusion LB model for the constitutive equation that includes an auxiliary source term, avoiding both spatial and temporal gradient computations. Third, we introduce a novel conservative non-equilibrium bounce-back (CNEBB) scheme that ensures proper conservation properties of the conformation tensor at solid boundaries. This conservation property is achieved by maintaining the physical constraint that the trace of the conformation tensor should remain constant along streamlines, even at boundary nodes. Through systematic validation on benchmark problems including the four-roll mill flow, force-driven Poiseuille flow, and flow past a circular cylinder, we demonstrate that this approach enables stable simulations at Weissenberg numbers substantially higher than previously reported in LBM studies for the Oldroyd-B model, while maintaining computational efficiency and accuracy.
	
	The remainder of this paper is organized as follows. Section~\ref{sec2} presents the governing equations for Oldroyd-B viscoelastic fluids. Section~\ref{sec3} describes our improved LBM formulation, including the two-relaxation-time regularized models for hydrodynamic and polymer fields (Sections~\ref{sec3.1} and~\ref{sec3.2}), and the conservative boundary scheme for the conformation tensor evolution (Section~\ref{sec3.3}). Section~\ref{sec4} presents validation results through benchmark problems including the simplified four-roll mill flow (Section~\ref{sec4.1}), force-driven Poiseuille flow (Section~\ref{sec4.2}), and flow past a circular cylinder (Section~\ref{sec4.3}). Section~\ref{sec5} summarizes our findings and discusses future research directions.	
	
	\section{Governing equations}\label{sec2}
	In this work, we consider weakly compressible isothermal viscoelastic fluids described by the Oldroyd-B constitutive model. The weakly compressible assumption allows for small density variations while maintaining computational efficiency, making it particularly suitable for lattice Boltzmann implementations. The viscoelastic behavior arises from the presence of dissolved polymer chains, which contribute an additional stress component to the total stress tensor. This leads to a coupled system involving two distinct physical fields: the hydrodynamic field governing fluid motion and the polymer field characterizing the microstructural evolution of polymer chains. The coupling between these fields captures the essential physics of viscoelastic flows, where polymer deformation influences fluid stresses and, conversely, fluid motion drives polymer chain evolution.
	\subsection{Hydrodynamic Field Equations}
	The hydrodynamic field is governed by the weakly compressible Navier-Stokes equations (NSE) as follows:
	\begin{subequations}\label{eq_NSE}
		\begin{gather}
			\label{eq_ev_fa}
			\partial_t \rho+\partial_\alpha\left(\rho u_\alpha\right)=0,
			\\
			\label{eq_ev_fb}
			\partial_t\left(\rho u_\alpha\right)+\partial_\beta\left(\rho u_\alpha u_\beta+p \delta_{\alpha \beta}\right)=\partial_\beta\left[\mu_s \left(\partial_\alpha u_\beta+\partial_\beta u_\alpha\right)\right]
			+\partial_\beta T_{\alpha \beta}+F_\alpha,
		\end{gather}
	\end{subequations}
	where $\rho$ is the density, $u_\alpha$ is the fluid velocity at time $t$ and spatial location $x_{\alpha}$ with the subscript $\alpha$ being the generic index, $\delta_{\alpha \beta}$ is the Kronecker delta, $p$ is the pressure, $\mu_{s}$ is the dynamic viscosity of the solvent, $T_{\alpha \beta}$ is the viscoelastic stress tensor which describes the effect of the polymers on the solvent, and $F_{\alpha}$ is the body force acting on the fluid. The Einstein summation convention is employed throughout, with Greek indices ranging over spatial dimensions.
	\subsection{Polymer Field Evolution}
	The evolution of the polymer microstructure is described through the conformation tensor $A_{\alpha\beta}$, which represents the second moment of the polymer chain end-to-end vector distribution. The conformation tensor satisfies the transport equation~\cite{alves2021numerical}
	\begin{equation}\label{eq_Constitutive}
		\frac{\partial A_{\alpha\beta}}{\partial t}+\partial_{\gamma}\left(A_{\alpha\beta}u_{\gamma}\right)=\partial_\gamma\left(\kappa \partial_\gamma A_{\alpha\beta}\right)-\frac{1}{\lambda}\mathcal{F}(A_{\alpha\beta})+A_{\alpha\gamma} \partial_{\gamma} u_{\beta} +A_{\gamma\beta}\partial_{\gamma}u_{\alpha}+A_{\alpha\beta}\partial_{\gamma} u_{\gamma},
	\end{equation}
	where $\kappa$ is the centre-of-mass diffusion coefficient, $\lambda$ is the characteristic relaxation time, and $\mathcal{F}(A_{\alpha\beta})$ is a model-dependent function defining the constitutive relationship. The terms on the right-hand side represent, respectively, diffusion, relaxation toward equilibrium, and deformation due to velocity gradients. The final term accounts for weak compressibility effects and vanishes for incompressible flows.
	\subsection{Stress-Conformation Coupling}
	The hydrodynamic and polymer fields are coupled through the Kramers formula
	\begin{equation}\label{eq_kramer}
		\tau_{\alpha\beta}=\frac{\mu_{p}}{\lambda}\mathcal{F}(A_{\alpha\beta}),
	\end{equation}
	where $\mu_{p}$ is the polymer contribution to viscosity. This relationship connects the microscopic polymer configuration to the macroscopic stress field.
	
	In Eqs. (\ref{eq_Constitutive}) and (\ref{eq_kramer}), the functions $\mathcal{F}(A_{\alpha\beta})$ can take different forms in various constitutive models~\cite{alves2021numerical}, such as the Oldroyd-B model~\cite{oldroyd1950formulation}, the Giesekus model~\cite{giesekus1982simple}, the finitely extensible nonlinear elastic (FENE) type models~\cite{warner1972kinetic,bird1980polymer,chilcott1988creeping}, the Phan-Thien-Tanner (PTT) model~\cite{thien1977new}, and the upper-convected Maxwell (UCM) model~\cite{alves2021numerical}, among others. The choice of function $\mathcal{F}(A_{\alpha\beta})$ is not limited in this study, as it appears in our model in the form of a source term. To facilitate the verification of the reliability of the model, this study has opted to consider the most popular and simple Oldroyd-B model with
	\begin{equation}\label{eqb4}
		\mathcal{F}(A_{\alpha\beta})=A_{\alpha\beta}-\delta_{\alpha\beta}.
	\end{equation}
	Thus, the extension of the model presented in this study to various complex viscoelastic fluid models is straightforward.
	\subsection{Dimensionless Parameters}
	The behavior of the Oldroyd-B viscoelastic fluid described in the above model can be characterized by the four dimensionless quantities: the Reynolds number ($\mathrm{Re}$), the Weissenberg number ($\mathrm{Wi}$), the viscosity ratio ($\beta$), and the Schmidt number ($\mathrm{Sc}$). 
	
	The Reynolds number $\mathrm{Re}$ in viscoelastic fluid governs the relative significance of inertial forces versus viscous forces in fluid flow flow and is defined as
	\begin{equation}\label{eqb5}
		\mathrm{Re}=\frac{\rho U_c L_c}{\mu_0},
	\end{equation}
	where $U_c$ is the characteristic velocity, $L_c$ is the characteristic length, $\mu_{0}$ is the total dynamic viscosity with $\mu_{0}=\mu_{s}+\mu_{p}$. 
	
	The Weissenberg number $\mathrm{Wi}$, which describes the relative importance of elastic effects compared to shear effects in viscoelastic fluids, is defined as:
	\begin{equation}\label{eqb6}
		\mathrm{Wi}=\frac{\lambda U_c}{L_c}.
	\end{equation}
	
	The viscosity ratio $\mathrm{\beta}$ describes the ratio of the solvent viscosity to the total viscosity given as
	\begin{equation}\label{eqb7}
		\mathrm{\beta}=\frac{\mu_{s}}{\mu_{0}}.
	\end{equation}
	
	The Schmidt number $\mathrm{Sc}$ characterizes the relative rate of momentum diffusion to solute diffusion within a fluid. It is defined as follows:
	\begin{equation}\label{eqb8}
		\mathrm{Sc}=\frac{\nu_{s}}{\kappa},
	\end{equation}
	where $\nu_{s}=\frac{\mu_{s}}{\rho}$ is kinematic viscosity of solvent. Numerous studies have demonstrated the actual existence of centre-of-mass diffusion. Consequently, assigning a value of 
	$\kappa=0$ and $\mathrm{Sc}=\infty$ is considered to be unrealistic. Here, the diffusion coefficients of several common dilute polymer solutions are presented in \hyperref[table1]{Table~\ref{table1}}. It is evident from the table that, despite the small magnitude of the center-of-mass diffusion coefficients $\kappa$, they cannot be neglected due to the inherently small self-diffusion coefficient of water ($O(10^{-9})$). There is a broad range of variation in the Schmidt number $\mathrm{Sc}$, which can significantly decrease with increasing solute concentration or molecular weight of the polymer. The $\mathrm{Sc}$ values for the various common viscoelastic polymer solutions displayed in the table range between $O(1)$ and $O(10^4)$. It is noteworthy that the diffusion coefficient $\kappa$ decreases sharply with an increasing number of DNA base pairs (bp)~\cite{lukacs2000size}. For instance, a single human DNA molecule, containing tens of millions to several hundred million base pairs, implies the possibility of higher $\mathrm{Sc}$ numbers, such as $\mathrm{Sc}$ greater than or equal to $O(10^6)$. So a comparatively broad range of Schmidt numbers (Sc) will be discussed in this paper.
	\begin{table}[h]
		\centering
		\caption{Diffusion coefficients $\kappa$ of common substances in water and corresponding $\mathrm{Sc}$.}
		\label{table1}
		\begin{tabular}{p{5.5cm}<{\centering}p{3.5cm}<{\centering}p{2.5cm}<{\centering}}
			\toprule
			Diffusion coefficient (in water) & $\kappa~(m^2/s)$ &  $\mathrm{Sc}$	  \\
			\midrule
			Water  ~\cite{agmon1995grotthuss}& $2.3\times 10^{-9}$ & 1		 \\
			Human serum albumin ~\cite{vogel1988life}& $6.1 \times10^{-11}$ & $3.8\times10^{1}$ 		 \\
			30S particle of ribosome ~\cite{dobitchin1983correlation} & $1.8\times^{-11}$ & $1.3\times10^{2}$ 			 \\
			Single mRNA molecules ~\cite{yan2016dynamics}& $1.0\times 10^{-13}$ & $2.3\times10^{4}$ 			 \\
			DNA (21bp) ~\cite{lukacs2000size} & $5.3\times10^{-11}$  & $4.3\times10^{1}$			 \\
			DNA (6000bp) ~\cite{lukacs2000size} &  $8.1\times10^{-13}$ & $2.8\times10^{3}$			 \\
			\bottomrule
		\end{tabular}
	\end{table}
	\subsection{Numerical Parameters}
	In addition to the dimensionless numbers characterizing the physical properties of fluids, this paper also considers three additional dimensionless parameters to ensure the robustness of numerical simulations: the Mach number ($\mathrm{Ma}$), the viscous incompressibility parameter ($\mathcal{T}$), and the Péclet number ($\mathrm{Pe}$). The Mach number $\mathrm{Ma}$ is defined as: 
	\begin{equation} \label{eq_Mach}
		\mathrm{Ma}=\frac{U_c}{c_s}.
	\end{equation}
	In this work, it is crucial to ensure that $Ma\leq 0.2$ to satisfy the condition of weak compressibility of fluids. The definition of the viscous incompressibility parameter $\mathcal{T}$ is~\cite{gsell2021lattice}  
	\begin{equation} \label{eq_T}
		\mathcal{T}=\frac{\mathrm{Ma}^2}{\mathrm{Re}} ,
	\end{equation}
	which is introduced to enhance the robustness of the TRT-RLB model in low Reynolds number problems. For such low-Reynolds-number flows, Gsell et al.~\cite{gsell2021lattice} and our recent results~\cite{yu2025tworelaxationtime} recommend $\mathcal{T} \leq 10^{-2}$, which we adopt for problems without stress singularities such as Poiseuille flow. For problems containing stress singularities that demand high stability requirements, we set $\mathcal{T} = 10^{-4}$. The Péclet number is defined as
	\begin{equation}
		\mathrm{Pe}=\frac{L_c U_c}{\kappa}=\mathrm{Re} \mathrm{Sc}.
	\end{equation}
	It is an effective measure for assessing the robustness of methods used to solve the constitutive equations Eq.~(\ref{eq_Constitutive}).
	
	\section{Numerical methods}\label{sec3}
	
	\subsection{The two-relaxation-time regularized lattice Boltzmann model for the hydrodynamic field}\label{sec3.1}

	The accurate simulation of viscoelastic fluids requires a robust hydrodynamic solver capable of handling intense variations in stress fields, thereby maintaining numerical stability at high Weissenberg numbers. In our recent work~\cite{yu2025tworelaxationtime2}, we developed a two-relaxation-time regularized lattice Boltzmann (TRT-RLB) model for the Navier-Stokes equations that achieves second-order spatial accuracy and demonstrates superior numerical stability compared to standard TRT and RLB models, particularly for high Reynolds number flows. The TRT-RLB model combines the advantages of both TRT and RLB approaches: the TRT scheme provides enhanced stability through dual relaxation parameters, while the regularization procedure eliminates spurious high-frequency modes that can compromise accuracy. Building upon this foundation, we extend the TRT-RLB framework to incorporate viscoelastic effects through a novel local discretization approach that directly integrates the stress tensor without requiring spatial derivatives.
	
	Traditional implementations of viscoelastic effects in LBM involve computing the viscoelastic force $F_{p,\alpha}=\partial_\beta \tau_{\alpha \beta}$ and incorporating it through forcing schemes~\cite{guo2002discrete}. However, this approach introduces errors due to gradient calculations and compromises the local nature of LBM. To address this limitation, we present a completely local discretization of the viscoelastic stress tensor that directly incorporates $\tau_{\alpha \beta}$ into the TRT-RLB framework without requiring spatial derivatives.
	
	In the present model, the mesoscopic evolution equation is given as follows:  
	\begin{align}\label{eq_ev_f}
		f_i (x_\alpha+e_{i \alpha} \Delta t&, t +\Delta t) =
		f_i^{\text{eq}}+ w_i\left(1-\frac{1}{\tau_{s,1}}\right) \left(\mathcal{A}_{0}^{\text{neq}}+ \frac{\mathcal{H}_{i, \alpha}}{c_s^2} \mathcal{A}_\alpha^{\text{neq}}+ \frac{\mathcal{H}_{i, \alpha \beta}}{2 c_s^4} \mathcal{A}_{\alpha \beta}^{\text{neq}}\right)  \notag \\
		& +w_i\left(1-\frac{1}{\tau_{s,2}}\right)  \frac{\mathcal{H}_{i, \alpha \beta \gamma}}{6 c_s^6} \mathcal{A}_{\alpha \beta \gamma}^{\text{neq}}  +G_i \Delta t+F_i \Delta t+ T_i \Delta t,
	\end{align}
	where $f_{i}$ and $f_i^{\text{eq}}$ are the distribution function and the equilibrium distribution function associated with the discrete velocity $e_{i\alpha}$ in the $i$-th direction ranging from $0$ to $(q-1)$ for a given $d$-dimensional D$d$Q$q$ lattice model. The equilibrium of the third-order Hermite expansion to suppress Galilean invariance errors:
	\begin{align}\label{eq2}
		f_i^{\mathrm{eq}}=w_i \rho\left(\mathcal{H}_i+\frac{\mathcal{H}_{i, \alpha} u_\alpha}{c_s^2}+\frac{\mathcal{H}_{i, \alpha \beta} u_\alpha u_\beta}{2 c_s^4}+\frac{\mathcal{H}_{i, \alpha \beta \gamma} u_\alpha u_\beta u_\gamma}{6 c_s^6}\right),
	\end{align}
	where $w_{i}$ are the weight coefficients, $c_s$ is the lattice speed of sound depended on the discrete velocity model, $\mathcal{H}_i$, $\mathcal{H}_{i, \alpha}$, $\mathcal{H}_{i, \alpha \beta}$ and $\mathcal{H}_{i, \alpha \beta \gamma}$ denote Hermite polynomials of various orders. The macroscopic density and velocity are recovered through moment relations:
	\begin{align}\label{eq_ev_f0}
		\sum_i f_i=\rho, \quad \sum_i e_{i \alpha} f_i=\rho u_\alpha-\frac{\Delta t}{2} F_\alpha.
	\end{align}
	The off-equilibrium moments of orders 0 through 3 are defined as
	\begin{subequations}
		\begin{gather}
			\label{eq4a}
			\mathcal{A}_{0}^{\text{neq}}=\sum_i \left(f_i-f_i^{\text{eq}}\right), 
			\\
			\label{eq4b}
			\mathcal{A}_\alpha^{\text{neq}}=\sum_i \mathcal{H}_{i, \alpha}\left(f_i-f_i^{\text{eq}}\right), \\ 
			\label{eq4c}
			\mathcal{A}_{\alpha \beta}^{\text{neq}}=\sum_i \mathcal{H}_{i, \alpha \beta}\left(f_i-f_i^{\text{eq}}\right), \\ 
			\label{eq4d}
			\mathcal{A}_{\alpha \beta \gamma}^{\text{neq}}=\sum_i \mathcal{H}_{i, \alpha \beta \gamma}\left(f_i-f_i^{\text{eq}}\right),
		\end{gather}
	\end{subequations}
	where $\mathcal{A}_{xxx}^{\text{neq}}=\mathcal{A}_{yyy}^{\text{neq}}$ can be employed to reduce computational complexity.
	
	The relaxation times $\tau_{s,1}$ and $\tau_{s,2}$ are related to the solvent viscosity and stability parameter through
	\begin{equation}
		\mu_{s}=\rho c_s^2\left(\tau_{s,1}-\frac{1}{2}\right) \Delta t,
	\end{equation}
	and
	\begin{equation}
		\Lambda_s=\left(\tau_{s,1}-\frac{1}{2}\right)\left(\tau_{s,2}-\frac{1}{2}\right),
	\end{equation}
	where $\Lambda_s=1/6$ is the magic parameter for the hydrodynamic system~\cite{gsell2021lattice}.
	
	The additional discrete source term $G_{i}$ will be responsible for eliminating the cubic defects in Galilean invariance on standard lattice, $T_i$ will contribute to incorporate the viscoelastic effect, and $F_{i}$ will account for any body forces.
	
	Although the extension to three dimensions is straightforward and will be implemented in the future, we only consider the two-dimensional cases for simplicity in this article. Thus we chose the two-dimensional discrete velocity model on standard lattices, namely the D2Q9 model with a Gauss-Hermite quadrature of fifth-order degree ~\cite{nie2008galilean,shan2016mathematical}. In this model, the weight coefficients are given as $w_0=4 / 9, w_{1,2,3.4}=1 / 9, w_{5,6.7 .8}=1 / 36$, the lattice speed of sound is described as $c_s=\frac{1}{\sqrt{3}}c$, where the lattice speed $c=\Delta x/\Delta t$ with $\Delta x$ and $\Delta t$ being the spatial and temporal step respectively, and the discrete velocities are given by 
	\begin{equation}
		e_{i\alpha}=\left\{\begin{array}{lr}
			(0,0) & i=0, \\
			\left(\cos \frac{\pi(i-1)}{2}, \sin \frac{\pi(i-1)}{2}\right) c & \text { for } i=1 \text { to } 4, \\
			\sqrt{2}\left(\cos \frac{\pi(i-9 / 2)}{2}, \sin \frac{\pi(i-9 / 2)}{2}\right) c & \text { for } i=5 \text { to } 8,
		\end{array}\right. 
	\end{equation}
	Following the work of Feng et al.~\cite{feng2019hybrid}, the correction source term is given as
	\begin{equation} \label{eq5}
		G_i=-w_i \left(1-\frac{1}{2\tau_{s,1}}\right)\frac{\mathcal{H}_{i, \alpha \beta}}{6 c_s^6} \partial_\gamma \Phi_{\alpha \beta \gamma},
	\end{equation}
	where
	\begin{gather}
		\Phi_{\alpha \beta \gamma}=\left\{\begin{array}{cc}
			\rho u_x^3 & \alpha=\beta=\gamma=x, \\
			\rho u_y^3 & \alpha=\beta=\gamma=y, \\
			0 & \text { others },\end{array}\right.\\
		\partial_\gamma \Phi_{\alpha \beta \gamma}=\partial_x\left(\rho u_x^3\right) \delta_{\alpha x} \delta_{\beta x}+\partial_y\left(\rho u_y^3\right) \delta_{\alpha y} \delta_{\beta y}.
	\end{gather}
	The newly proposed discrete viscoelastic source term in this article is defined as
	\begin{equation}\label{eq6}
		T_i=-w_i \frac{\mathcal{H}_{i, \alpha \beta}}{2 c_s^4} \frac{\tau_{\alpha \beta}}{\tau_{s,1} \Delta t}
	\end{equation}
	where $\tau_{\alpha \beta}$ is the viscoelastic stress tensor which describes the effect of the polymers on the solvent. Furthermore, a first-order Hermite expansion of discrete forcing term is incorporated to improve the accuracy~\cite{postma2020force,bawazeer2021critical}, which can be expressed as
	\begin{equation}
		F_i =  \left(1-\frac{1}{2 \tau_{s, 1}}\right) w_i \frac{\mathcal{H}_{i, \alpha}}{c_s^2} F_\alpha
	\end{equation}
	
	With the aid of the equation of state $p=\rho c_s^2$, the aforementioned mesoscopic model can be readily recovered to the NSE as shown in Eqs.~(\ref{eq_NSE}) via the Chapman-Enskog analysis technique~\cite{yu2025tworelaxationtime}.
	
	\subsection{The improved two-relaxation-time regularized lattice Boltzmann model for the polymer field}\label{sec3.2}
	
	To model the evolution of the viscoelastic stress tensor $\tau_{\alpha\beta}$, a common approach is to solve the continuum-level differential constitutive equations with respect to the configuration tensor $A_{\alpha\beta}$. Once these equations are solved, the stress tensor $\tau_{\alpha\beta}$ can be calculated from the configuration tensor $A_{\alpha\beta}$ by Eqs.~(\ref{eq_kramer}) and (\ref{eqb4}). It should be noted that we assume the viscoelastic fluid to be isotropic, which means that for 2D problems, the viscoelastic configuration tensor has only three mutually independent components: $A_{xx}$, $A_{yy}$, and $A_{xy}$. Consequently, we must solve the following convection-diffusion equation (CDE) three times concurrently and independently:
	\begin{equation} \label{eq_cde}
		\partial_t \phi+\partial_\alpha\left(\phi u_\alpha\right)=\partial_\alpha\left(\kappa \partial_\alpha \phi\right)+S,
	\end{equation}
	where the variables $\phi$ and $S$ represent the three different components of $A_{\alpha \beta}$ and $\Phi_{\alpha \beta}$, respectively, in practice, with
	\begin{equation}
		\Phi_{\alpha\beta}=-\frac{1}{\lambda}(A_{\alpha\beta}-\delta_{\alpha\beta})+A_{\alpha\gamma} \partial_{\gamma} u_{\beta} +A_{\gamma\beta} \partial_{\gamma}u_{\alpha}+A_{\alpha\beta}\partial_{\gamma} u_{\gamma}.
	\end{equation} 
	
	To solve Eq.~(\ref{eq_cde}), we propose an improved regularized lattice Boltzmann model in which the mesoscopic distribution functions $g_{i}\left(x_{\alpha},t\right)$ satisfy the following evolution process:
	\begin{align} \label{eq_g0}
		g_i\left(x_\alpha+e_{i \alpha} \Delta t, t +\Delta t\right)
		& =g_i^{\text{eq}} +w_i \left(1-\frac{1}{\tau_{p,1}}\right) \left(\mathcal{B}_{0}^{\text{neq}}+\frac{\mathcal{H}_{i, \alpha}}{c_s^2} \mathcal{B}_\alpha^{\text{neq}}\right) \notag \\
		&\quad+ w_i \left(1-\frac{1}{\tau_{p,2}}\right) \frac{\mathcal{H}_{i, \alpha\beta}}{2c_s^4} \mathcal{B}_{\alpha\beta}^{\text{neq}} \notag \\
		& \quad+\widetilde{G}_i \Delta t+\widetilde{F}_i \Delta t +\frac{\Delta t^2}{2} \partial_t \widetilde{F}_i,
	\end{align}
	where the field quantities on the right-hand side of the equation, such as $g_{i}^{\text{eq}}$, $\mathcal{B}_{0}^{\text{neq}}$, $\mathcal{B}_{\alpha}^{\text{neq}}$, $\mathcal{B}_{\alpha\beta}^{\text{neq}}$, $\widetilde{G}_{i}$, and $\widetilde{F}_{i}$, all depend on $(x,t)$, although we have omitted this dependency for simplicity in our notation. We use the D$2$Q$9$ instead of the D$2$Q$5$ lattice model because Li et al.~\cite{li2017lattice} have shown that the D$2$Q$9$ lattice model offers better accuracy in dealing with convection-dominant problems, albeit with increased computational overhead. Therefore, the parameters related to the discrete velocity model, including $e_{i \alpha}$, $w_i$, and $c_s$, remain consistent with the settings described in the previous section. 
	
	A second-order Hermite expansion is used to improve the accuracy of the equilibrium distribution function $g_i^{\text{eq}}$, as shown below:
	\begin{equation} \label{eq_geq}
		g_i^{\text{eq}}=w_i \phi\left(1+\frac{\mathcal{H}_{i, \alpha}}{c_s^2} u_\alpha+\frac{\mathcal{H}_{i, \alpha \beta}}{2 c_s^4} u_\alpha u_\beta\right).
	\end{equation}
	
	The regularization procedure requires the computation of off-equilibrium moments of different orders. The zero-order off-equilibrium moment $\mathcal{B}_{0}^{\text{neq}}$ is derived as
	\begin{equation}
		\mathcal{B}_{0}^{\mathrm{neq}}=\sum_i \left(g_i-g_i^{\mathrm{eq}}\right)=0,
	\end{equation}
	which vanishes due to the zero-order moment conservation relation
	\begin{equation}
		\sum_i g_i=\sum_i g_i^{\mathrm{eq}}=\phi.
	\end{equation}
	Similarly, the first-order and second-order off-equilibrium moments are defined as
	\begin{subequations}\label{eq_p_neq_m}
		\begin{align}
			\mathcal{B}_\alpha^{\mathrm{neq}} &= \sum_i \mathcal{H}_{i, \alpha}\left(g_i-g_i^{\mathrm{eq}}\right), \label{eq_p_neq_m_a}\\
			\mathcal{B}_{\alpha \beta}^{\mathrm{neq}} &= \sum_i \mathcal{H}_{i, \alpha \beta}\left(g_i-g_i^{\mathrm{eq}}\right). \label{eq_p_neq_m_b}
		\end{align}
	\end{subequations}
	
	In contrast to the approach taken by Malaspinas et al.~\cite{malaspinas2010lattice}, a higher-order discrete source term is employed to enhance the accuracy. Additionally, we propose a new correction term $\widetilde{G}_i$ that considers the NSE-CDE coupling. Unlike the formulation proposed by Shi et al.~\cite{shi2008new}, the correction term presented here circumvents the calculation of velocity gradients $\partial_{\alpha} u_{\alpha}$ and instead involves calculations of nearly negligible density gradients $\partial_{\alpha} \rho$, which is particularly applicable to weakly compressible fluids. The auxiliary term $\widetilde{G}_i$ is defined as
	\begin{equation}
		\label{eq_p_G}
		\widetilde{G}_i=w_i \left(1-\frac{1}{2 \tau_{p,1}}\right)\frac{e_{i \alpha}}{c_s^2} \frac{\phi}{\rho}\left(F_\alpha-c_s^2 \partial_\alpha \rho\right).
	\end{equation}
	For the completely incompressible case, this reduces to
	\begin{equation}
		\label{eq_p_G_1}
		\widetilde{G}_i=w_i \left(1-\frac{1}{2 \tau_{p,1}}\right)\frac{e_{i \alpha}}{c_s^2} \frac{\phi}{\rho}F_\alpha,
	\end{equation}
	where $F_\alpha$ is the force term in the NSE as introduced in the previous subsection.
	
	The force term $\widetilde{F}_i$ is given by
	\begin{equation}
		\label{eq_p_F}
		\widetilde{F}_i=w_i S+ \left(1-\frac{1}{2 \tau_{p,1}}\right) w_i \frac{e_{i \alpha}}{c_s^2} u_\alpha S.
	\end{equation}
	
	Through Chapman-Enskog multiscale expansion analysis, it can be proven that the above model correctly recovers Eq.~(\ref{eq_cde}) and exhibits the following transport coefficient relationship:
	\begin{equation}
		\kappa=\left(\tau_{p,1}-\frac{1}{2}\right) \Delta t c_s^2.
	\end{equation}
	
	The second relaxation time $\tau_{p,2}$ is determined by the magic parameter $\Lambda_p$~\cite{yu2025tworelaxationtime2,ginzburg2012truncation}:
	\begin{equation}
		\tau_{p,2}=\frac{\Lambda_p}{\tau_{p,1}-0.5}+0.5,
	\end{equation}
	where the optimal value of $\Lambda_p$ depends on the convection-diffusion scaling and will be specified for each test case. Through multiscale analysis, it has been shown that $\tau_{p,2}$ does not affect the macroscopic equation recovery, making $\Lambda_p$ a free parameter that can be optimized for numerical stability and accuracy. The determination of appropriate $\Lambda_p$ values has been subject to preliminary numerical analysis in our recent work~\cite{yu2025tworelaxationtime2}, and its selection will be detailed in the validation section for different flow configurations.
	
	The temporal derivative $\partial_t \widetilde{F}_i$ in Eq.~(\ref{eq_g0}) is calculated using the following finite difference scheme:
	\begin{equation}
		\partial_t \widetilde{F}_i\left(x_{\alpha},t\right)=\frac{\widetilde{F}_i\left(x_{\alpha},t\right)-\widetilde{F}_i\left(x_{\alpha}, t-\Delta t\right)}{\Delta t}.
	\end{equation}
	
	This model adopts the standard D2Q9 lattice, and the macroscopic quantity $\phi$ satisfies the following relationship:
	\begin{equation}
		\sum_i g_i=\sum_i g_i^{\mathrm{eq}} = \phi.
	\end{equation}
	
	\subsection{Conservative boundary scheme for the convection-diffusion system governing constitutive equation evolution}\label{sec3.3}
	
	This section presents a novel boundary condition scheme for the convection-diffusion lattice Boltzmann method that ensures strict conservation of the conformation tensor at solid boundaries. As will be demonstrated in subsequent analyses, this method proves essential for achieving numerical stability in wall-bounded viscoelastic flows at high Weissenberg numbers.
	
	In viscoelastic flow simulations, the conformation tensor exhibits sharp gradients near solid boundaries that undergo rapid temporal evolution. Existing boundary treatment methods typically employ zeroth- or first-order extrapolation schemes when imposing boundary conditions for the conformation tensor (or equivalently, the polymeric stress tensor). These approximations inevitably introduce numerical errors that manifest as non-physical "leakage" of the conformation tensor components across the boundary. This spurious loss of conformation tensor conservation is analogous to the well-documented mass conservation errors encountered in discrete kinetic methods when implementing no-slip boundary conditions. Such conservation violations become particularly problematic at high Weissenberg numbers, where they can trigger numerical instabilities and compromise solution accuracy.
	
	To address this fundamental challenge, we develop a boundary treatment method that maintains strict conservation of all conformation tensor components while simultaneously enforcing the no-slip condition without numerical artifacts. Drawing inspiration from the mass-conserving boundary schemes developed by Yu et al.~\cite{yu2020modified} for multiphase lattice Boltzmann methods, we formulate a conservative non-equilibrium bounce-back (CNEBB) method tailored specifically for the convection-diffusion equation governing conformation tensor evolution.
	
	The proposed method operates as follows. When implementing the non-equilibrium bounce-back procedure to reconstruct the unknown post-streaming distribution functions, we lack access to the exact macroscopic quantities at time \( t+\Delta t \). To address this challenge, following the approach pioneered by Yu et al.~\cite{yu2020modified}, we compute the conformation tensor component at the subsequent time step using:
	
	\begin{equation}
		\phi(\mathbf{x}_b,t+\Delta t) = \sum_{\gamma \in \Gamma}g_{\gamma}(\mathbf{x}_b,t+\Delta t)+\sum_{\eta \in H}g_{\eta}^{\dagger}(\mathbf{x}_b, t)
	\end{equation}
	where $\mathbf{x}_b$ denotes the outermost fluid nodes adjacent to the solid boundary, $\Gamma$ represents the set of lattice directions with known post-streaming distribution functions, $H$ denotes the set of velocity directions pointing toward the solid boundary at each boundary node $\mathbf{x}_b$, and $g_{\eta}^{\dagger}$ denotes the post-collision distribution functions prior to streaming.
	
	With the conserved quantity $\phi$ thus determined, the equilibrium distribution function is subsequently computed using Eq.~(\ref{eq_geq}). The unknown distribution functions are then reconstructed by utilizing the non-equilibrium information from the opposite lattice directions:
	
	\begin{equation}
		g_i = g_i^{eq} + g_{\bar{i}}^{neq} = g_i^{eq} + (g_{\bar{i}} - g_{\bar{i}}^{eq})
	\end{equation}
	where $i \in H$ and $\bar{i}$ denotes the lattice direction opposite to $i$. This reconstruction assumes that the non-equilibrium components exhibit anti-symmetric behavior at the boundary, consistent with the physical bounce-back mechanism.
	
	Finally, to ensure exact conservation of \( \phi \), the rest distribution function is updated according to:
	
	\begin{equation}
		g_0 = \phi(\mathbf{x}_b,t+\Delta t) - \sum_{i \neq 0} g_i(\mathbf{x}_b,t+\Delta t)
	\end{equation}
	
	The proposed conservative NEBB method constitutes a significant enhancement of the classical non-equilibrium bounce-back approach. Consequently, it preserves the versatility of the classical method, demonstrating equal efficacy for straight wall boundaries and staircase-approximated curved boundaries while offering seamless extension to three-dimensional configurations. Moreover, the method's exclusive reliance on post-collision distribution functions, without direct involvement in the collision process itself, ensures broad compatibility across diverse collision models and lattice Boltzmann formulations, including those incorporating external force terms. This generality, combined with its conservation properties, positions the CNEBB method as a robust boundary treatment scheme for high-fidelity viscoelastic flow simulations.

	\section{Results and discussions}\label{sec4}
	\subsection{The simplified four-roll mill problem}\label{sec4.1}
\begin{figure}[htpb]
		\centering
		\includegraphics[width=0.4\textwidth]{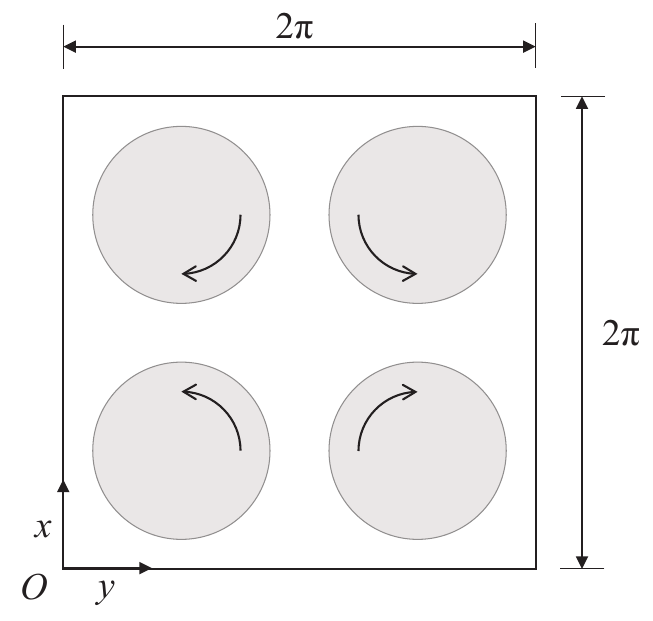}
		\caption{Schematic diagram of the four-roll mill problem. The computational domain is within a space of $[0, 2\pi]^2$.}
		\label{fig_SFRM}
	\end{figure}

	The simplified four-roll mill problem without initial disturbances is a classic benchmark problem for validating the accuracy of numerical models in simulating steady-state Oldroyd-B fluid flows~\cite{thomases2007emergence,malaspinas2010lattice,thomases2011analysis,osmanlic2016lattice,gutierrez2019proper,sedaghat2021hybrid,dzanic2022hybrid,zhang2025latticea}. As demonstrated by Renardy et al.~\cite{renardy2006comment}, this configuration exhibits a viscoelastic stress singularity that is fundamentally different from boundary layer singularities. This singularity emerges even in the absence of geometric discontinuities in the flow domain, which explains why simulating this problem at high Weissenberg numbers remains particularly challenging. The rotation of four cylinders, illustrated in Figure~\ref{fig_SFRM}, induces a stretching flow along both horizontal and vertical symmetry axes while establishing a stagnation point at the domain center.
	
	To establish a consistent unit system, we adopt the convention that physical quantities marked with a tilde (\textasciitilde) are expressed in physical units, while all other quantities are given in lattice units~\cite{kruger2017lattice} unless otherwise specified. Rather than explicitly modeling four rotating cylinders, the fluid motion is driven by an equivalent external force $F_{\alpha}$, expressed as
	\begin{equation} \label{eq-force}
		\left(F_x,F_y\right)=F_0 \left(\sin{\tilde{x}}\cos{\tilde{y}},-\cos{\tilde{x}}\sin{\tilde{y}}\right),
	\end{equation}
	where $\tilde{x},\tilde{y}\in \left[0,\tilde{L}_d\right]$ represent the spatial coordinates in a square domain with side length $\tilde{L}_d=2\pi$. The corresponding lattice domain side length is $L_d=N \Delta x$, where $N$ denotes the number of grid points in each direction and $\Delta x$ is the lattice grid spacing. The characteristic length is defined as $\tilde{L}_c=1$ in physical units, or equivalently $L_c=L_d\tilde{L}_c/\tilde{L}_d$ in lattice units. The characteristic time is defined by $T_c=L_c/U_c$. The force amplitude in Eq.~\eqref{eq-force} is then given by
	$$F_0=2\mu_{s}U_c\left(\frac{2\pi}{N}\right)^2.$$
	This forcing generates a steady-state velocity field in Newtonian fluids: 
	\begin{equation}\label{eq-steady}
		\left(u_x,u_y\right)=U_c\left(\sin{\tilde{x}}\cos{\tilde{y}},-\cos{\tilde{x}}\sin{\tilde{y}}\right).
	\end{equation}

	When implementing the wet-node bi-periodic boundary scheme~\cite{kruger2017lattice},the spatial coordinates are discretized as $\tilde{x}=\frac{2\pi}{N}x_i$ and $\tilde{y}=\frac{2\pi}{N}y_i$, where $x_i=i-1$ and $y_j=j-1$ represent the lattice coordinates of the $i$-th and $j$-th nodes in the horizontal and vertical directions, respectively, with $i,j=1,2,...,N+1$. The simulation procedure begins by applying the continuous force field, as given by Eq.~(\ref{eq-force}), to a Newtonian fluid. Once the flow field reaches the steady-state solution as Eq.~(\ref{eq-steady}) according to the convergence criterion defined in Eq.~(\ref{eq_err}), viscoelastic effects are introduced. The convergence criterion is specified as
	\begin{equation}\label{eq_err}
		\max(\left|\frac{u_\alpha\left(x_{\alpha},t_n+T_c\right)-u_\alpha\left(x_{\alpha},t_n\right)}{U_c}\right|)<10^{-8}.
	\end{equation}
	where $t_n=n \Delta t$ with $n$ denoting the number of iterations.
	
	For the Oldroyd-B fluids with viscosity ratio $\beta=2/3$ in the Stokes limit, Thomases and Shelley~\cite{thomases2007emergence} demonstrated analytically that when the flow reaches steady state, the components of the conformation tensor can be approximated as
	\begin{subequations}
		\label{eq_Aab_ana}
		\begin{align}
			\label{eq_Aab_ana1}
			A_{x x}(\pi, \tilde{y})&=\frac{1}{1-2 Wi^{\mathrm{eff}}}+C|\tilde{y}-\pi|^{\frac{\left(1-2 Wi^{\mathrm{eff}}\right)}{Wi^{\mathrm{eff}}}},\\
			\label{eq_Aab_ana2}
			A_{y y}(\pi, \tilde{y})&=\frac{1}{1+2 Wi^{\mathrm{eff}}}+|\tilde{y}-\pi|^{\frac{\left(1+2 Wi^{\mathrm{eff}}\right)}{Wi^{\mathrm{eff}}}},\\
			\label{eq_Aab_ana3}
			A_{x y}(\pi, \tilde{y})&=0.
		\end{align}
	\end{subequations}
	where $C$ is a constant, and $\mathrm{Wi}^{\mathrm{eff}}$ is the effective Weissenberg number at the stagnation point (domain center), defined by
	\begin{equation} \label{eq_Wieff}
		Wi^{\mathrm{eff}}=\dot{\varepsilon} Wi,
	\end{equation}
	where $\dot{\varepsilon}$ is the dimensionless local elongational rate given by
	\begin{equation}
		\dot{\varepsilon}=\frac{L_c}{U_c}\partial_{x}u_{x}(\pi,\pi)=-\frac{L_c}{U_c}\partial_{y}u_{y}(\pi,\pi).
	\end{equation}
	To ensure the accuracy of $\dot{\varepsilon}$ in practical numerical computations, an averaged form is adopted as follows:
	\begin{equation} \label{eq_LER}
		\dot{\varepsilon}_\mathrm{num}=\frac{L_c}{U_c}\frac{\partial_{x}u_{x}(\pi,\pi)-\partial_{y}u_{y}(\pi,\pi)}{2}.
	\end{equation}
	
	In our simulations, the Reynolds number is set to 1 to ensure low Reynolds number conditions while maintaining reasonable computational efficiency. Unless otherwise specified, we employ the following default parameters: grid resolution $N=256$	grid points in each direction, providing sufficient accuracy for capturing the stress singularities while keeping computational costs manageable; viscosity ratio $\beta=2/3$	to match the analytical solution derived by Thomases and Shelley~\cite{thomases2007emergence}; Schmidt number $\mathrm{Sc}=10^5$ representing typical polymer solution conditions while minimizing artificial diffusion effects; Mach number $Ma=0.01$	to satisfy the weak compressibility constraint; magic parameters $\Lambda_s=1/4$ for the hydrodynamic field and $\Lambda_p=1.0\times10^{-6}$ for the polymer field, where the small value of $\Lambda_p$ enhances stability for the convection-dominated constitutive equations at high Weissenberg numbers. These default values may be adjusted for specific test cases as detailed in the respective subsections, particularly when investigating grid convergence effects or exploring different flow regimes.
	
	\subsubsection{Grid convergence tests}
	\begin{figure}[htpb] 
		\centering 
		\includegraphics[width=0.4\textwidth]{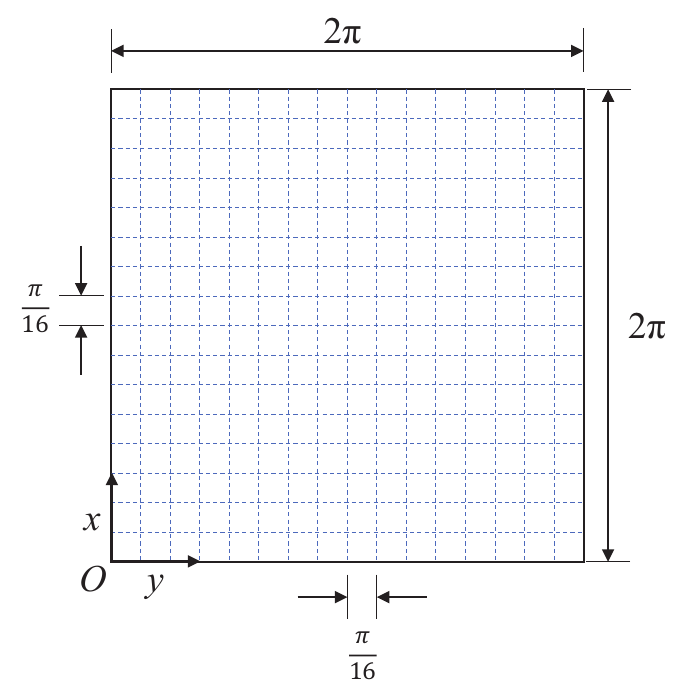} 
		\caption{Distribution of monitoring points for grid convergence analysis. The points are uniformly spaced at intervals of \( \pi/16 \) throughout the computational domain.} 
		\label{fig1_SFRM2}
	\end{figure}
	
	We first assessed the grid convergence characteristics of the proposed model using seven different mesh resolutions: \( N = 32, 64, 128, 256, 512, 1024 \), and \( 2048 \). After reaching steady state, we recorded the velocity components $u_\alpha$ and conformation tensor components $A_{\alpha \beta}$ at the monitoring points illustrated in Figure~\ref{fig1_SFRM2}. Our preliminary tests revealed that increasing the number of monitoring points significantly improves the reliability of convergence rate estimates. Therefore, we selected \( m = (32+1)^2 = 1089 \) monitoring points \( \boldsymbol{x}_k \) (where \( k = 1, ..., m \)), uniformly distributed with a spacing of \( \pi/16 \). This configuration provides sufficient spatial sampling while enabling consistent comparisons across all grid resolutions, including the coarsest mesh (\( N = 32 \)).
	
	Using the finest grid (\( N = 2048 \)) as the reference solution, we computed the relative errors for each mesh resolution according to:
	
	\begin{equation} 
		E_{u_\alpha} = \sqrt{\frac{1}{m}\sum_{k=1}^{m}{\left(\frac{u_{\alpha}(\boldsymbol{x}_{k}) - u_{\alpha,\mathrm{ref}}(\boldsymbol{x}_{k})}{U_c}\right)^2}}
	\end{equation}
	
	\begin{equation} 
		E_{A_{\alpha\beta}} = \sqrt{\frac{1}{m}\sum_{k=1}^{m}{(A_{\alpha\beta}(\boldsymbol{x}_{k}) - A_{\alpha\beta,\mathrm{ref}}(\boldsymbol{x}_{k}))^2}}
	\end{equation}
	
	\begin{figure}[htpb]
		\centering
		\begin{subfigure}[b]{0.48\textwidth}
			\includegraphics[width=\textwidth]{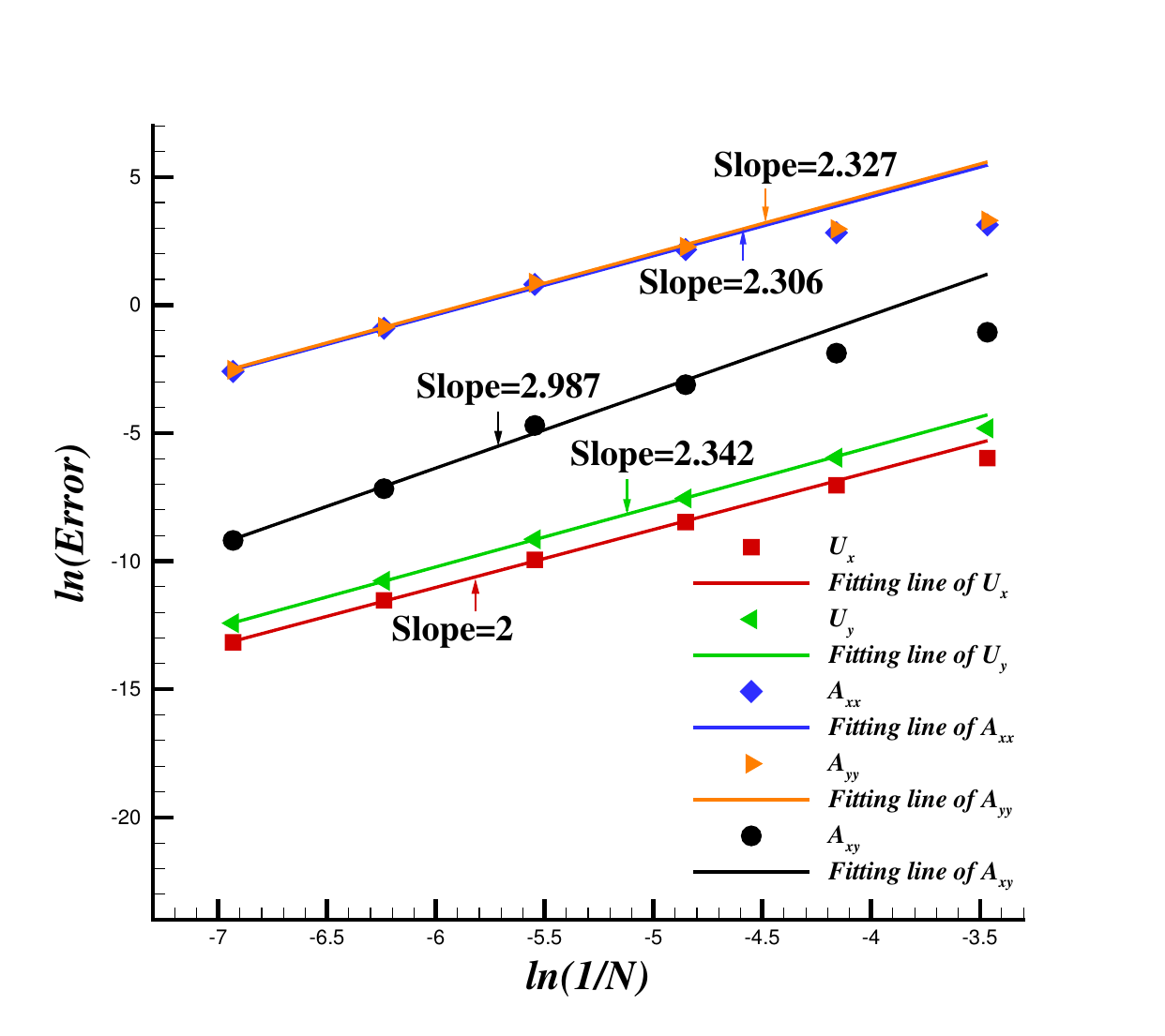}
			\caption{Wi=5}
			\label{fig2:sub1}
		\end{subfigure}
		\hfill
		\begin{subfigure}[b]{0.48\textwidth}
			\includegraphics[width=\textwidth]{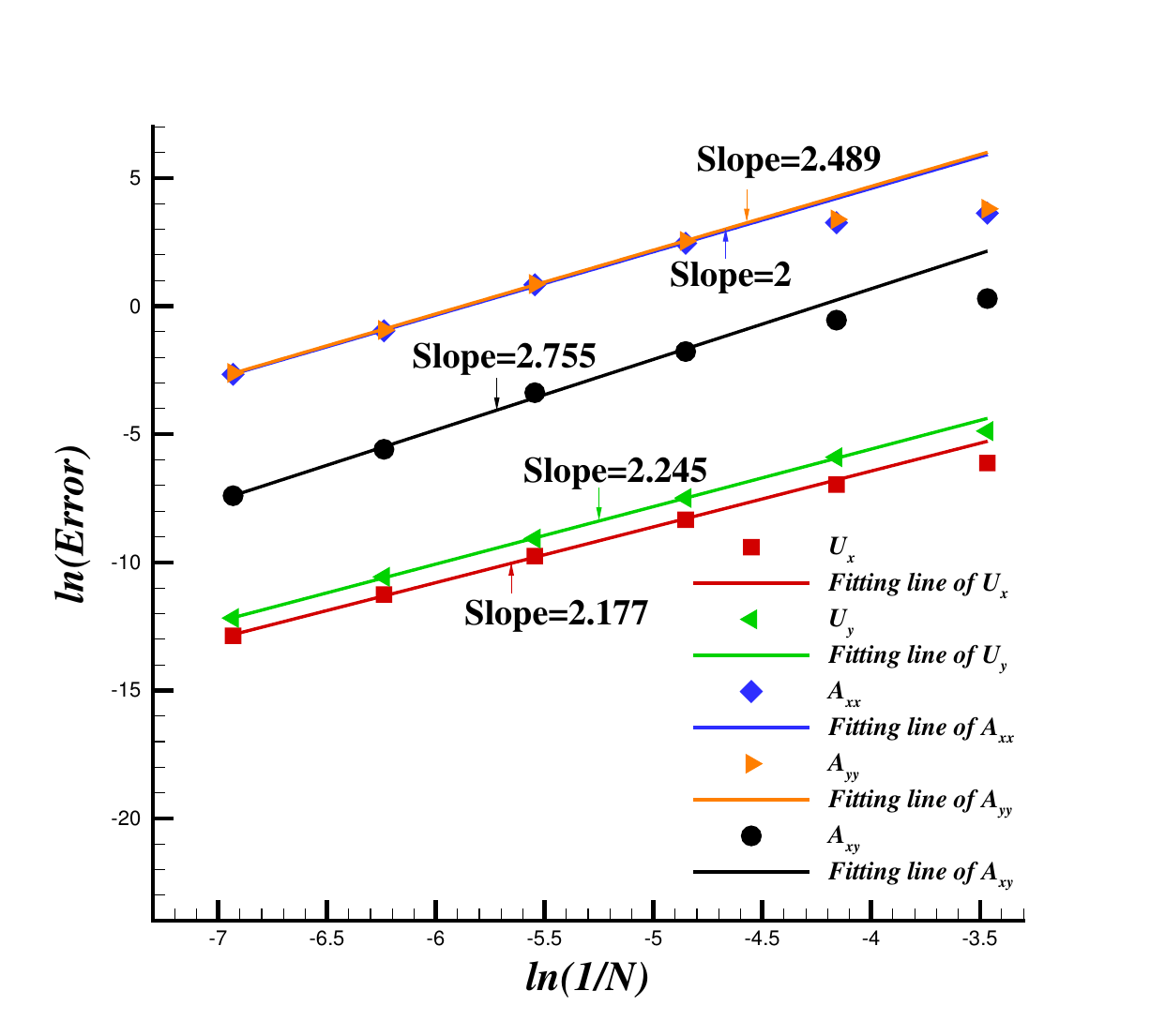}
			\caption{Wi=10}
			\label{fig2:sub2}
		\end{subfigure}
		\caption{Grid convergence analysis showing relative errors versus mesh resolution on log-log scale. (a) $\mathrm{Wi}=5$; (b) $\mathrm{Wi}=10$. The solid lines represent linear regression fits with indicated slopes.}
		\label{fig2_Wi}
	\end{figure}

	Figure~\ref{fig2_Wi} presents the relationship between relative error and grid resolution. We performed linear regression analysis on the data from $N = 128, 256, 512, 1024$, and $2048$, excluding the coarser meshes ($N = 32$ and $64$) which exhibited pre-asymptotic behavior. The regression analysis reveals convergence rates exceeding 2.17 for $u_\alpha$, at least 2.30 for $A_{xx}$ and $A_{yy}$, and above 2.75 for $A_{xy}$. The coefficient of determination ($R^2$) for these regressions exceeds 0.9991, 0.9974, and 0.9938, respectively, indicating excellent goodness of fit and confirming the reliability of the computed convergence rates.
	
	While conventional numerical methods, including standard lattice Boltzmann schemes, typically achieve second-order accuracy, our coupled method demonstrates super-convergence behavior for all macroscopic quantities examined. The convergence rate for \( A_{xy} \) reaches 2.75, which is particularly noteworthy. We attribute this enhanced convergence to the strong coupling strategy employed in our model, which manifests in two key aspects. First, the hydrodynamic force term in Eq.~\eqref{eq_p_G_1} strengthens the coupling between the Navier-Stokes and constitutive equations, thereby avoiding discretization errors inherent in traditional auxiliary source term methods that require explicit spatial derivative calculations. Second, the direct incorporation of viscoelastic stress without spatial discretization, as shown in Eq.~\eqref{eq6}, not only improves numerical accuracy but also enhances the overall convergence order of the method. These features collectively contribute to the superior accuracy and robust numerical stability observed in our simulations.
	
	\subsubsection{Validation}
\begin{figure}[!ht]
		\centering
		
		\begin{subfigure}{0.32\textwidth}
			\includegraphics[width=\linewidth]{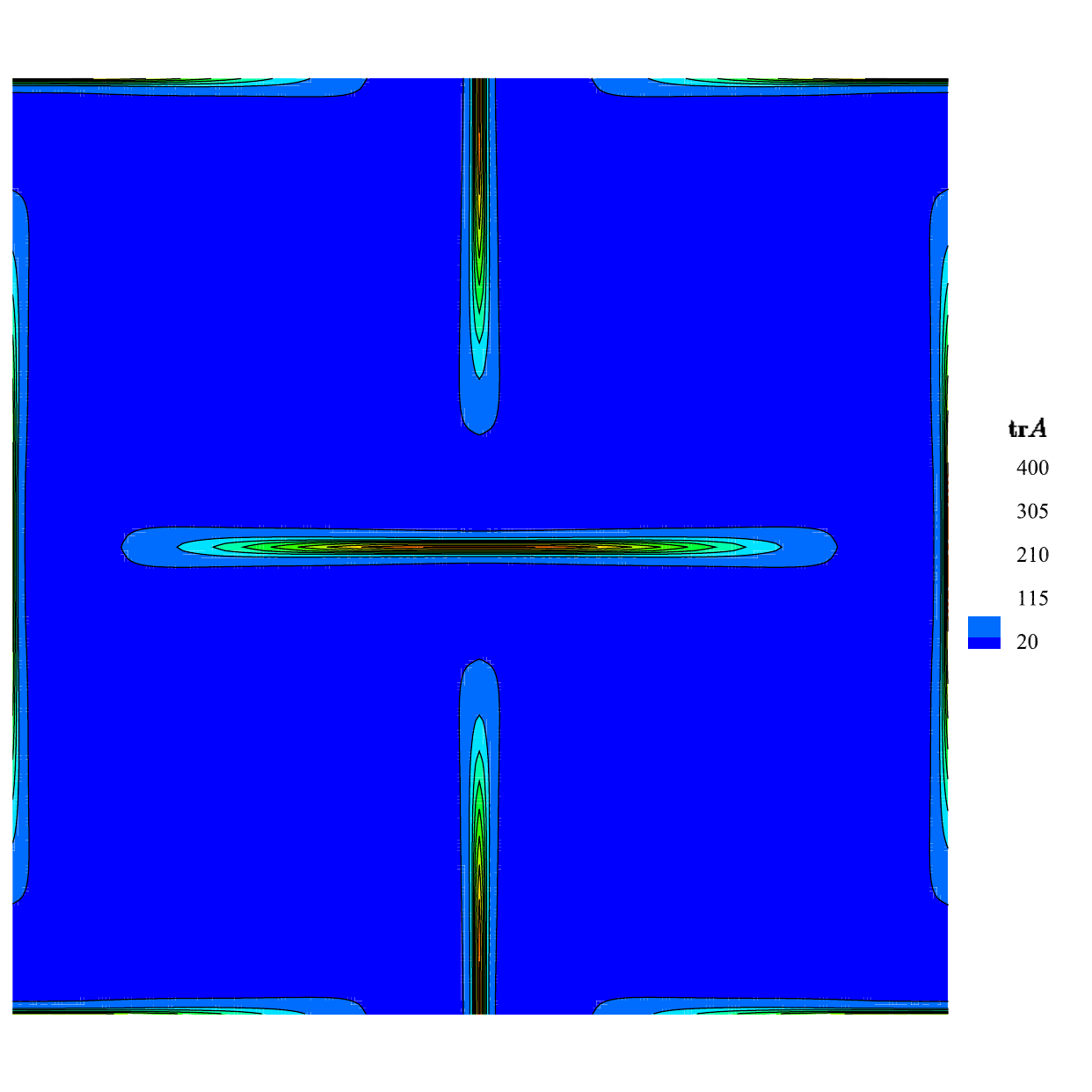}
			\caption{}
			\label{fig_SFRM_VAL_1a}
		\end{subfigure}
		\hfill
		\begin{subfigure}{0.32\textwidth}
			\includegraphics[width=\linewidth]{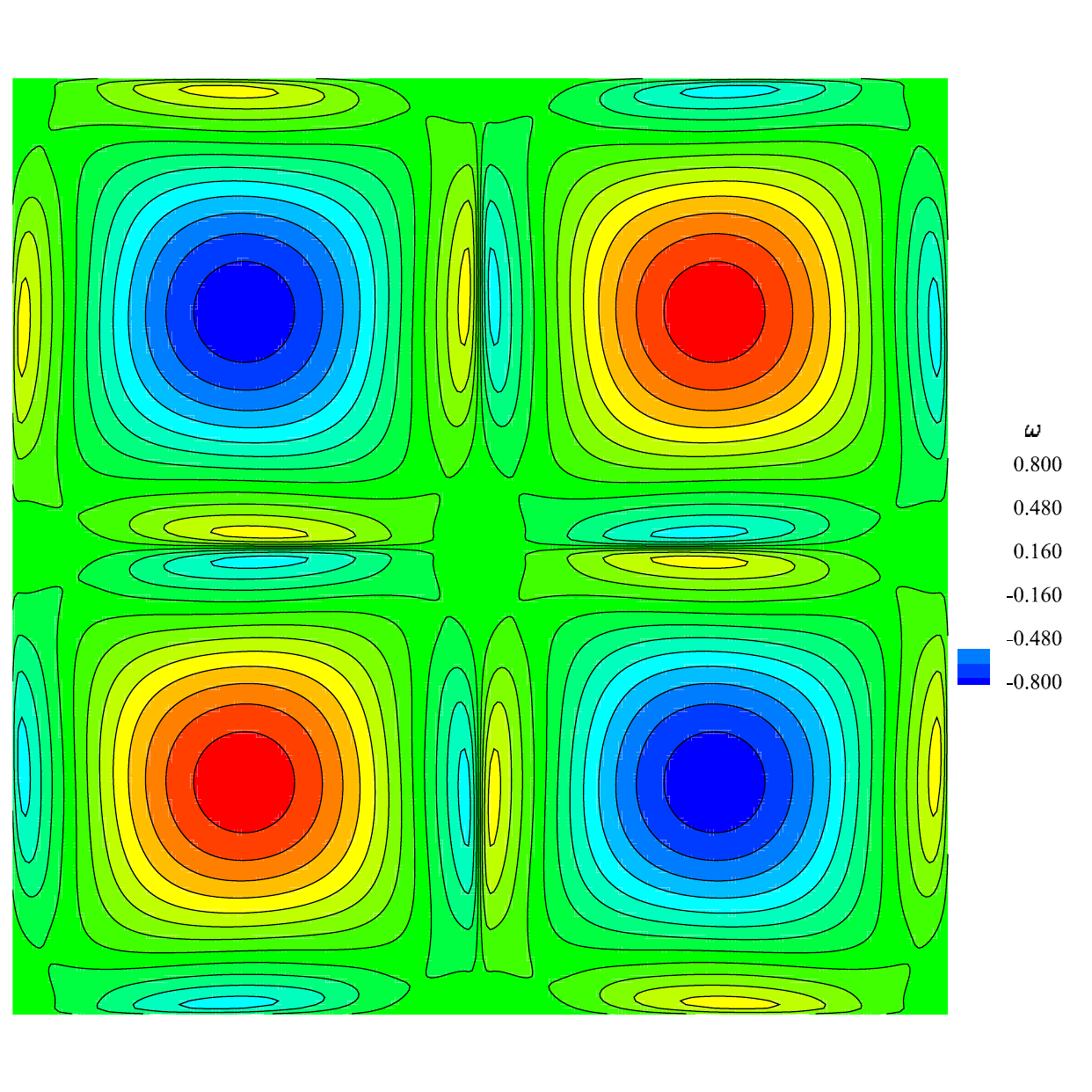}
			\caption{}
			\label{fig_SFRM_VAL_1b}
		\end{subfigure}
		\hfill
		\begin{subfigure}{0.32\textwidth}
			\includegraphics[width=\linewidth]{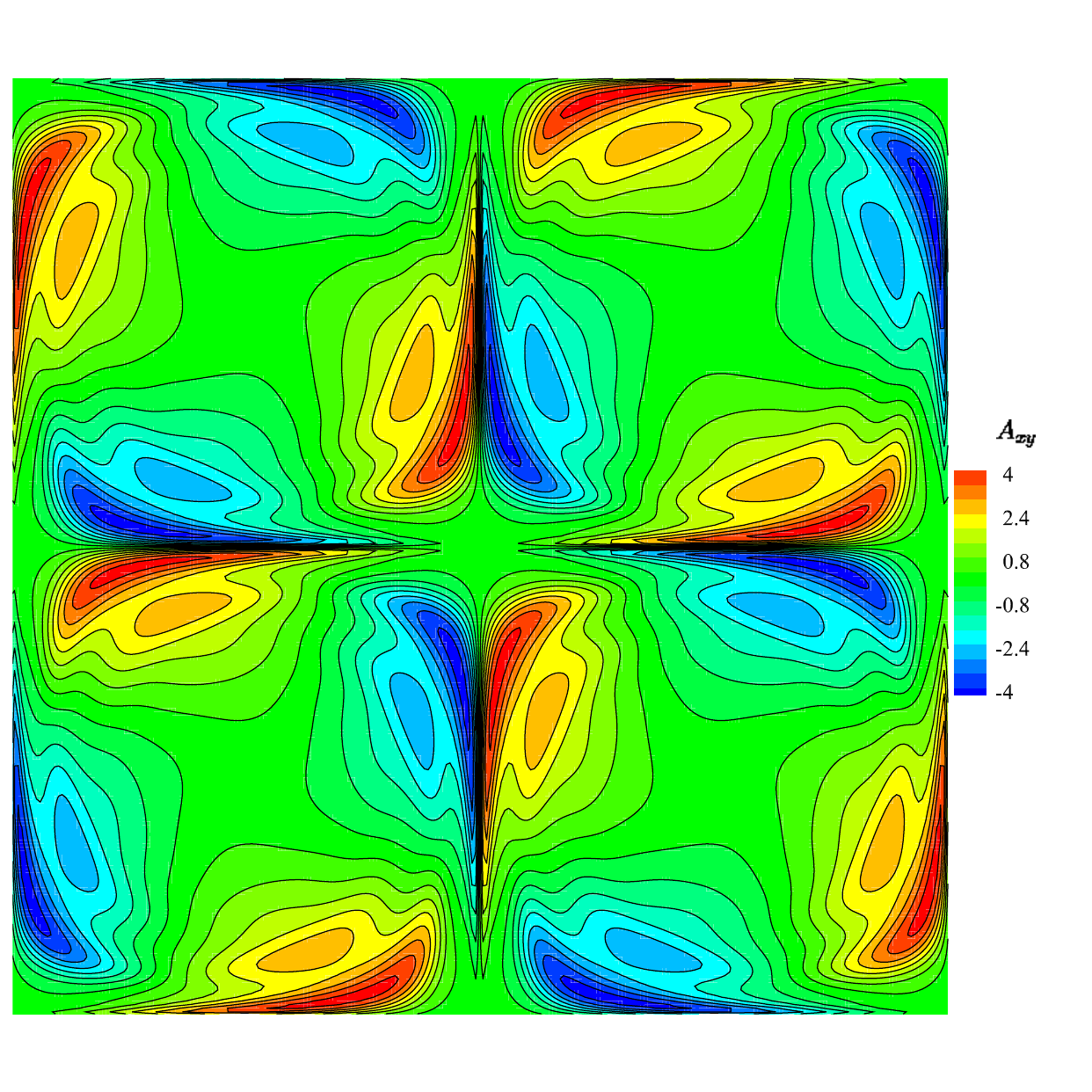}
			\caption{}
			\label{fig_SFRM_VAL_1c}
		\end{subfigure}

		\begin{subfigure}{0.32\textwidth}
			\includegraphics[width=\linewidth]{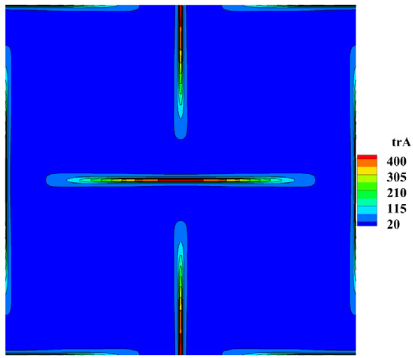}
			\caption{}
			\label{fig_SFRM_VAL_1d}
		\end{subfigure}
		\hfill
		\begin{subfigure}{0.32\textwidth}
			\includegraphics[width=\linewidth]{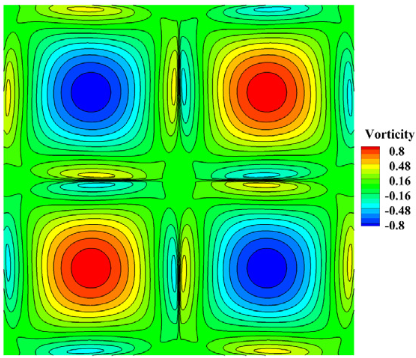}
			\caption{}
			\label{fig_SFRM_VAL_1e}
		\end{subfigure}
		\hfill
		\begin{subfigure}{0.32\textwidth}
			\includegraphics[width=\linewidth]{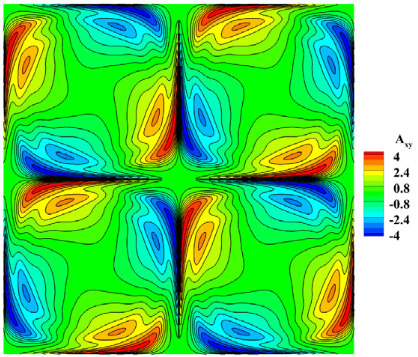}
			\caption{}
			\label{fig_SFRM_VAL_1f}
		\end{subfigure}
		\caption{Comparison of distribution contours for $\mathrm{tr}(\mathbf{A})$, $A_{xy}$, and $\omega$ at $Wi=5$ and $t^*=10$: (a-c) results from the present model; (d-f) results from Zhang et al.~\cite{zhang2025latticea}.}
		\label{fig_SFRM_VAL_1}
	\end{figure}

	To validate the accuracy of the proposed method, a series of assessments were conducted on the simulation results. Firstly, the spatial distributions of the trace of the conformation tensor $\mathrm{tr}(\mathbf{A})$, the off-diagonal component of the conformation tensor $A_{xy}$, and the vorticity magnitude $\omega$ at dimensionless time $t^*=10$ under $Wi=5.0$ were plotted, and these are presented in Figure~\ref{fig_SFRM_VAL_1}. (Note that quantities marked with an asterisk (*) throughout this work denote dimensionless variables.) The results demonstrate excellent agreement between the current results and those obtained by Zhang et al.~\cite{zhang2025latticea}. 
	
	Previous studies by Thomases et al.~\cite{thomases2007emergence}, Osmanlic et al.~\cite{osmanlic2016lattice}, and Zhang et al.~\cite{zhang2025latticea} have investigated not only the case of $Wi=5$ but also lower Weissenberg numbers such as $Wi=0.3$ and $Wi=0.6$. Since the low Weissenberg number cases closely resemble Newtonian fluid behavior and pose minimal methodic challenges, only the $Wi=5$ case is presented here for brevity.
	
	A comprehensive quantitative analysis was subsequently performed to rigorously validate the accuracy of the proposed method. Simulations were conducted for Weissenberg numbers ranging from $Wi = 0.1$ to $5.0$ (fourteen values in total) while maintaining the parameter settings described previously. For each simulation, the conformation tensor component $A_{xx}$ was extracted along the vertical centerline at the dimensionless time $t^* = 6$. 
	
	These numerical results were systematically compared against the benchmark solutions of the Stokes-Oldroyd-B equation system obtained by Thomases and Shelley~\cite{thomases2007emergence} using a pseudospectral method. The comparison revealed excellent agreement between the present lattice Boltzmann results and the reference pseudospectral solutions across the entire range of Weissenberg numbers investigated. This strong correspondence validates both the accuracy and reliability of the proposed method for simulating viscoelastic flows over a broad parameter space.
	
	\begin{figure}[!ht]
		\centering
		\begin{subfigure}{0.32\textwidth}
			\centering
			\includegraphics[width=\textwidth]{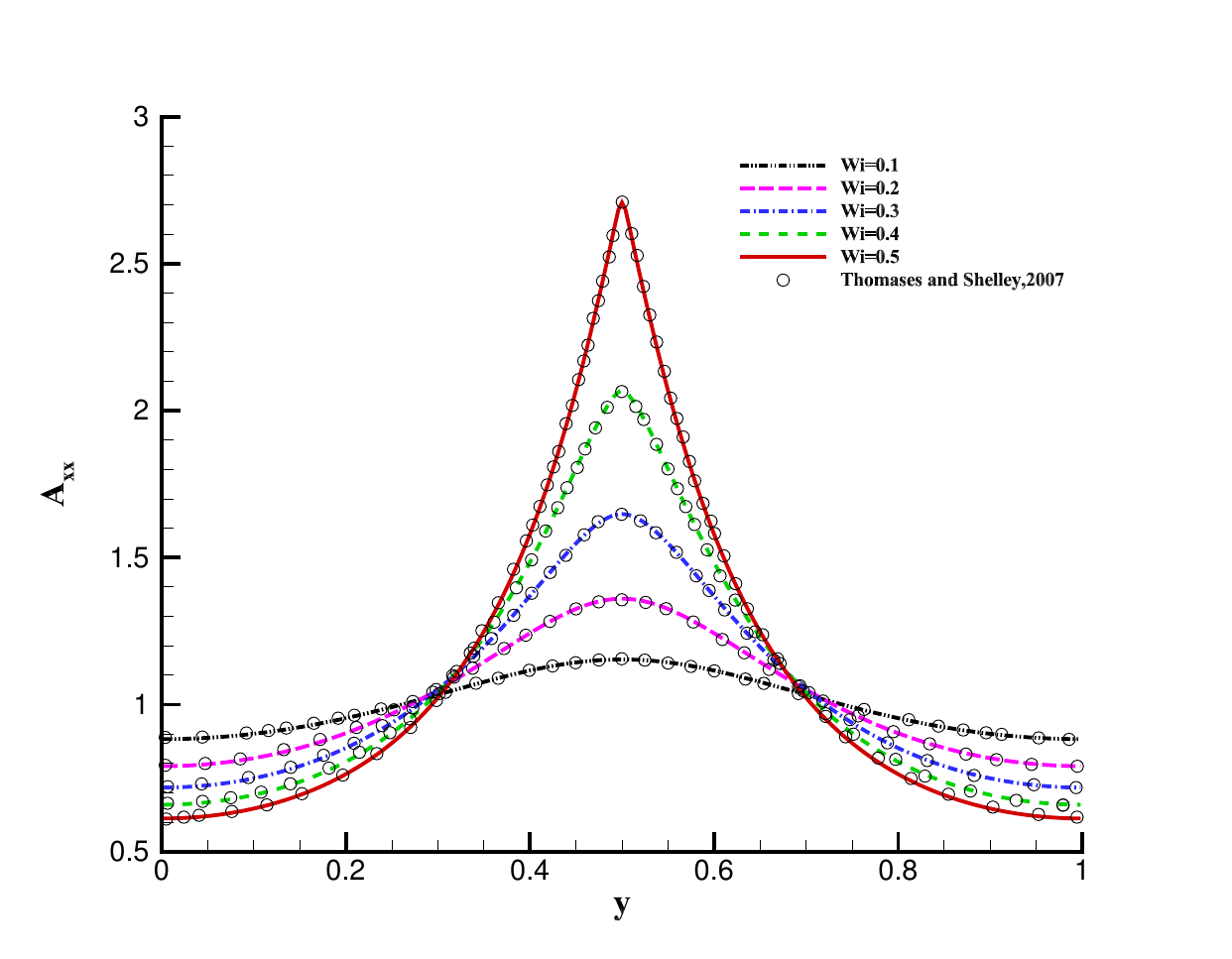}
			\caption{}
			\label{fig:subfig_j}
		\end{subfigure}
		\hfill
		\begin{subfigure}{0.32\textwidth}
			\centering
			\includegraphics[width=\textwidth]{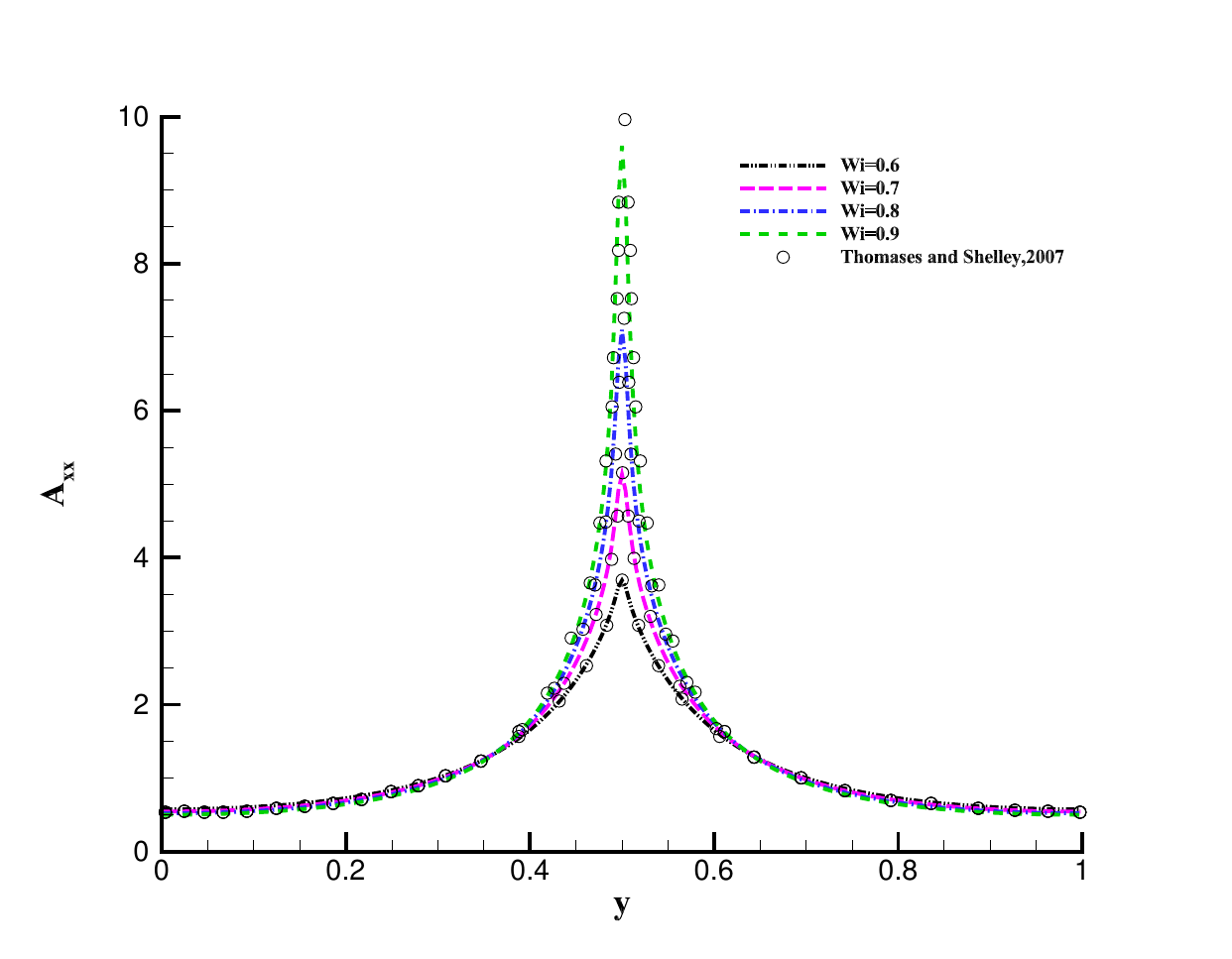}
			\caption{}
			\label{fig:subfig_k}
		\end{subfigure}
		\hfill
		\begin{subfigure}{0.32\textwidth}
			\centering
			\includegraphics[width=\textwidth]{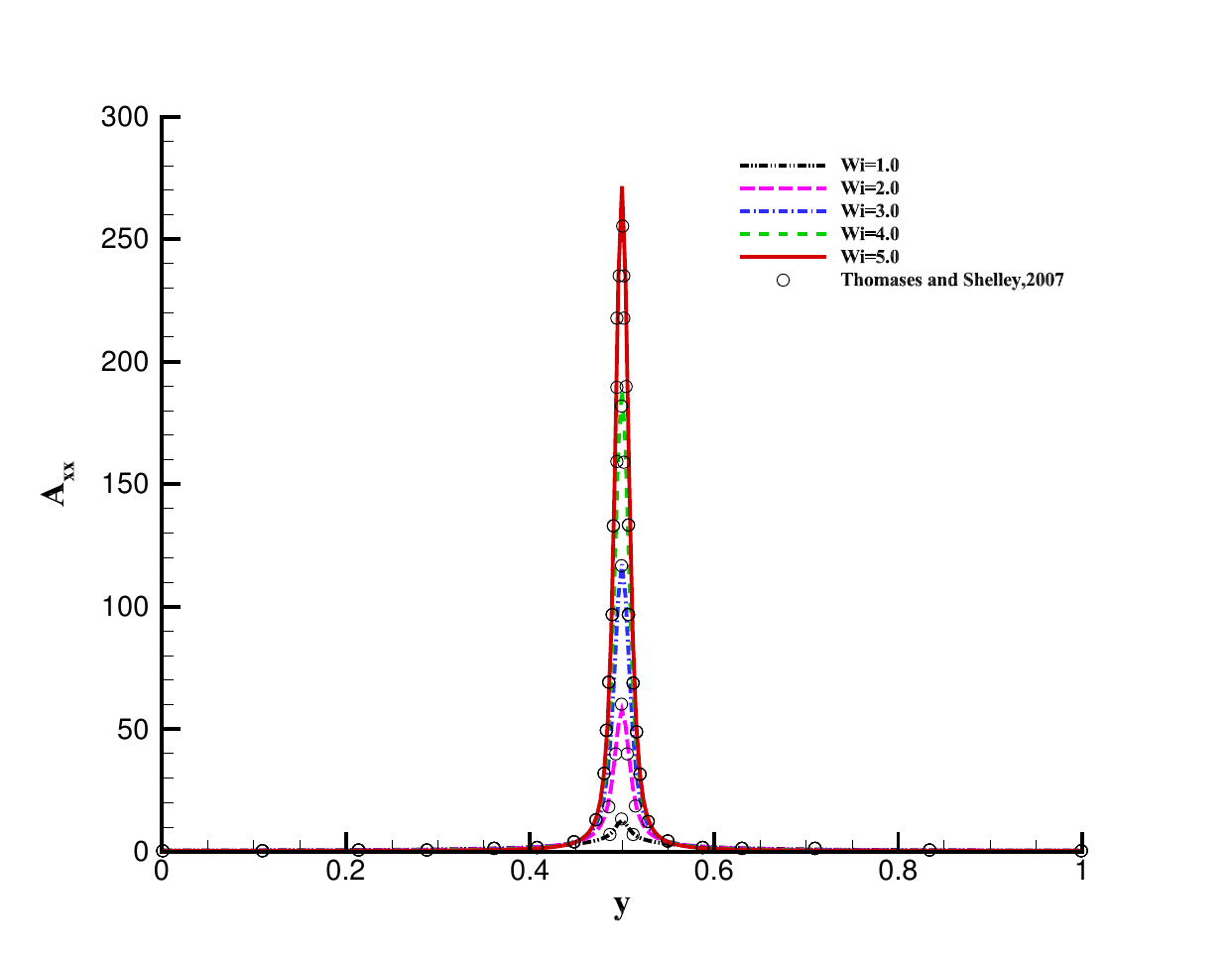}
			\caption{}
			\label{fig:subfig_l}
		\end{subfigure}
		\caption{Comparison of the conformation tensor component $A_{xx}$ along the vertical centerline at dimensionless time $t^* = 6$ for various Weissenberg numbers: (a) 0.1-0.5; (b) 0.6-0.9; (c) 1.0-5.0. The present lattice Boltzmann results are compared with the pseudospectral solutions of Thomases and Shelley~\cite{thomases2007emergence} for the Stokes-Oldroyd-B system.}
		\label{fig:axx_sc5}
	\end{figure}
	
	\subsubsection{Numerical Performance at High Weissenberg Numbers}

\begin{figure}[!ht]
		\centering
		\begin{subfigure}{0.45\textwidth}
			\includegraphics[width=\textwidth]{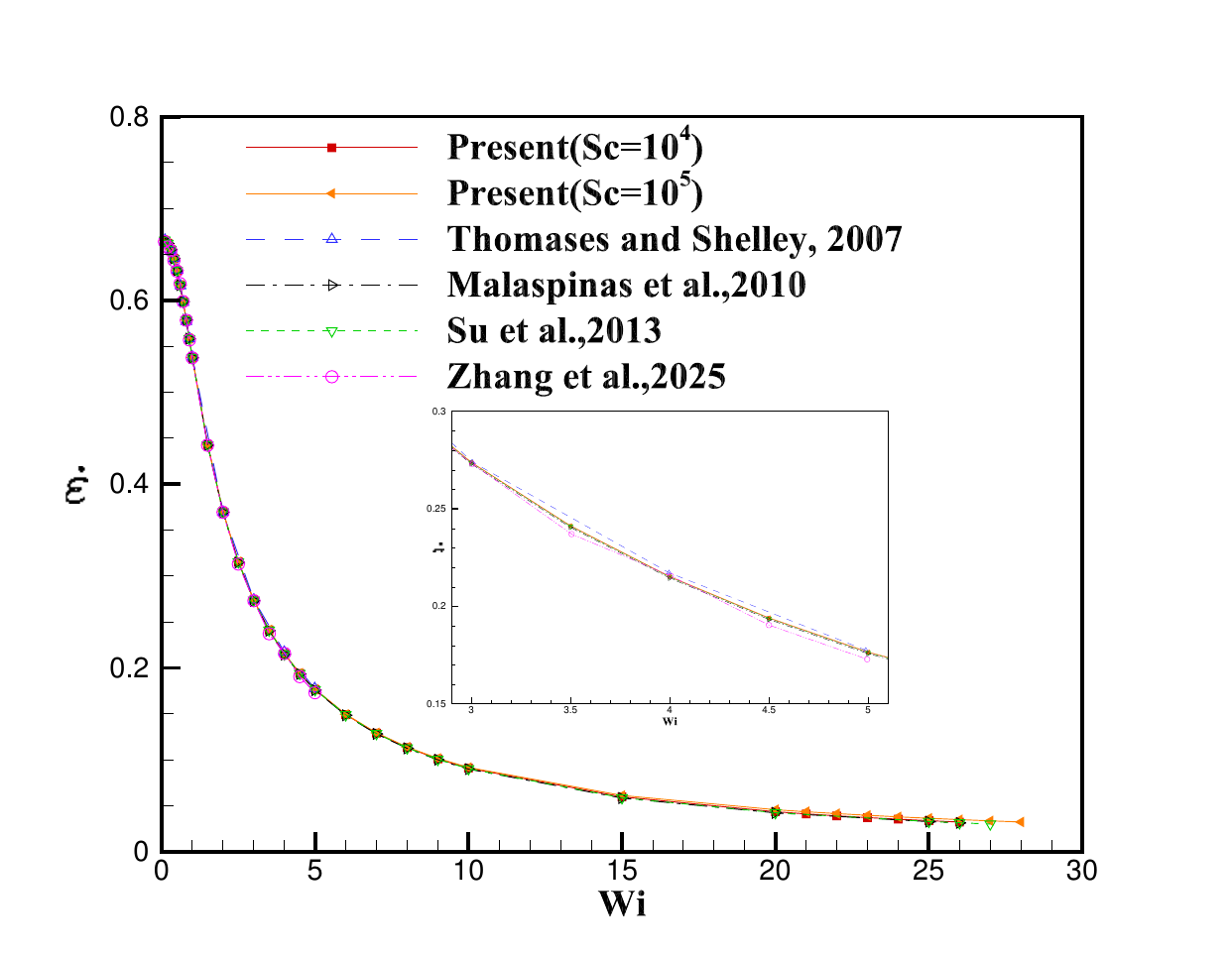}
			\caption{}
		\end{subfigure}
		\hfill
		\begin{subfigure}{0.45\textwidth}
			\includegraphics[width=\textwidth]{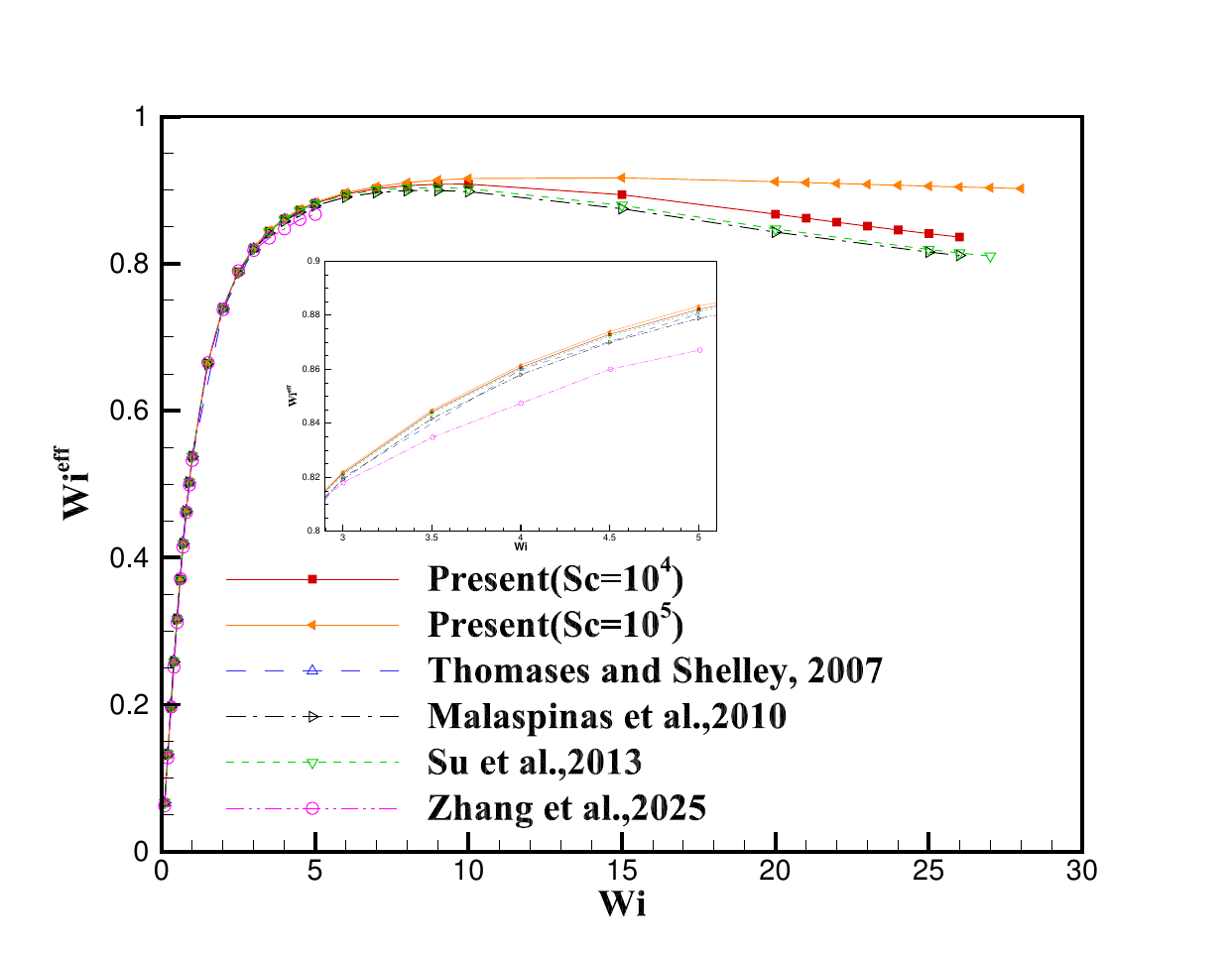}
			\caption{}
		\end{subfigure}
		\caption{Comparison of (a) local elongation rate $\dot{\varepsilon}$ and (b) effective Weissenberg number $\mathrm{Wi}^{\mathrm{eff}}$ versus nominal Weissenberg number $\mathrm{Wi}$ for different methods. Present method uses $\mathrm{Sc}=10^5$ and $\mathrm{Sc}=10^4$; Malaspinas et al.~\cite{malaspinas2010lattice} and Su et al.~\cite{su2013lattice} methods use $\mathrm{Sc}=10^4$ (the latter two methods encounter numerical breakdown at $\mathrm{Wi}<1$ when $\mathrm{Sc}=10^5$). Reference data from Thomases et al.~\cite{thomases2007emergence} (pseudospectral method) and Zhang et al.~\cite{zhang2025latticea} are included for comparison.}
		\label{fig_LER_Wieff}
	\end{figure}

	This section evaluates the performance of the proposed method at high Weissenberg numbers and provides a comprehensive comparison with existing methods within the LBM framework. All computational parameters remain consistent with those specified in the previous section. The simulations continue until reaching $t^* = 1000$ or satisfying the convergence criterion defined in Eq.~\eqref{eq_err}. Subsequently, the dimensionless local elongation rate $\dot{\varepsilon}$ is calculated using Eq.~(\ref{eq_LER}), and the effective Weissenberg number $\mathrm{Wi}^{\mathrm{eff}}$ at the stagnation point is determined through Eq.~(\ref{eq_Wieff}). The obtained results are compared with data reported by Thomases et al.~\cite{thomases2007emergence} and Zhang et al.~\cite{zhang2025latticea}. For comprehensive evaluation, we also implemented two well-established viscoelastic fluid LBM formulations: the methods proposed by Malaspinas et al.~\cite{malaspinas2010lattice} and Su et al.~\cite{su2013lattice}. A comparison of all results is shown in Figure~\ref{fig_LER_Wieff}.
	
	It should be noted that the numerical results from Thomases et al. are considered highly accurate reference solutions for the range $\mathrm{Wi} \leq 5$. This is because they employ a pseudospectral method to solve the Stokes–Oldroyd-B equation system, where the stress field evolution uses a second-order Adams-Bashforth method, the velocity field is obtained through Fourier space inversion of the Stokes equation, and a smoothing filter is introduced to enhance stability and improve accuracy beyond standard dealiasing rules. As the singular structure of the stress field intensifies, spatial discretization is correspondingly doubled, reaching a maximum resolution of 4096, and simulations are terminated before exceeding thresholds in high wavenumber regions, ensuring resolution in critical areas. The time step is verified to have second-order accuracy, and positive definiteness of the stress tensor is maintained throughout all simulations. More importantly, local analytical solutions are in excellent agreement with numerical results, such as the exponential form of stress singular structures and evolution rates, which are consistent with theoretical predictions, further confirming the reliability of the numerical results.
	
	Our results demonstrate that within the conventionally tested range of $\mathrm{Wi} = 0.1$–$5$ (studied by Thomases et al.~\cite{thomases2007emergence}, Malaspinas et al.~\cite{malaspinas2010lattice}, and Zhang et al.~\cite{zhang2025latticea}), our method's numerical predictions are closest to the pseudospectral results of Thomases et al. These findings indicate that our method exhibits higher accuracy in the moderate Weissenberg number range compared to other representative LBM methods.
	
	For higher Weissenberg numbers with $Wi>5$, to the authors' knowledge, no previous studies have reported successful computations. We further present results for high Wi up to 30 and reproduce the methods of Malaspinas et al.\cite{malaspinas2010lattice} and Su et al.~\cite{su2013lattice} for comparison, as shown in Figure~\ref{fig_LER_Wieff}. Our method demonstrates results for both $Sc = 10^4$ and $10^5$, while the reference methods were tested at $Sc = 10^4$. When we attempted to set $Sc = 10^5$ to achieve higher accuracy with these reference methods, both approaches encountered numerical breakdown at $Wi< 1$, preventing any meaningful high-$Wi$ computations and thus are not shown in the figure.
	
	The results reveal that when artificial dissipation is relatively high (smaller $Sc$ values), all three methods exhibit a decreasing trend in effective Weissenberg number ($Wi^{eff}$) at high $Wi$. However, when $Sc = 10^5$, our method maintains numerical stability and shows that $Wi^{eff}$ stabilizes around approximately $0.907$ for $Wi \geq 20$, with only a very gradual decay rate. This demonstrates that our method can achieve excellent numerical stability at high Weissenberg numbers without requiring excessive artificial dissipation, representing a significant advancement in computational capability for viscoelastic flow simulation.

	To provide further quantitative comparison, we employ the present model, Su's model, and Malaspinas's model to compute this problem and record the effective Weissenberg number at the center position. The magic parameter is set to $\Lambda_p = 10^{-6}$ for all calculations. Two Schmidt numbers ($Sc = 10^4$ and $10^5$) are considered for comparison, with results presented in Table~\ref{tab:qq}. The results demonstrate that when $Sc = 10^4$, all three methods can avoid numerical breakdown within the considered Weissenberg number range ($Wi \leq 30$). However, when artificial dissipation is reduced to $Sc = 10^5$, both Su's and Malaspinas's methods experience premature breakdown, while the present method maintains stability without breakdown across all tested Weissenberg numbers.
	
	\begin{table}[htpb]
		\centering
		\caption{Stability comparison of different LBM approaches for the four-roll mill problem showing effective Weissenberg numbers ($Wi^{eff}$) at the domain center. Grid resolution: $L_x = L_y = 257$, Reynolds number: $Re = 1$, magic parameter: $\Lambda_p = 10^{-6}$. "NaN" indicates numerical breakdown/divergence; "unsteady" denotes time-oscillatory solutions when steady state is not achieved by $t^* = 1000$.}
		\label{tab:qq}
		\resizebox{\textwidth}{!}{
			\begin{tabular}{c c c c c c c}
				\toprule
				\multicolumn{1}{c}{\multirow{2}{*}{$Wi$}} 
				& \multicolumn{3}{c}{$Wi^{eff}$ ($Sc = 10^4$)} 
				& \multicolumn{3}{c}{$Wi^{eff}$ ($Sc = 10^5$)} \\
				\cmidrule(lr){2-4}  \cmidrule(lr){5-7}
				& Present & Su & Malaspinas  
				& Present & Su & Malaspinas \\
				\midrule
				0.1   & 0.067 & 0.067 & 0.067 & 0.067 & 0.067 & 0.067 \\
				0.2   & 0.132 & 0.132 & 0.132 & 0.132 & 0.132 & 0.132 \\
				0.3   & 0.197 & 0.197 & 0.197 & 0.197 & 0.197 & 0.197 \\
				0.4   & 0.258 & 0.258 & 0.258 & 0.258 & NaN    & 0.258 \\
				0.5   & 0.316 & 0.316 & 0.316 & 0.316 & NaN    & 0.316 \\
				0.6   & 0.370 & 0.370 & 0.370 & 0.370 & NaN    & 0.370 \\
				0.7   & 0.419 & 0.419 & 0.419 & 0.419 & NaN    & 0.419 \\
				0.8   & 0.463 & 0.463 & 0.463 & 0.463 & NaN    & 0.463 \\
				0.9   & 0.503 & 0.503 & 0.502 & 0.503 & NaN    & NaN \\
				1.0   & 0.538 & 0.538 & 0.537 & 0.538 & NaN    & NaN \\
				5.0   & 0.882 & 0.881 & 0.879 & 0.883 & NaN    & NaN \\
				10.0  & 0.908 & 0.902 & 0.898 & 0.916 & NaN    & NaN \\
				15.0  & 0.894 & 0.879 & 0.875 & 0.917 & NaN    & NaN \\
				20.0  & 0.867 & 0.847 & 0.843 & 0.911 & NaN    & NaN \\
				25.0  & 0.841 & 0.819 & 0.816 & 0.905 & NaN    & NaN \\
				26.0  & 0.836 & 0.815 & 0.811 & 0.904 & NaN    & NaN \\
				27.0  & unsteady & 0.810 & unsteady & 0.903 & NaN & NaN \\
				28.0  & unsteady & unsteady & unsteady & 0.902 & NaN & NaN \\
				29.0  & unsteady & unsteady & unsteady & unsteady & NaN & NaN \\
				30.0  & unsteady & unsteady & unsteady & unsteady & NaN & NaN \\
				\bottomrule
			\end{tabular}
		}
	\end{table}
	\subsubsection{Optimal Selection of the Magic Parameter $\Lambda_p$ for Bulk Flow Regions}

\begin{figure}[htbp]
		\centering
		\begin{subfigure}[b]{0.32\textwidth}
			\centering
			\includegraphics[width=\textwidth]{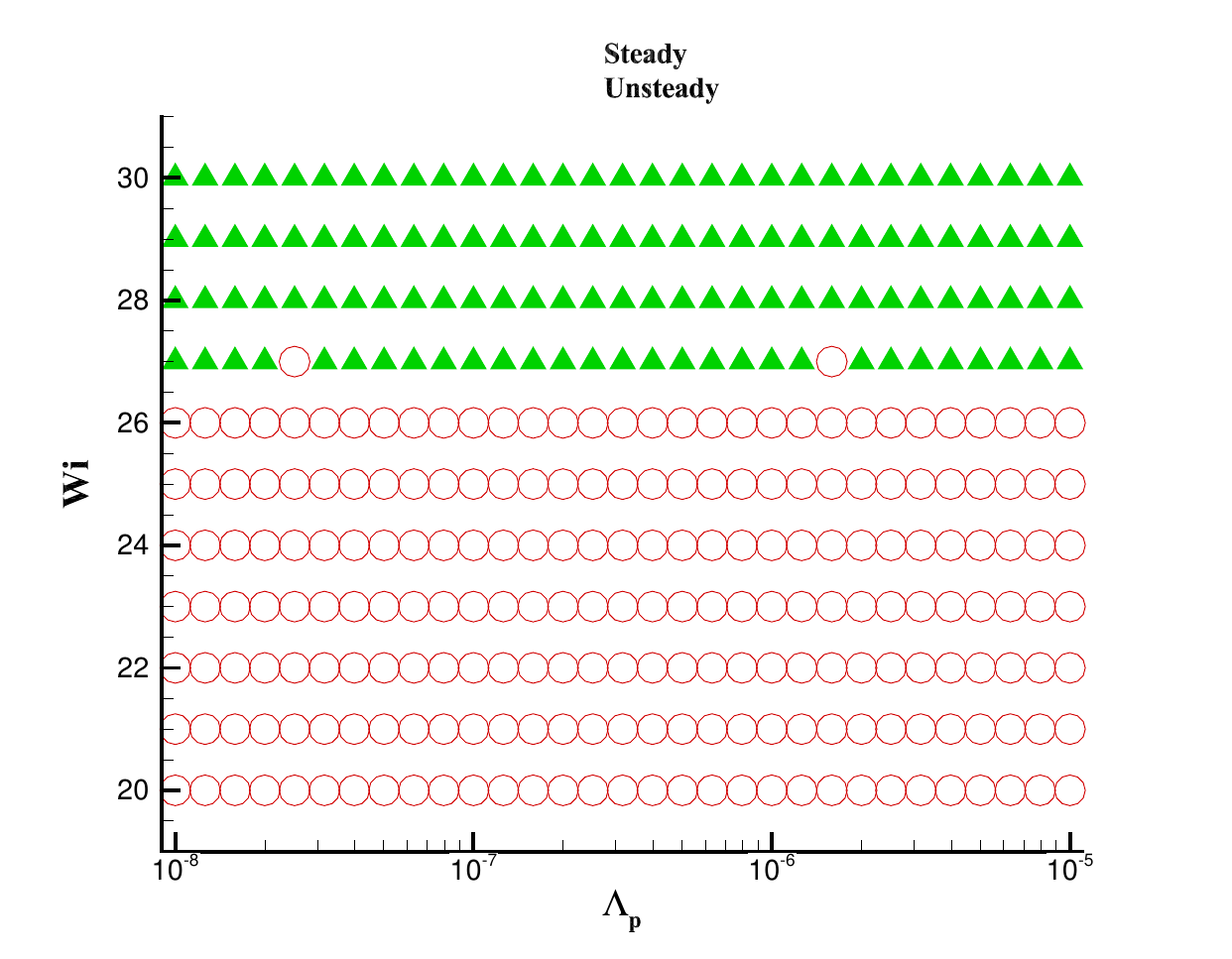}
			\caption{$Sc = 10^4$}
			\label{fig:max_wi_sc4}
		\end{subfigure}
		\hfill
		\begin{subfigure}[b]{0.32\textwidth}
			\centering
			\includegraphics[width=\textwidth]{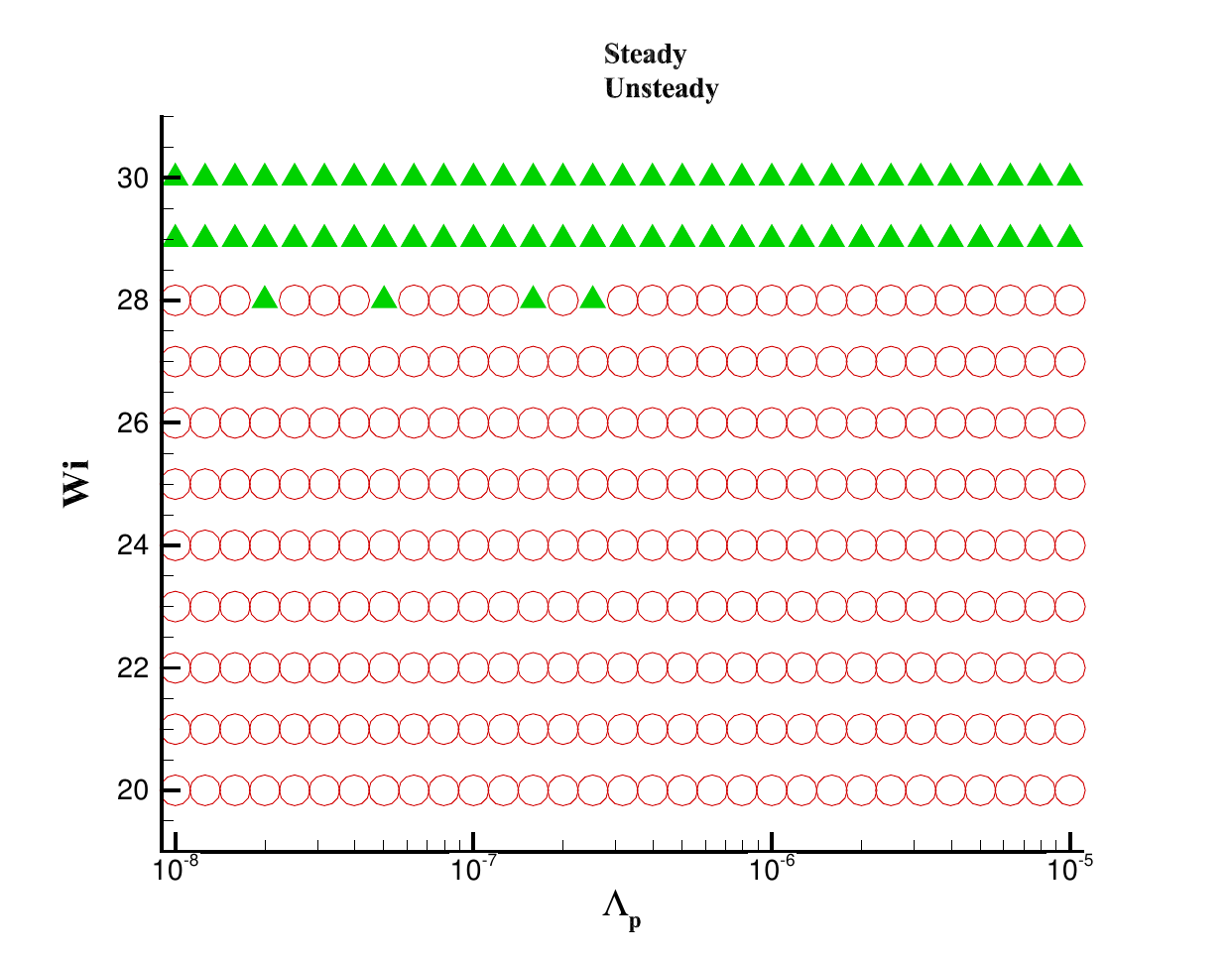}
			\caption{$Sc = 10^5$}
			\label{fig:max_wi_sc5}
		\end{subfigure}
		\hfill
		\begin{subfigure}[b]{0.32\textwidth}
			\centering
			\includegraphics[width=\textwidth]{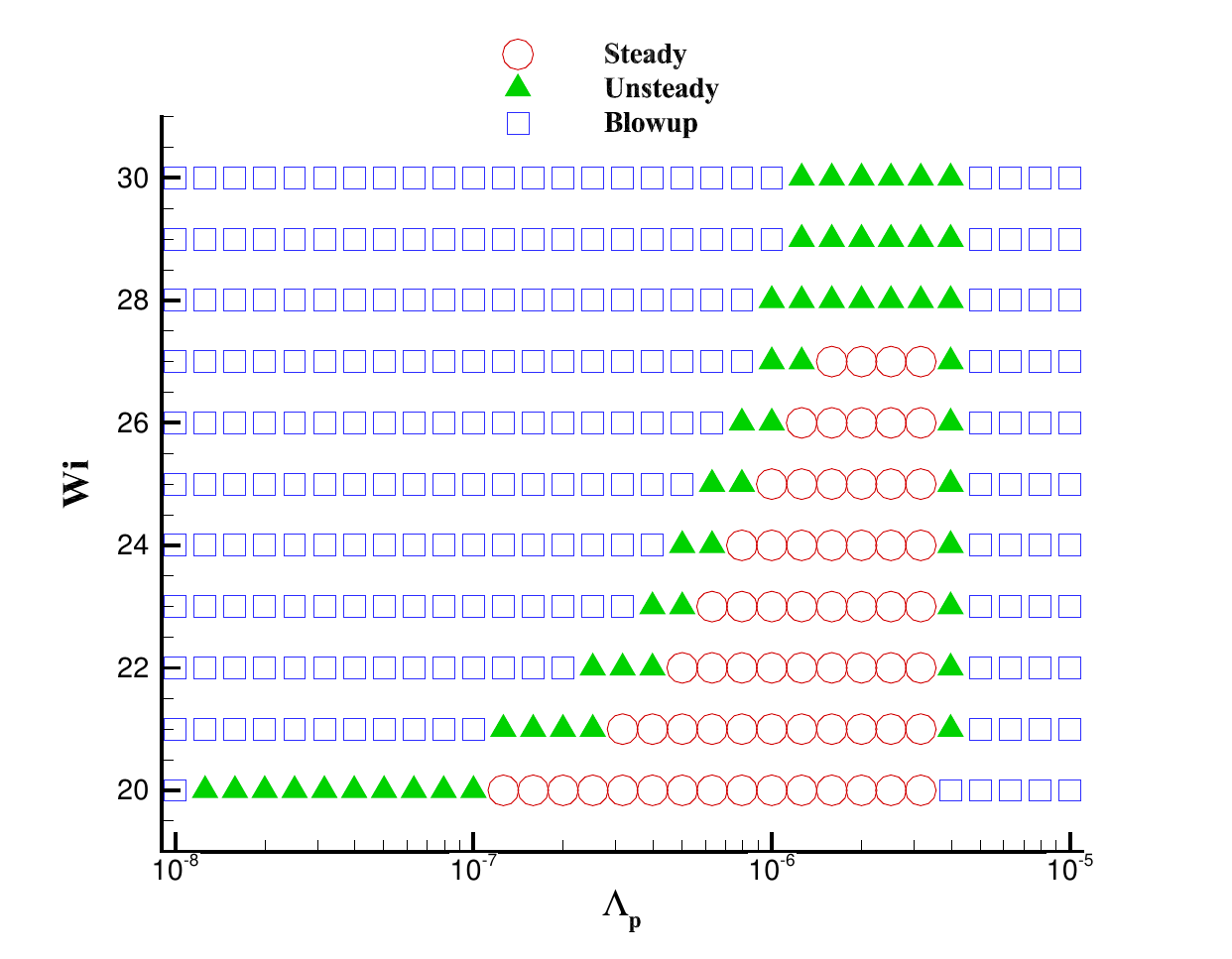}
			\caption{$Sc = 10^6$}
			\label{fig:max_wi_sc6}
		\end{subfigure}
		
		\caption{Phase diagrams showing computational stability regions in the $\Lambda_p$-Wi parameter space for the four-roll mill problem (bulk flow without solid boundaries) under different Schmidt numbers: (a) $Sc = 10^4$, (b) $Sc = 10^5$, and (c) $Sc = 10^6$. Three computational states are distinguished: steady-state solutions, unsteady solutions, and numerical breakdown. Lower artificial viscosity (higher Sc) requires more careful selection of the magic parameter $\Lambda_p$ to maintain stability.}
		\label{fig:max_wi_lambda_p}
	\end{figure}

	Our previous work has demonstrated that for TRT-RLB models solving ADE, the selection of the magic parameter is crucial for model stability~\cite{yu2025tworelaxationtime}. Currently, existing literature cannot analytically specify the optimal free relaxation parameter $\tau_{p,2}$ or magic parameter $\Lambda_p$. Therefore, in this subsection, we employ numerical methods to select the optimal free relaxation parameter for fine-tuning algorithm stability. We consider three different Schmidt numbers ($Sc = 10^4$, $10^5$, $10^6$), with Weissenberg numbers ranging from 20 to 30 in increments of 1. Simulations are run until convergence, or remain unconverged at $t^* = 1000$, or experience numerical breakdown, which are categorized as "steady", "unsteady", and "blowup", respectively. The results are shown in Figure~\ref{fig:max_wi_lambda_p}.
	
	The results indicate that when $Sc = 10^4$, the maximum Weissenberg number achieving convergence is approximately 26, with stable unsteady solutions obtainable at higher values. Similar behavior occurs when $Sc = 10^5$, where the maximum Weissenberg number for convergent solutions reaches approximately 28. For both Schmidt number cases mentioned above, unsteady solutions can still be obtained when Wi is further increased to 30. When further increased to $Sc = 10^6$, the significantly reduced artificial viscosity poses greater challenges to the numerical model. Therefore, when artificial viscosity is moderately high, $\Lambda_p$ can be selected from a wide range with minimal impact on the model, whereas when artificial viscosity is very low, model stability becomes closely related to the choice of $\Lambda_p$.
	
	At $Sc = 10^6$, if an inappropriate $\Lambda_p$ is selected, the program fails prematurely as the Weissenberg number increases. However, when $\Lambda_p$ approaches the optimal range of $1.6\times10^{-6}$ to $3.2\times10^{-6}$, the maximum Wi producing stable solutions reaches 27, and stable computations can be maintained even when Wi is further increased to 30. Based on our understanding of TRT collision models and TRT-RLB models~\cite{yu2025tworelaxationtime}, different boundary conditions also affect the optimal magic parameter. Considering this, for problems with periodic boundary conditions, we recommend an optimal magic parameter value of approximately $10^{-6}$, while the selection for problems involving solid wall boundaries will be discussed further in subsequent sections.

	\subsection{Force-driven Poiseuille flow}\label{sec4.2}
	
	Force-driven Poiseuille flow represents one of the few viscoelastic fluid simulation problems with available time-dependent analytical solutions. The presence of solid walls makes this benchmark crucial for validating algorithmic accuracy, particularly regarding temporal precision and the implementation of boundary condition schemes. As the Weissenberg number increases, viscoelastic stress tensors and their gradients near boundaries become increasingly large, presenting significant computational challenges. Therefore, this benchmark serves as an effective test for algorithm stability under high-Wi conditions.
	
	Numerous numerical methods have been applied to simulate this problem, with several studies explicitly reporting numerical stability limitations at high Weissenberg numbers. For instance, Ellero and Tanner~\cite{ellero2005sph} employed the SPH method to achieve stable UCM model simulations up to $Wi = 1.0$, noting that while calculations remained stable at $Wi = 0.1$, small spatial oscillations began to appear at higher Wi, explicitly acknowledging the numerical challenges associated with elevated Weissenberg number regimes. Fiétier and Deville~\cite{fietier2003time} developed a time-dependent viscoelastic algorithm based on the spectral element method, systematically demonstrating progressive capability improvements: from unstabilized approaches achieving $Wi \approx 0.6-2.7$, to pure DEVSS techniques reaching $Wi \approx 5.0$, and finally to DEVSS combined with filtering techniques extending to $Wi = 10.0$. They clearly stated that pure Galerkin methods face significant numerical instability limitations, while pure DEVSS techniques require excessively high artificial viscosity to maintain stability at $Wi = 5.0$. Jafari et al.~\cite{jafari2012simulation} made notable contributions to high-Wi viscoelastic flow simulation using a filter-based stabilized spectral element method. They systematically analyzed spurious mode growth in Poiseuille flow, investigating factors including time evolution, mesh refinement, boundary conditions, Wi, and extensibility parameters. Their proposed mesh-transfer technique enhanced filter effectiveness in multi-element configurations, improving stability for both Oldroyd-B and FENE-P models. For Poiseuille flow, they achieved simulations up to $Wi = 100$ using the more stable FENE-P model, though simulations encountered severe instability due to spurious mode amplification and the conformation tensor trace approaching critical values, limiting further Wi increases.
	
	However, many studies primarily focus on low-Wi analytical solutions as references for time-dependent problems and do not actively report performance at high Wi numbers.  Without reproducing their work, it remains difficult to definitively conclude whether their methods encounter numerical challenges when solving Poiseuille flow problems beyond certain Wi thresholds. Nevertheless, we can enumerate the highest Wi numbers reported in their studies, though these may not necessarily represent their maximum computational capabilities. For example, Xu and Yu~\cite{xu2016multiscale} developed a multiscale SPH method that achieved stable numerical simulations of transient viscoelastic flows, successfully computing cases up to $Wi = 5.0$. Zou \textit{et al.} ~\cite{zou2014integrated} proposed an integrated lattice Boltzmann-finite volume framework (ILFVE), employing a hybrid strategy where LBM solves the Navier-Stokes equations while FVM handles the constitutive equations, achieving approximately 5-fold computational efficiency improvement while maintaining accuracy, with maximum Weissenberg numbers reaching $Wi = 1.0$ in Poiseuille flow.  Lee \textit{et al.} ~\cite{lee2017time} systematically applied the advection-diffusion lattice Boltzmann method to study time-dependent viscoelastic characteristics of Oldroyd-B fluids, validating the method's accuracy through realistic rheological tests including step-shear and oscillatory shear, with maximum Weissenberg numbers of Wi = 0.1 in Poiseuille flow.  Su \textit{et al.} ~\cite{su2018lattice} extended the computational range to Wi = 15.0 through an asynchronous coupling strategy between solvent and polymer phases. Notably, in the literature surveyed, we have not observed reported results for the Oldroyd-B model in Poiseuille flow simulations exceeding $Wi = 20$.
	
	It is important to note that although stress tensor magnitudes and their gradients continuously increase with rising Wi numbers, no stress singularities or their evolution occur in the Poiseuille flow problem.  This absence of singularities means that for certain algorithms, particularly LBM methods, high-Wi conditions in this problem do not pose fundamental computational challenges.  We successfully reproduced Malaspinas et al.'s LB algorithm and confirmed its capability to simulate Poiseuille flow at extremely high Wi numbers (up to 100,000) with stable and accurate results.  While our improved algorithm demonstrates similar robustness for high-Wi simulations, this section will focus on demonstrating the performance of our own method.

	\begin{figure}[!htp]
		\centering
		\includegraphics[width=0.4\textwidth]{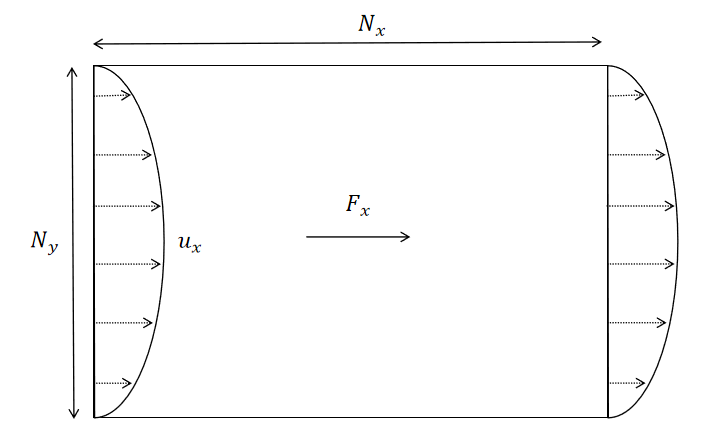}
		\caption{Configuration of plane Poiseuille flow with periodic boundary conditions in the streamwise direction and no-slip conditions at the walls.}
		\label{poiseuille}
	\end{figure}
	
	The computational domain for plane Poiseuille flow is illustrated in Figure~\ref{poiseuille}, where the channel length and height are denoted as $N_x$ and $N_y$, respectively. The flow is driven by a constant body force $F_{\alpha}$, with periodic boundary conditions imposed at the inlet and outlet, and no-slip boundary conditions applied at the upper and lower walls. The body force components are defined as:
	\begin{equation}
		(F_x,F_y) = \left(\frac{8}{L_c^2}(\nu_s+\nu_p)U_c, 0\right),
	\end{equation}
	where $\nu_s$ and $\nu_p$ represent the solvent and polymer viscosities, respectively, $U_c$ is the characteristic velocity, and $L_c=N_y$ is the characteristic length. For this configuration, Waters and King \cite{waters1970unsteady} derived the analytical solution for transient plane Poiseuille flow of an Oldroyd-B fluid. The dimensionless velocity field is expressed as:
	\begin{equation}
		u_x^*(y^*,t^*)=-4y^*(y^*-1)-32\sum_{n=1}^\infty\frac{\sin Ny^*}{N^3}G_N\left(\frac{\mathrm{Wi}}{\mathrm{Re}},t^*\right),
	\end{equation}
	where $u_x^*$, $y^*$, and $t^*$ represent the dimensionless velocity, coordinate, and time, respectively, normalized by the characteristic velocity $U_c$, characteristic length $L_c$, and characteristic time $T_c$. Here, $N = (2n-1)\pi$, and the time-dependent function $G_N$ is given by
	\begin{equation}
		\begin{aligned}
			G_N\left(\frac{\mathrm{Wi}}{\mathrm{Re}},t^*\right) =
			\begin{cases}
				e^{-\frac{\alpha_Nt^*}{2\mathrm{Wi}}}\left\{\cosh\left(\frac{\beta_Nt^*}{2\mathrm{Wi}}\right)+\frac{\gamma_N}{\beta_N}\sinh\left(\frac{\beta_Nt^*}{2\mathrm{Wi}}\right)\right\}, & \beta_N^2\geq0, \\[0.5em]
				e^{-\frac{\alpha_Nt^*}{2\mathrm{Wi}}}\left\{\cos\left(\frac{\beta_Nt^*}{2\mathrm{Wi}}\right)+\frac{\gamma_N}{\beta_N}\sin\left(\frac{\beta_Nt^*}{2\mathrm{Wi}}\right)\right\}, & \beta_N^2<0,
			\end{cases}
		\end{aligned}
	\end{equation}
	where the auxiliary parameters in the above expressions are defined as
	\begin{align}
		&\alpha_N=1+\beta\frac{\mathrm{Wi}}{\mathrm{Re}}N^{2},\quad\beta_{N}^{2}=\alpha_{N}^{2}-4N^{2}\frac{\mathrm{Wi}}{\mathrm{Re}},\\
		&\beta_N=\sqrt{|\beta_N^2|},\quad\gamma_N=1+(\beta-2)N^2\frac{\mathrm{Wi}}{\mathrm{Re}},
	\end{align}

	For the steady-state solution, as demonstrated by Zou et al. \cite{zou2014integrated}, when $t^* \rightarrow \infty$, the velocity profile reduces to the classical parabolic distribution
	\begin{equation}\label{eq_ux_ana}
		u_x^*(y^*) = 4y^*(1-y^*),
	\end{equation}
	and the corresponding steady-state viscoelastic stress components are given by
	\begin{align} \label{eq_Tab_ana}
		&\tau_{xx}^* = 2\mathrm{Wi}(1-\beta)\left(\frac{\partial u^*}{\partial y^*}\right)^2,\\
		&\tau_{xy}^* = (1-\beta)\left(\frac{\partial u^*}{\partial y^*}\right),\\
		&\tau_{yy}^* = 0,    
	\end{align}
	where $\tau_c = (\mu_0 U_c)/L_c$ is the characteristic stress scale, and $\tau_{\alpha \beta}^* = \tau_{\alpha \beta}/\tau_c$ denotes the dimensionless stress tensor components.
	
	The convergence criterion employed in this study is defined as:
	\begin{equation}\label{2eq4}
		\max\left(\left|\frac{u_{\alpha}(x_{\alpha}, t_n+T_c) - u_{\alpha}(x_{\alpha}, t_n)}{U_c}\right|\right) \leq 10^{-8},
	\end{equation}
	where $t_n=n \Delta t$ and $n$ represents the iteration number. This criterion ensures that the temporal variation of the velocity field is sufficiently small to consider the solution as steady-state.
	
	Unless otherwise specified, simulations are initialized with unit density, zero velocity and stress fields, identity conformation tensor, and equilibrium distribution functions. Default parameters are $\mathrm{Re} = 1$, $\mathrm{Ma} = 0.1$, and $\mathrm{Sc} = 10^6$, ensuring negligible compressibility and adequate diffusion resolution. These settings may be modified for specific validation cases as detailed in the following subsections.
	
	\subsubsection{Comparison with Analytical Solutions}

	In this subsection, we conduct a comprehensive comparison between the numerical results and analytical solutions. We first examine the steady-state solutions under three representative Weissenberg numbers: $Wi = 1$, $100$, and $10000$, corresponding to moderate, high, and ultra-high Wi regimes, respectively. The viscosity ratio is fixed at $\beta = 0.5$ throughout these simulations. Three distinct Schmidt numbers ($Sc = 10^6$, $10^7$, $10^8$) are investigated to assess the influence of artificial diffusion on solution accuracy.
	\begin{figure}[!htbp]
		\centering
		\begin{subfigure}[b]{0.32\textwidth}
			\centering
			\includegraphics[width=\textwidth]{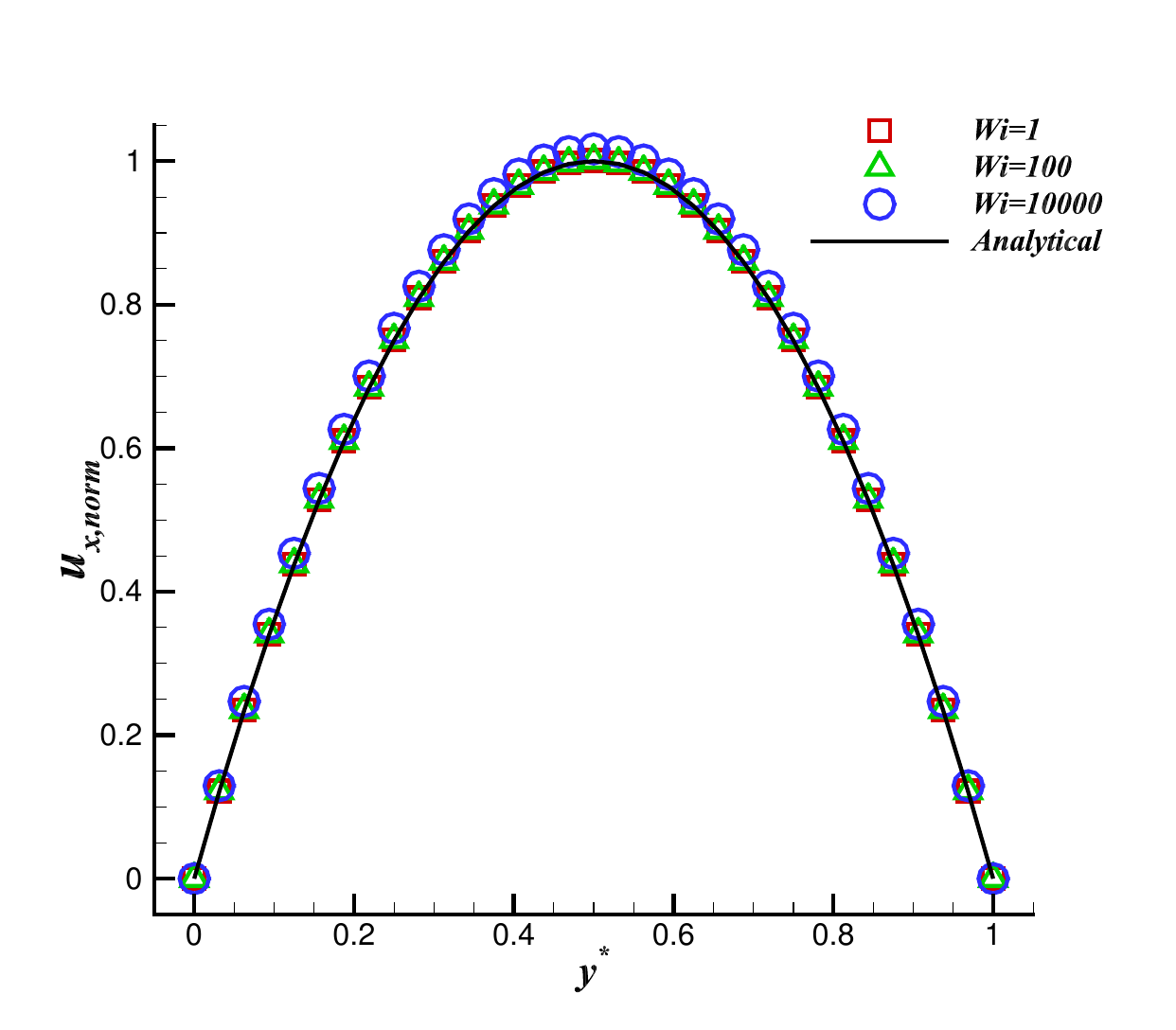}
			\caption{ $u_{x,\text{norm}}$, $Sc = 10^6$}
			\label{fig:ux_sc6}
		\end{subfigure}
		\hfill
		\begin{subfigure}[b]{0.32\textwidth}
			\centering
			\includegraphics[width=\textwidth]{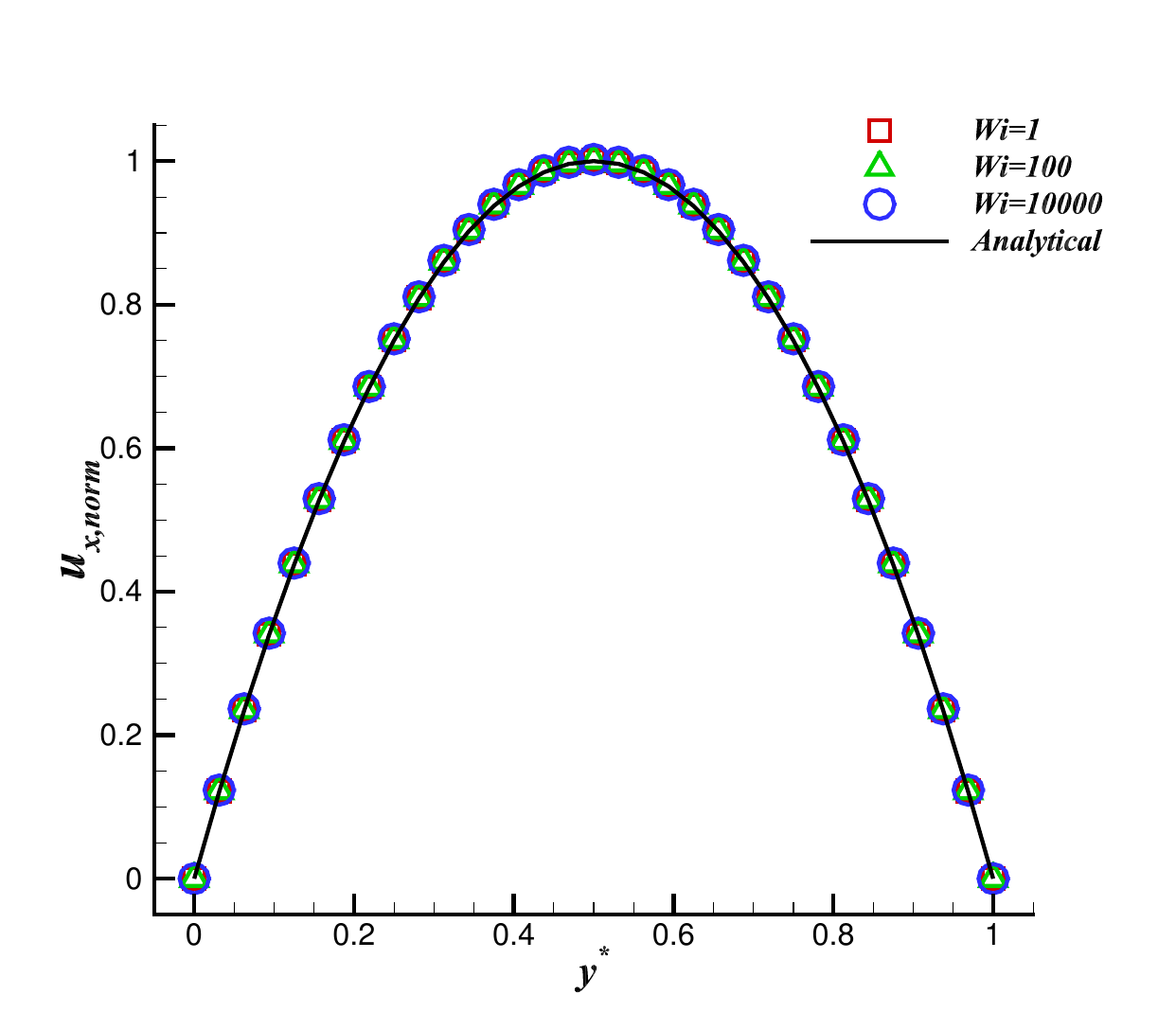}
			\caption{$u_{x,\text{norm}}$, $Sc = 10^7$}
			\label{fig:ux_sc7}
		\end{subfigure}
		\hfill
		\begin{subfigure}[b]{0.32\textwidth}
			\centering
			\includegraphics[width=\textwidth]{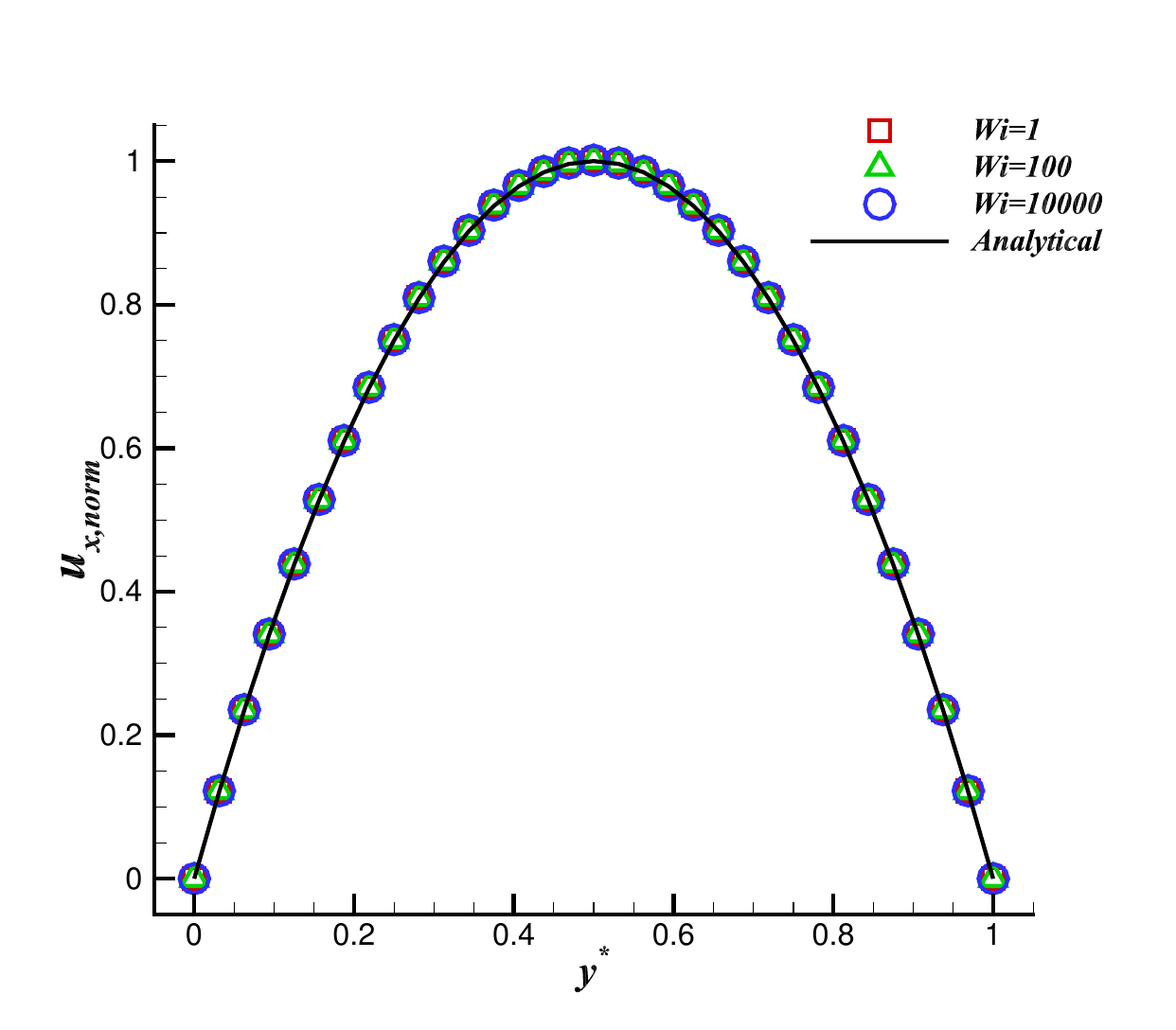}
			\caption{$u_{x,\text{norm}}$, $Sc = 10^8$}
			\label{fig:ux_sc8}
		\end{subfigure}
		
		\vspace{0.3cm}
		
		\begin{subfigure}[b]{0.32\textwidth}
			\centering
			\includegraphics[width=\textwidth]{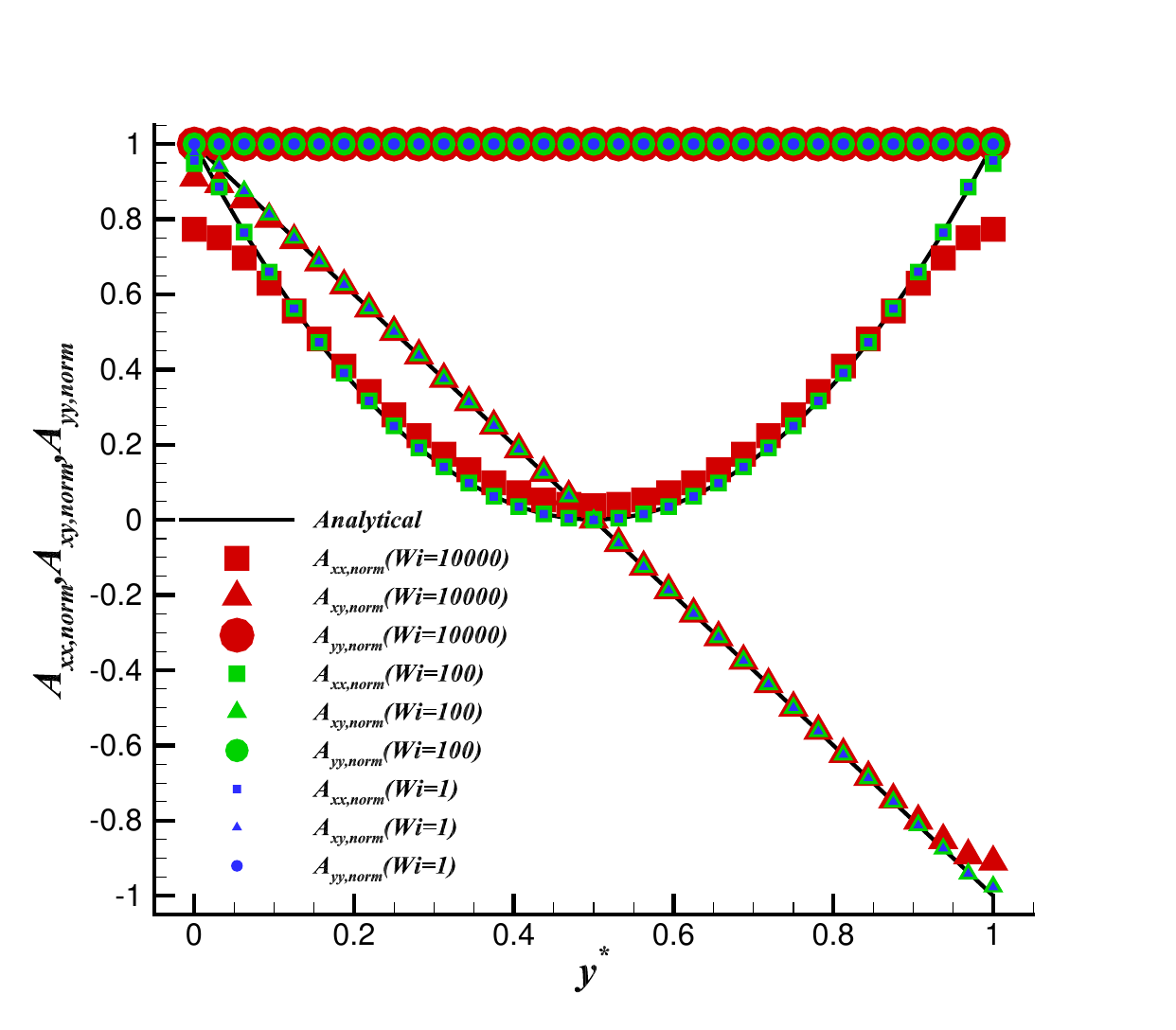}
			\caption{$A_{\alpha\beta,\text{norm}}$, $Sc = 10^6$}
			\label{fig:A_sc6}
		\end{subfigure}
		\hfill
		\begin{subfigure}[b]{0.32\textwidth}
			\centering
			\includegraphics[width=\textwidth]{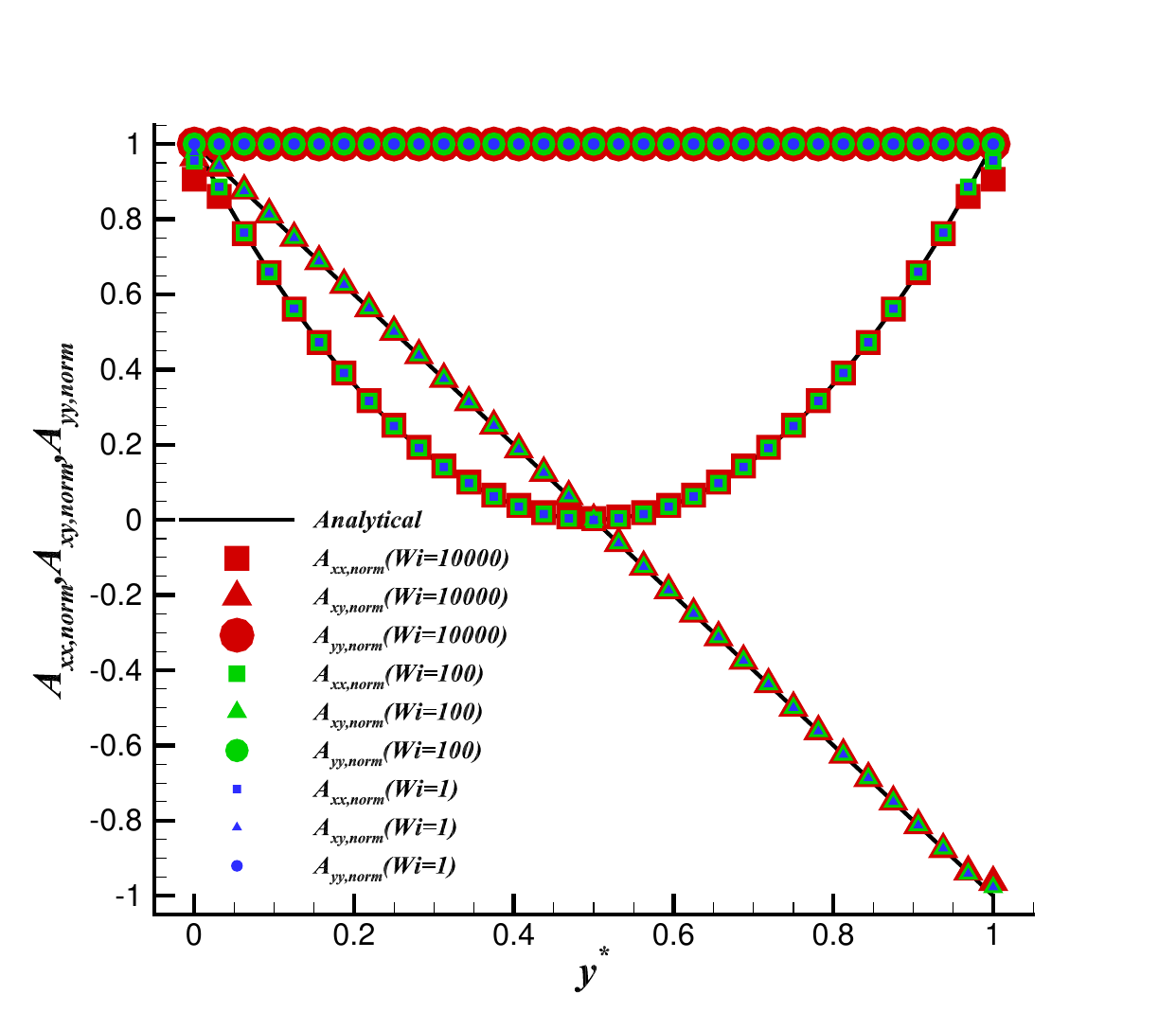}
			\caption{$A_{\alpha\beta,\text{norm}}$, $Sc = 10^7$}
			\label{fig:A_sc7}
		\end{subfigure}
		\hfill
		\begin{subfigure}[b]{0.32\textwidth}
			\centering
			\includegraphics[width=\textwidth]{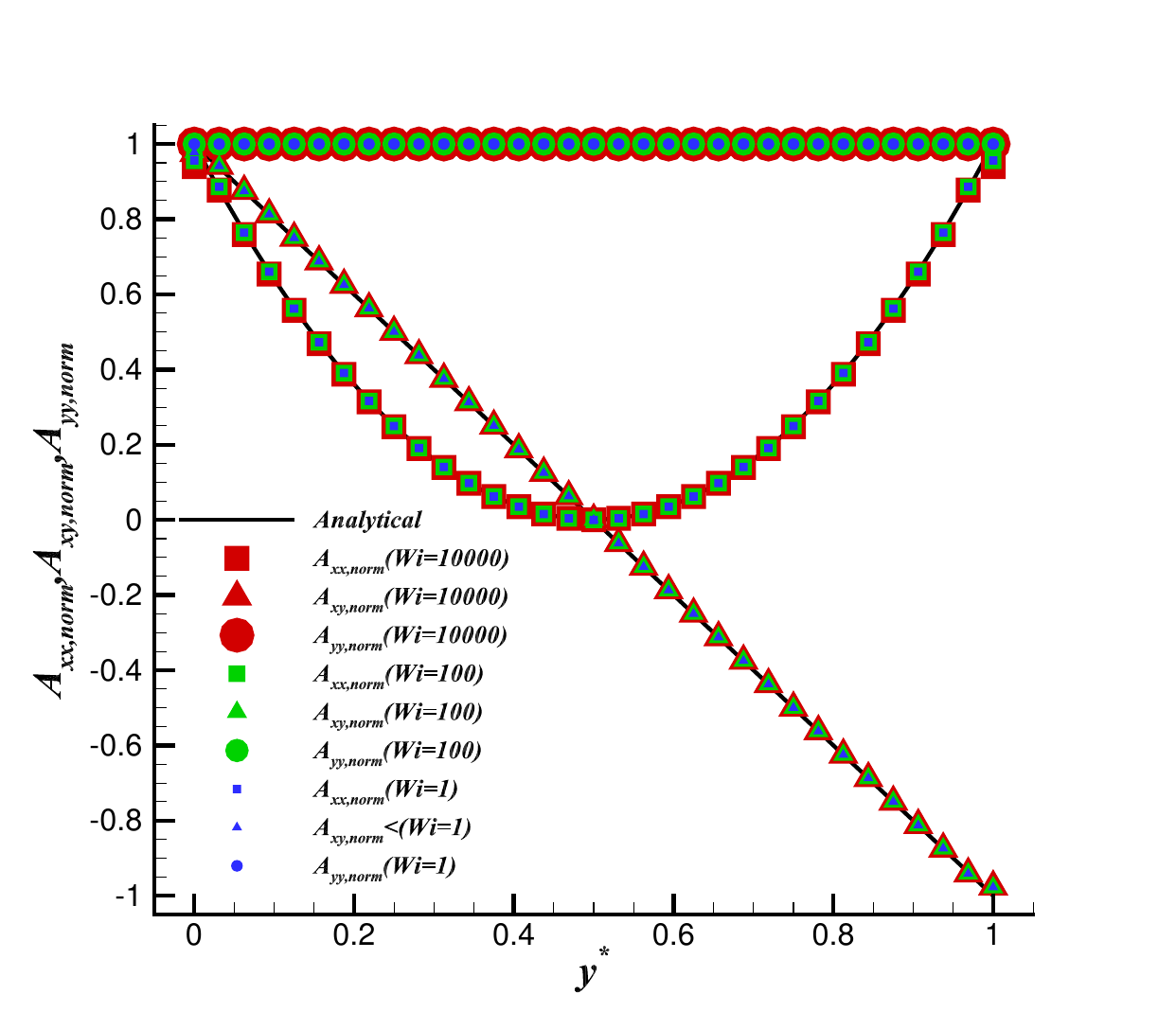}
			\caption{$A_{\alpha\beta,\text{norm}}$, $Sc = 10^8$}
			\label{fig:A_sc8}
		\end{subfigure}
		
		\caption{Comparison of numerical results with analytical solutions for normalized velocity $u_{x,\text{norm}}$ (a-c) and conformation tensor components $A_{\alpha\beta,\text{norm}}$ (d-f) under different Schmidt numbers. Panels (a,d): $\mathrm{Sc} = 10^6$; panels (b,e): $\mathrm{Sc} = 10^7$; panels (c,f): $\mathrm{Sc} = 10^8$. Three Weissenberg numbers ($Wi = 1, 100, 10000$) are considered with $N_y=32$, $\beta = 0.5$ and $\mathrm{Re} = 1$.}
		\label{fig:steady_state_comparison}
	\end{figure}
	
	Upon reaching steady state, we extract the $y$-directional profiles of the velocity component $u_x$ and conformation tensor components $A_{\alpha\beta}$ for quantitative comparison with analytical solutions. To facilitate direct comparison, all quantities are normalized according to: $u_{x,\text{norm}}= \frac{u_x}{\max(u_{x,\text{ana}})}$, $A_{xx,\text{norm}}= \frac{A_{xx} - 1}{\max(A_{xx,\text{ana}} - 1)}$, $A_{xy,\text{norm}}= \frac{A_{xy}}{\max(A_{xy,\text{ana}})}$, and $A_{yy,\text{norm}}=\frac{A_{yy}}{\max(A_{yy,\text{ana}})}$, where the subscript $\text{norm}$ indicates normalized quantities, and $\max(\cdot)$ denotes the maximum value of the corresponding analytical solution along the $y$-direction. For the $A_{xx}$ component, we subtract unity to account for its equilibrium value before normalization, ensuring that the normalized range reflects the polymer-induced deformation. Both numerical and analytical solutions undergo identical normalization procedures, with the comparative results presented in Figure~\ref{fig:steady_state_comparison}. The results demonstrate excellent agreement between numerical and analytical solutions across all cases, with one notable exception: at ultra-high Weissenberg number ($Wi = 10000$), the conformation tensor components $A_{xx}$ and $A_{xy}$ exhibit noticeable deviations from the analytical solutions near the boundary nodes when $Sc = 10^6$.
	
	This deviation arises from the extremely large absolute values of conformation tensor components at boundary nodes under ultra-high Wi conditions. When $Sc = 10^6$, the prescribed artificial diffusion becomes sufficiently significant to compromise the accurate resolution of the conformation tensor field. However, a systematic comparison across different Schmidt numbers reveals that as $Sc$ is progressively increased (thereby reducing artificial diffusion), these boundary-region deviations observed under ultra-high Wi conditions are effectively suppressed. This behavior underscores the critical role of minimizing artificial diffusion for maintaining solution accuracy in the challenging ultra-high Wi regime.
	
	\begin{figure}[htbp]
		\centering
		\begin{subfigure}[b]{0.32\textwidth}
			\centering
			\includegraphics[width=\textwidth]{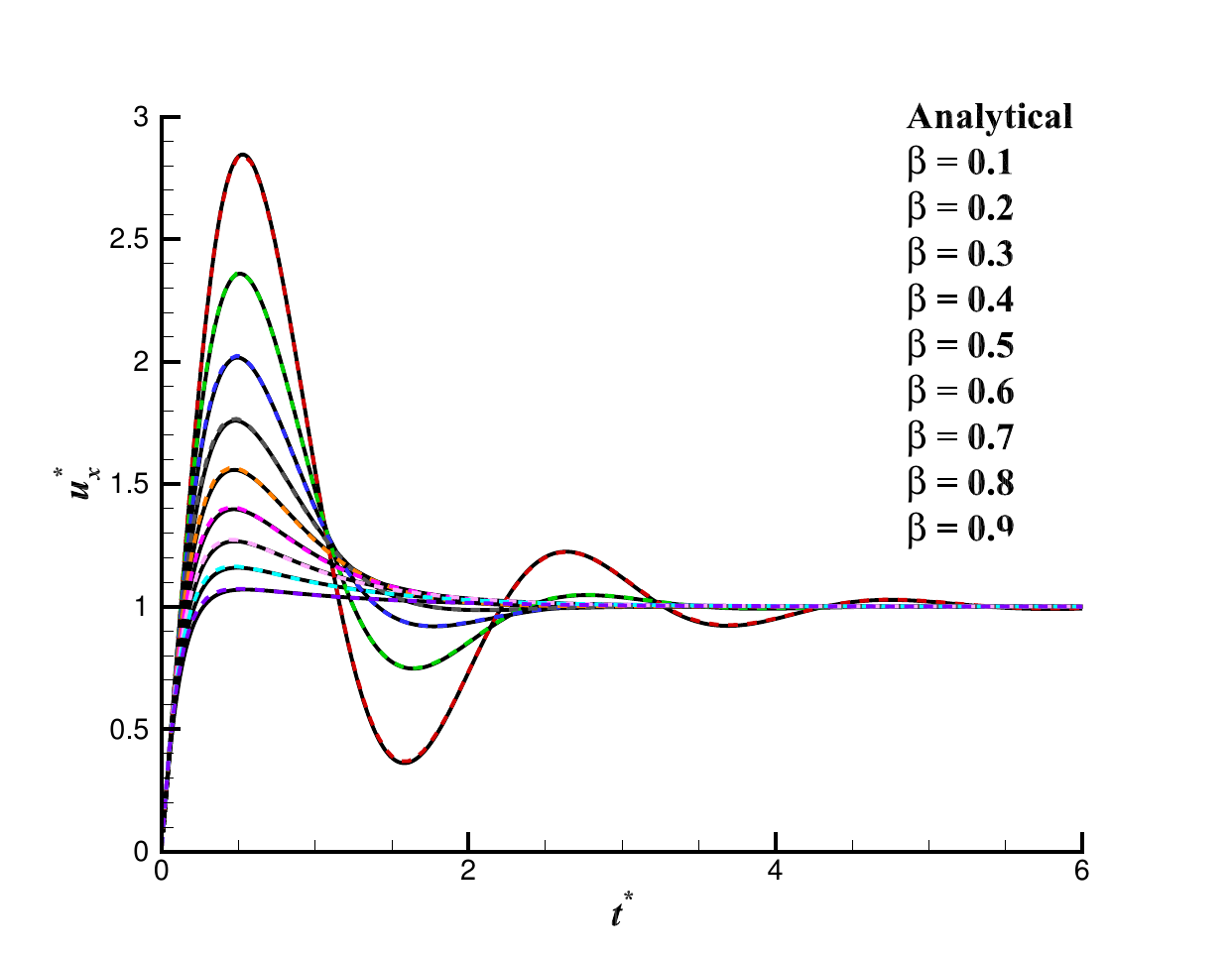}
			\caption{$Wi = 1$}
			\label{fig:ux_center_wi1}
		\end{subfigure}
		\hfill
		\begin{subfigure}[b]{0.32\textwidth}
			\centering
			\includegraphics[width=\textwidth]{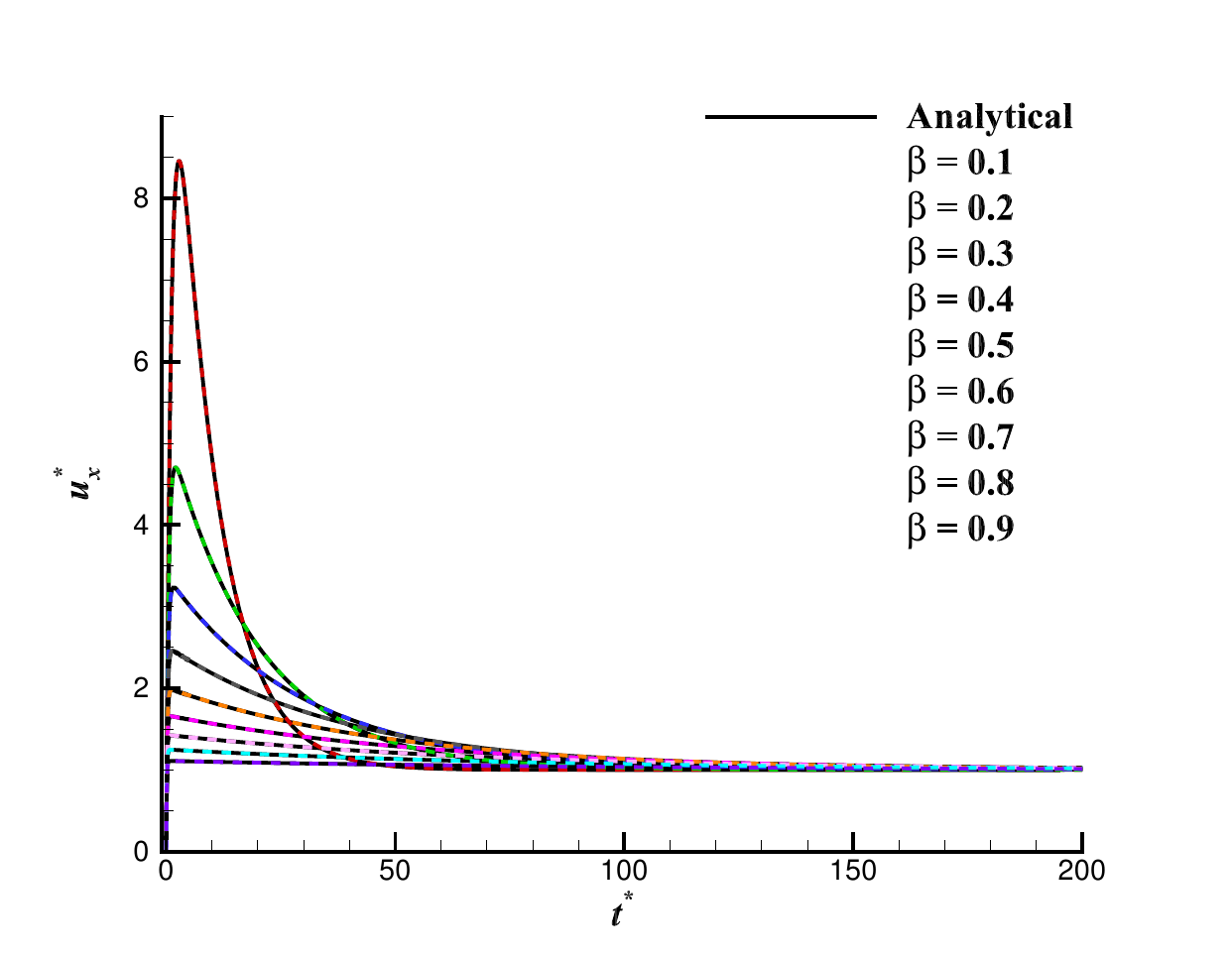}
			\caption{$Wi = 100$}
			\label{fig:ux_center_wi100}
		\end{subfigure}
		\hfill
		\begin{subfigure}[b]{0.32\textwidth}
			\centering
			\includegraphics[width=\textwidth]{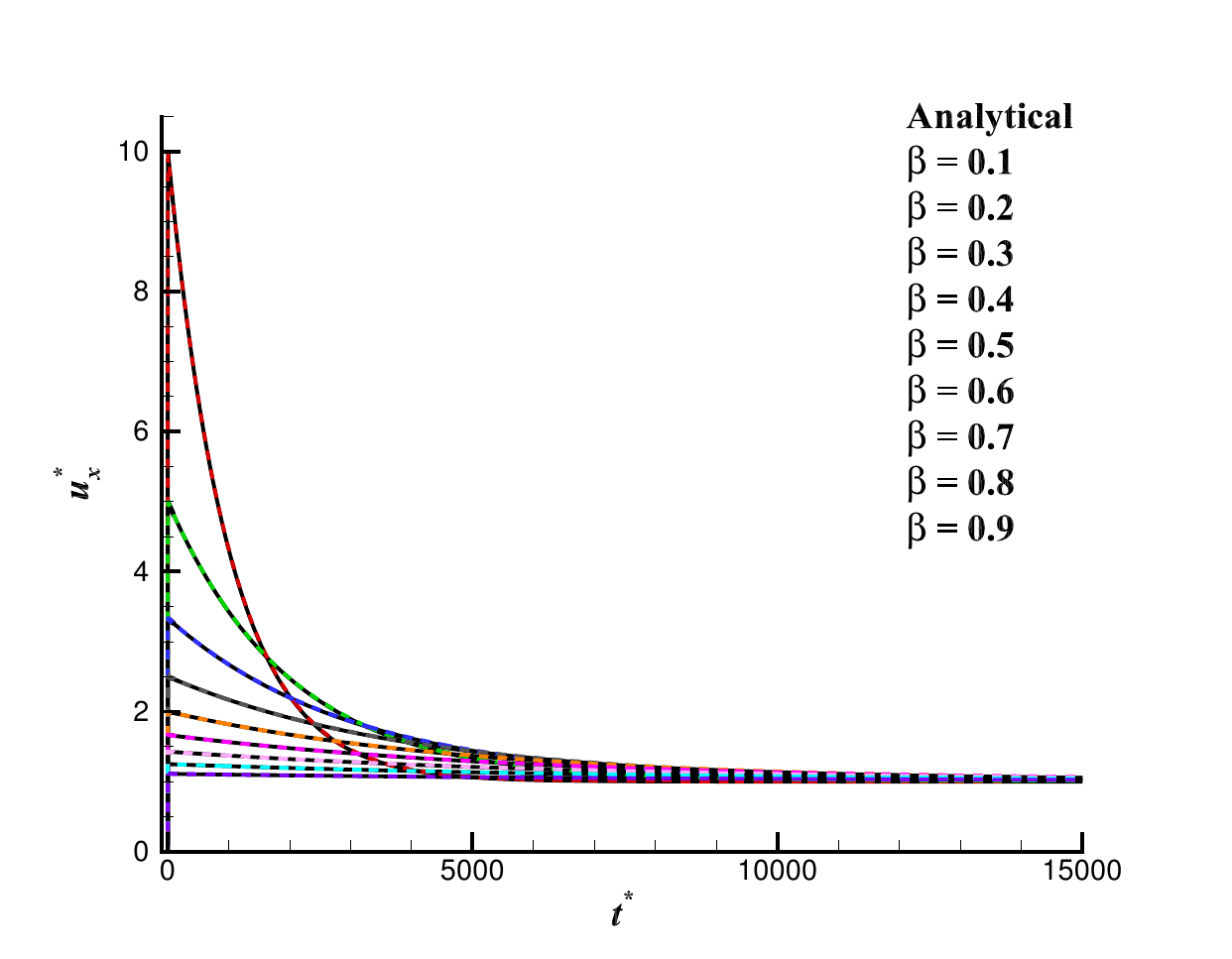}
			\caption{$Wi = 10000$}
			\label{fig:ux_center_wi10000}
		\end{subfigure}
		\caption{Time evolution of dimensionless velocity $u_x^{*}$ at channel center for different viscosity ratios ($\beta = 0.1$ to $0.9$) under three Weissenberg numbers: (a) $Wi = 1$, (b) $Wi = 100$, and (c) $Wi = 10000$. Schmidt number is fixed at $Sc = 10^6$ with $Re = 1$ and $N_y=32$.}
		\label{fig:ux_center_comparison}
	\end{figure}
	
	Subsequently, we turn our attention to the time-dependent transient behavior. We compute the temporal evolution of the dimensionless velocity component $u_x^{*}$ at the channel center as a function of dimensionless time $t^{*}$ for the three aforementioned Weissenberg numbers across nine different viscosity ratios ranging from $\beta = 0.1$ to $0.9$. The numerical results are compared against the corresponding analytical solutions and presented in Figure~\ref{fig:ux_center_comparison}. The comparison demonstrates that the present method accurately captures the transient dynamics across the entire parameter space, encompassing various Weissenberg numbers and viscosity ratios investigated.
	
	\subsubsection{Spatial Grid Convergence Study}
	
	Although our previous analysis of the four-roll mill problem demonstrated super-second-order convergence in the bulk region, the implementation of solid boundary conditions significantly influences the overall convergence behavior for Poiseuille flow problems. Therefore, this section focuses on examining the spatial convergence characteristics of the proposed method.
	
	We compare four different boundary condition schemes for implementing the no-slip boundary condition of the conformation tensor at solid walls: (1) our proposed Conservative Non-Equilibrium Bounce-Back (CNEBB) method, (2) the Non-Equilibrium Extrapolation (NEQE) method, (3) a method adapted from the multiphase wetting wall scheme proposed by Yu, Li, and Wen et al.~\cite{yu2020modified}, referred to as the YLW algorithm, and (4) the boundary algorithm from Malaspinas et al. It should be noted that the Navier-Stokes equations employ the non-equilibrium bounce-back scheme for all cases, with the sole difference among these methods being the boundary treatment for the advection-diffusion equation (ADE) system at solid walls. 
	
	Grid convergence studies are conducted using various mesh resolutions with $N_y = 16, 32, 64, 128, 256, 512, 1024$. The dimensionless velocity and conformation tensor components are recorded at the channel center and compared against analytical solutions. The global relative error (GRE) is computed using the following formula:
	\begin{equation}
		\text{GRE}(\theta)=\frac{\sum\limits_{k=1}^M |\theta(\textbf{x}_k)-\theta_{ana}(\textbf{x}_k)|}{\sum\limits_{k=1}^M |\theta_{ana}(\textbf{x}_k)|},
	\end{equation}
	where $\theta$ represents $u_x$, $A_{xx}$, or $A_{xy}$, $k$ is the node index, $M$ is the total number of computational nodes, and the subscript "ana" denotes the corresponding analytical solution obtained from Eqs.  (\ref{eq_ux_ana}), (\ref{eq_Tab_ana}), (\ref{eq_kramer}), and (\ref{eqb4}).
	
\begin{figure}[htbp]
		\centering
		\begin{subfigure}[b]{0.32\textwidth}
			\centering
			\includegraphics[width=\textwidth]{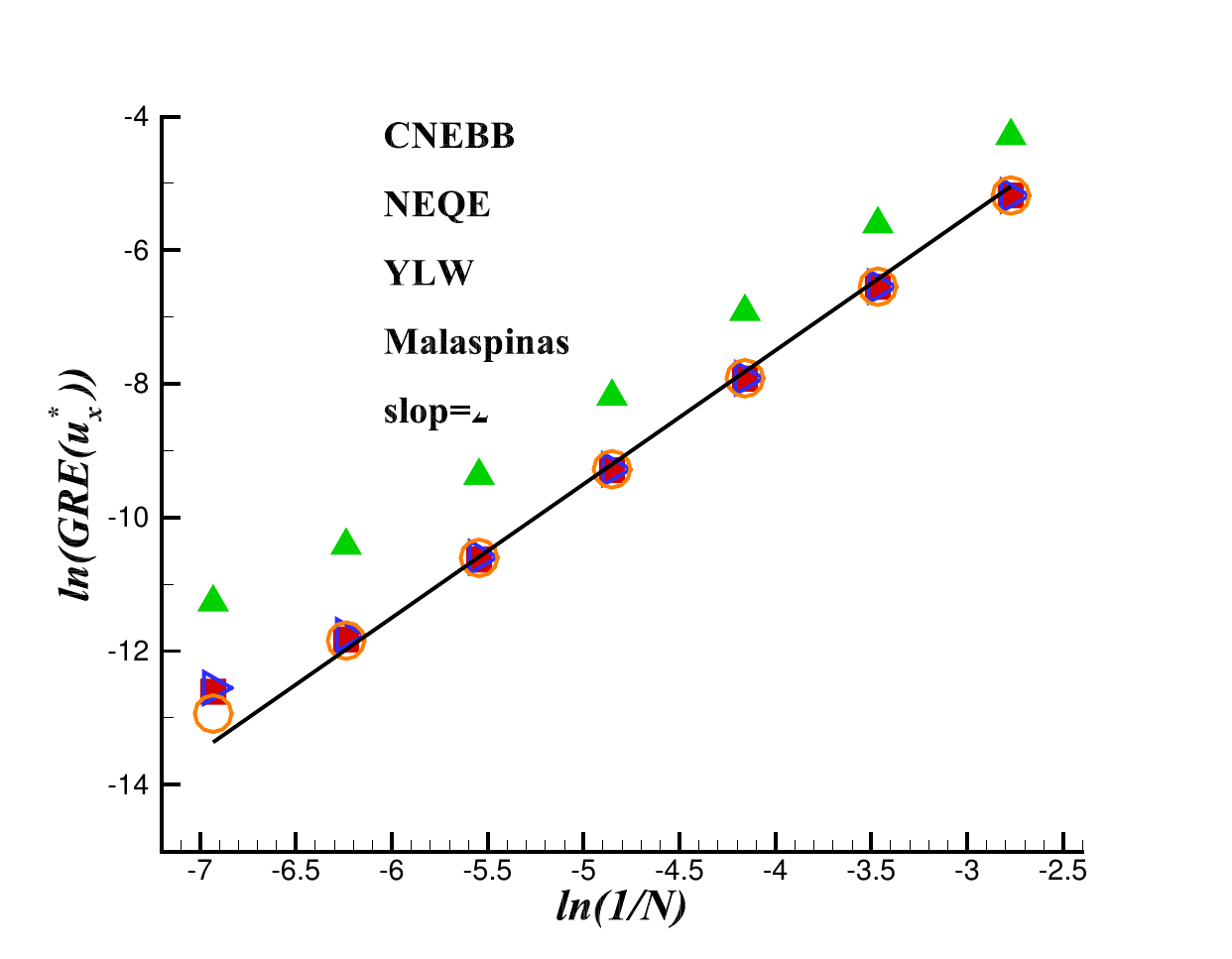}
			\caption{$u_x^{*}$}
			\label{fig:ux_grid}
		\end{subfigure}
		\hfill
		\begin{subfigure}[b]{0.32\textwidth}
			\centering
			\includegraphics[width=\textwidth]{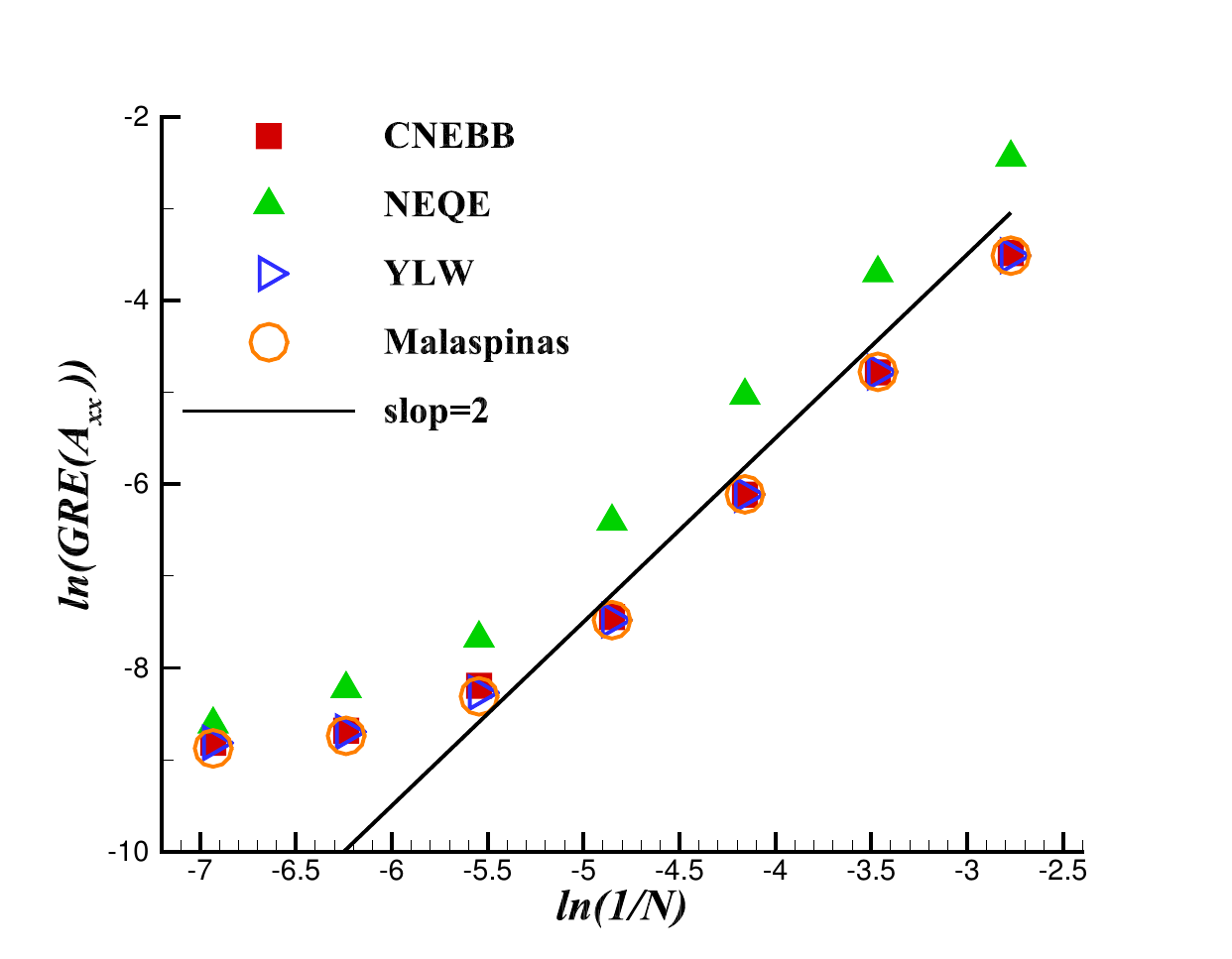}
			\caption{$A_{xx}$}
			\label{fig:axx_grid}
		\end{subfigure}
		\hfill
		\begin{subfigure}[b]{0.32\textwidth}
			\centering
			\includegraphics[width=\textwidth]{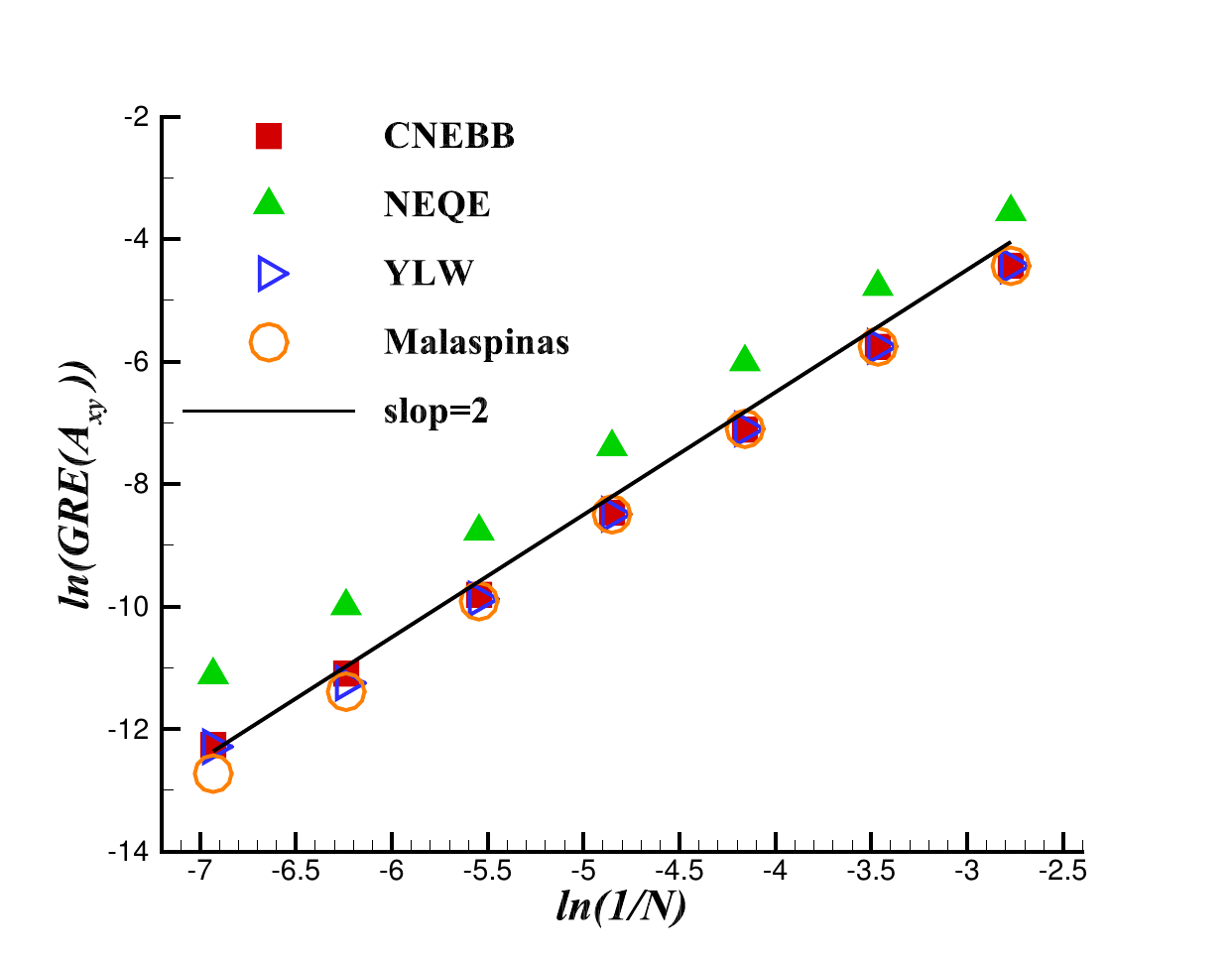}
			\caption{$A_{xy}$}
			\label{fig:axy_grid}
		\end{subfigure}
		
		\caption{Grid convergence analysis comparing different boundary schemes for dimensionless velocity $u_x^{*}$ and conformation tensor components $A_{xx}$ and $A_{xy}$ in Poiseuille flow. Grid resolution varies as $N_y = 16, 32, 64, 128, 256, 512, 1024$ with $Wi = 1$, $\beta = 0.5$, $Sc = 10^6$, and $Re = 1$. CNEBB (red solid squares), NEQE (blue solid triangles), YLW (blue hollow triangles), and Malaspinas (orange hollow circles) methods are compared. The black line indicates second-order convergence slope for reference.}
		\label{fig:grid_convergence}
	\end{figure}

	The comparative results are presented in Figure~\ref{fig:grid_convergence}. At low to moderate resolutions ($N_y \leq 256$), all schemes exhibit approximately second-order accuracy. The NEQE scheme demonstrates relatively higher errors compared to the other three methods, while the remaining three schemes yield comparable error levels. At high resolutions, the Malaspinas boundary scheme shows slightly lower errors for $u_x$ and $A_{xy}$. These results confirm the reliability of the proposed CNEBB scheme.
	
\begin{figure}[htbp]
		\centering
		\includegraphics[width=0.5\textwidth]{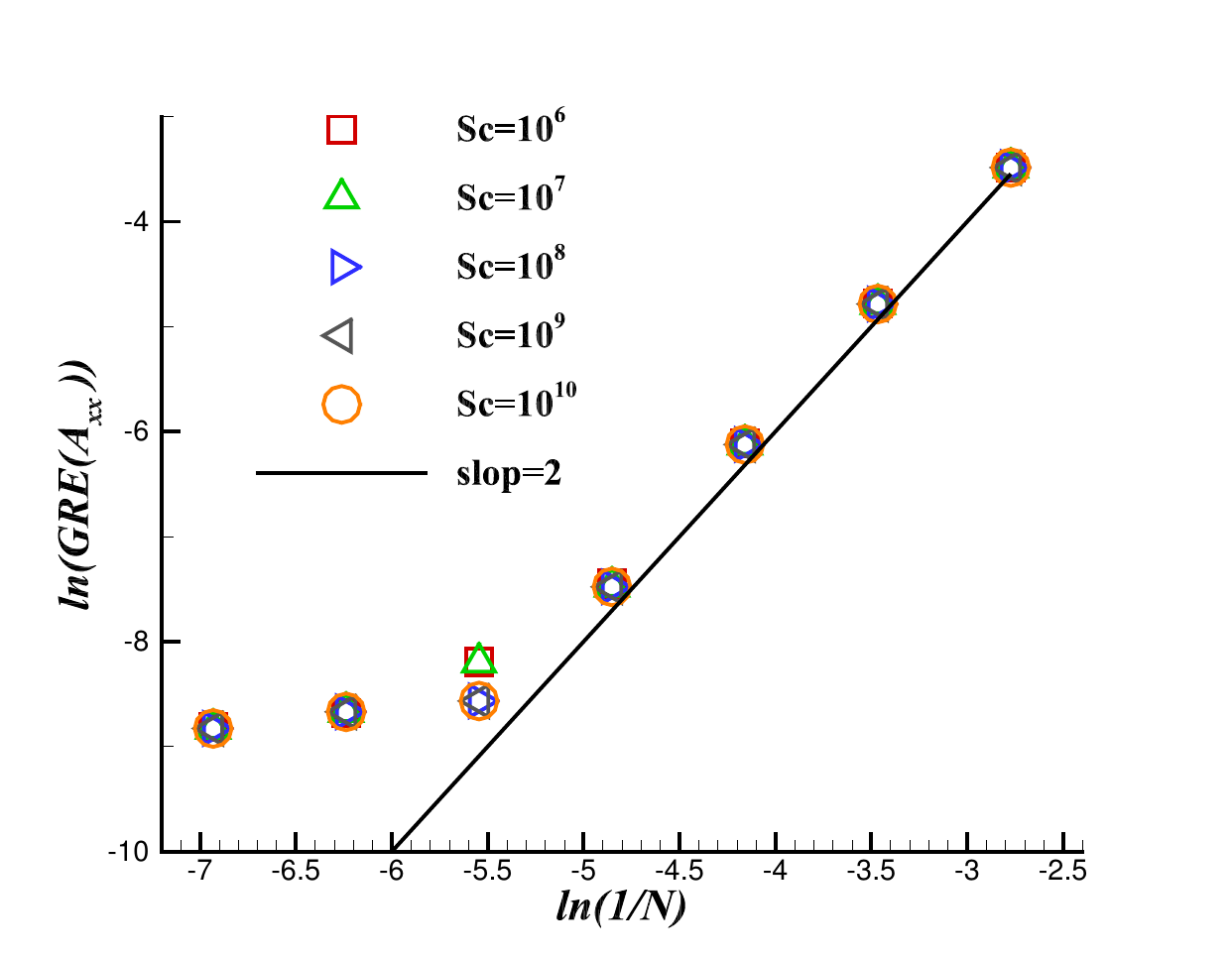}
		\caption{Grid convergence analysis for conformation tensor component $A_{xx}$ under different Schmidt numbers ranging from $Sc = 10^6$ to $Sc = 10^{10}$. The convergence order deterioration at high resolutions persists despite reduced artificial viscosity, with $Wi = 1$, $\beta = 0.5$, and $Re = 1$.}
		\label{fig:convergence_sc_effect}
	\end{figure}

	An important and intriguing observation emerges from this analysis: for all methods, the convergence order begins to deteriorate significantly at high resolutions, approaching zero in some cases, indicating that errors no longer decrease with grid refinement. This convergence order degradation with increasing resolution is a well-documented phenomenon in viscoelastic flow computations at high Weissenberg numbers. Previous studies have commonly attributed this behavior to excessive artificial viscosity, suggesting that artificial diffusion dominates the solution accuracy when numerical schemes attempt to resolve high-stress regions with insufficient stabilization, similar to the phenomenon illustrated in Figure~\ref{fig:convergence_sc_effect}.	However, our investigation reveals a surprising finding. To test whether artificial viscosity is indeed the cause of this convergence deterioration, we conducted additional simulations with higher Schmidt numbers (thereby significantly reducing artificial viscosity), as shown in Figure~\ref{fig:steady_state_comparison}. Contrary to the conventional explanation, increasing the Schmidt number does not alleviate this convergence order degradation. This observation suggests that the convergence order deterioration with grid refinement is not primarily caused by artificial viscosity effects, representing a novel finding that challenges the commonly accepted interpretation in viscoelastic flow simulations.

	\subsection{Flow Past a Circular Cylinder}\label{sec4.3}
	Flow past a circular cylinder represents a canonical benchmark problem in computational fluid dynamics that poses significant challenges for viscoelastic fluid simulations. Unlike previous validation cases involving simple geometries with straight boundaries, this configuration introduces curved solid surfaces that test the robustness and accuracy of boundary condition implementations. The presence of flow separation, recirculation zones, and complex stress distributions around the cylinder makes this problem particularly demanding for numerical methods, especially at elevated Weissenberg numbers where elastic effects become prominent. For viscoelastic fluids, cylinder flow exhibits rich physics including stress boundary layers, elastic wake modifications, and potential flow instabilities that are absent in Newtonian cases. The curved geometry necessitates careful treatment of conformation tensor evolution at the cylinder surface, where the proposed conservative boundary scheme faces its most rigorous test. The ability to maintain proper conservation properties along curved boundaries while accurately capturing complex stress distributions validates the method's applicability to realistic engineering flows.
	
	\begin{figure}[htp]
		\centering
		\includegraphics[width=0.8\textwidth]{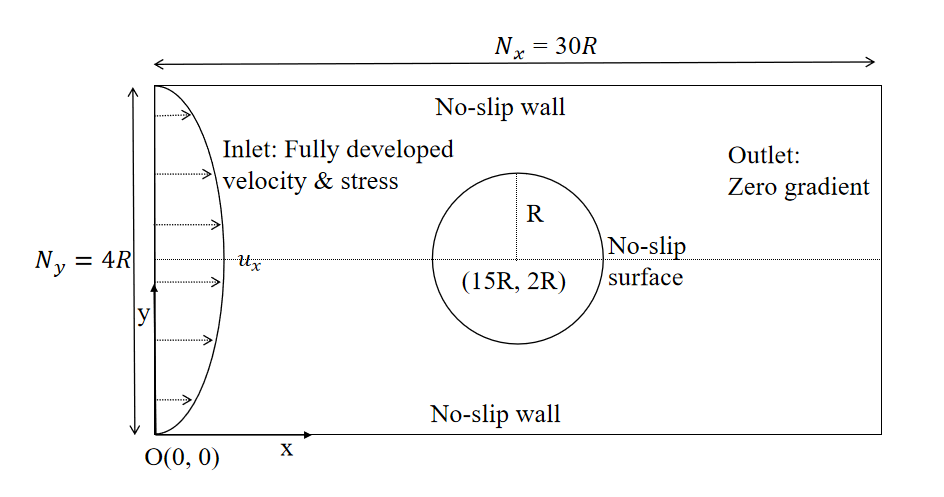}
		\caption{Computational domain and boundary conditions for flow past a circular cylinder. The rectangular domain has dimensions $30R \times 4R$ with the cylinder positioned at $(15R, 2R)$. Inlet: fully developed velocity and stress profiles; Outlet: zero-gradient conditions; Walls and cylinder surface: no-slip boundary conditions.}
		\label{fig_cylinder}
	\end{figure}
	
	The computational setup is illustrated in Figure~\ref{fig_cylinder}, consisting of a rectangular domain with dimensions $L_x \times L_y = 30R \times 4R$. The cylinder is positioned at the domain center with coordinates $(15R, 2R)$. The characteristic length is specified as $L_c = R$, and the characteristic velocity is $U_c = 2U_{\max}/3$, where $U_{\max}$ is the maximum value of the parabolic velocity profile at the inlet. The characteristic velocity $U_c$ is initialized through Mach number calculation with Ma = 0.01. Boundary conditions are implemented as follows: fully developed velocity and stress profiles at the inlet, zero-gradient (convective) conditions at the outlet, and no-slip conditions on the walls and cylinder surface.
	
	The fully developed inlet conditions are prescribed according to the analytical solution for plane Poiseuille flow:
	\begin{gather}
		u_x(y) = 1.5 \times 4U_c y(N_y-y)/N_y^2, \quad u_y = 0,\\
		A_{xx} = 1 + 2 \left(\frac{\lambda}{N_y}\right)^2\left(\frac{\partial u_x}{\partial y}\right)^2, \quad A_{yy} = 1, \quad A_{xy} = \frac{\lambda}{N_y} \frac{\partial u_x}{\partial y}
	\end{gather}
	where $A_{ij}$ represents the components of the conformation tensor at the inlet.
	
	The convergence criterion follows the condition specified in Eq.~(\ref{2eq4}). When this criterion is satisfied, the simulation is classified as "steady". If convergence is not achieved by dimensionless time $t^* > 1000$, the case is marked as "unsteady". Simulations that experience numerical breakdown are indicated by "NaN". For the NSE system, the inlet and outlet boundaries employ the NEBB scheme, while the walls utilize the same scheme. The cylinder surface adopts the YLW scheme~\cite{yu2020modified}, which has been proven to possess second-order accuracy for curved boundary problems. However, when the cylinder surface uses the half-way bounce-back (HWBB) scheme, a link-wise type scheme, for the CDE system, the cylinder surface correspondingly adopts the HWBB scheme for the NSE system to maintain consistency, since the previous algorithms are all wet node type schemes.
	
	For the CDE system, the inlet and outlet boundaries use the NEQE scheme, the walls employ our proposed CNEBB scheme, and the cylinder surface implements several different boundary conditions for systematic comparison, including CNEBB, YLW, HWBB, the scheme proposed by Malaspinas et al.~\cite{malaspinas2010lattice}, and NEQE. The CNEBB, YLW, and HWBB schemes satisfy the local conservation properties of the stress tensor, which is crucial for maintaining physical consistency.
	
	The default parameter configuration is as follows: $\mathrm{Re} = 1$, $\beta = 0.59$, $\Lambda_p = 2.5 \times 10^{-7}$ (the selection criterion for this parameter is discussed subsequently), $\Lambda_s = 1/4$, and $\mathrm{Sc} = 10^6$. The choice of $\mathrm{Re} = 1$ requires clarification: while most previous studies~\cite{ma2023high,hulsen2005flow,chai2021efficient,castillo2015first} use $\mathrm{Re} = 0.01$, our TRT regularized method, although stable and accurate for low Reynolds number cases, exhibits slow convergence rates. Moreover, for viscoelastic fluids at very low Reynolds numbers, the method becomes susceptible to numerical instabilities. Consequently, significant deviations are expected between $\mathrm{Re} = 1$ and $\mathrm{Re} = 0.01$ results, making direct quantitative comparison with previous studies challenging. However, the qualitative behavior and relative performance assessment among different boundary schemes remain valid.
	
	To quantitatively assess the accuracy and convergence behavior of different boundary schemes, the drag coefficient $C_d$ is computed for all simulations. This parameter serves as a sensitive indicator of numerical solution quality and provides a basis for comparing the performance of various boundary implementations. The total drag force $F_x$ acting on the cylinder surface is calculated using the momentum exchange method~\cite{ladd1994numerical}, which evaluates momentum transfer between fluid and solid through the distribution functions at boundary nodes:
	\begin{equation}\label{eq_F_x}
		F_x=\sum_{\textbf{x}_f \in \Gamma_b} \sum_{i \in I_w}  \frac{\Delta x^2}{\Delta t}e_{ix}(f_i^{\dagger}(\textbf{x}_f,t) + f_{\bar{i}}(\textbf{x}_f,t+\Delta t))
	\end{equation}
	where $\Gamma_b$ represents the set of fluid nodes adjacent to the cylinder surface that interface with the solid boundary, $I_w$ denotes the set of directions pointing toward the wall at boundary node $\textbf{x}_f$, $\bar{i}$ is the direction opposite to $i$, $f_i^{\dagger}$ is the post-collision distribution function, and $e_{ix}$ is the x-component of discrete velocity $\textbf{e}_i$. The drag coefficient is calculated as:
	\begin{equation}
		C_d = \frac{F_x}{\frac{1}{2}\rho U_{\text{avg}}^2 D}
	\end{equation}
	where $\rho$ is the fluid density, $U_{\text{avg}}$ is the average inlet velocity, and $D = 2R$ is the cylinder diameter.
	
	\subsubsection{Grid Independence Analysis of Different Conformation Tensor Boundary Implementation Schemes}
	A grid independence study is conducted to evaluate the performance and stability characteristics of different boundary schemes. The analysis employs the aforementioned comparative algorithms for simulations with cylinder radii ranging from $R = 5$ to $40$ (in increments of 5), three Weissenberg numbers ($\mathrm{Wi} = 0.1, 0.5, 1.0$), and systematic examination of Schmidt number effects ($\mathrm{Sc} = 10^4, 10^5, 10^6$). The results are presented in Tables~\ref{Wi_Sc_1.d4_Cd}, \ref{Wi_Sc_1.d5_Cd}, and \ref{Wi_Sc_1.d6_Cd}, respectively.
	
	The results reveal that implementing cylinder surface CDE boundaries using the NEQE algorithm yields poor performance among all tested schemes. At $\mathrm{Wi} = 0.5$ and $1.0$, regardless of grid density, the method essentially fails to converge. At the lowest Weissenberg number ($\mathrm{Wi} = 0.1$), convergence is achieved only when $R \leq 20$, but further grid refinement leads to numerical breakdown. The HWBB and Malaspinas schemes demonstrate marginally better performance but show no substantial improvement, converging across all considered grid densities only at $\mathrm{Wi} = 0.1$, while failing completely at higher Weissenberg numbers.
	
	Two schemes emerge as competitive methods: our newly proposed CNEBB method and the YLW algorithm extended to viscoelastic flows, both possessing identical local macroscopic quantity conservation strategies. The YLW scheme demonstrates stability improvements compared to the three aforementioned schemes. At $\mathrm{Wi} = 0.1$, convergence is achieved across all three Schmidt numbers as the grid is refined ($R$ increases). However, at $\mathrm{Wi} = 0.5$, while convergence is maintained under high artificial viscosity conditions ($\mathrm{Sc} = 10^4$), abrupt numerical breakdown occurs at $R = 35$ when artificial viscosity is reduced. The instability and breakdown phenomena become more pronounced at $\mathrm{Wi} = 1.0$. 
	
	In contrast, our newly proposed CNEBB method demonstrates good stability, achieving stable convergence across all tested conditions. This performance highlights the effectiveness of the specific conservative formulation employed in the CNEBB scheme for handling complex stress distributions at curved interfaces.
	
	Further examination of results computed with the CNEBB scheme reveals that at $\mathrm{Wi} = 0.1$ and $0.5$, drag coefficient values exhibit monotonic convergence to stable asymptotic values as the grid is refined. At $\mathrm{Wi} = 1.0$, although steady-state solutions can be consistently obtained, grid convergence becomes less apparent. This phenomenon of reduced grid convergence at higher Weissenberg numbers is commonly observed in existing viscoelastic cylinder flow studies~\cite{hulsen2005flow,alves2021numerical}. The underlying causes are multifaceted and potentially related to the interplay between artificial viscosity, boundary scheme accuracy, and the inherently complex physics at high elasticity numbers. Complete resolution of this challenging issue extends beyond the scope of the present work and remains an active area of research.
	
	\begin{table}[htp]
		\caption{Steady-state drag coefficients for viscoelastic flow past a circular cylinder at different grid densities (cylinder radius $R$) and Weissenberg numbers, comparing five different boundary schemes for the conformation tensor at the cylinder surface ($\mathrm{Sc}=10^4$). "Unsteady" indicates no steady state achieved at $t^*=1000$; "NaN" denotes numerical breakdown.}
		\label{Wi_Sc_1.d4_Cd}
		\centering
		\scriptsize
		\setlength{\tabcolsep}{2pt}
		\begin{tabular}{@{}c c c c c c c c c c c c c c c c@{}}
			\toprule
			\multicolumn{1}{c}{\multirow{2}{*}{$R$}} 
			& \multicolumn{3}{c}{$C_D$ (CNEBB)} 
			& \multicolumn{3}{c}{$C_D$ (YLW)}
			& \multicolumn{3}{c}{$C_D$ (HWBB)}
			& \multicolumn{3}{c}{$C_D$ (Malaspinas)}
			& \multicolumn{3}{c}{$C_D$ (NEQE)} \\
			\cmidrule(lr){2-4}  \cmidrule(lr){5-7}  \cmidrule(lr){8-10}  \cmidrule(lr){11-13}  \cmidrule(lr){14-16}
			& \rotatebox{270}{Wi=0.1} & \rotatebox{270}{Wi=0.5} & \rotatebox{270}{Wi=1.0}  
			& \rotatebox{270}{Wi=0.1} & \rotatebox{270}{Wi=0.5} & \rotatebox{270}{Wi=1.0}  
			& \rotatebox{270}{Wi=0.1} & \rotatebox{270}{Wi=0.5} & \rotatebox{270}{Wi=1.0}  
			& \rotatebox{270}{Wi=0.1} & \rotatebox{270}{Wi=0.5} & \rotatebox{270}{Wi=1.0}  
			& \rotatebox{270}{Wi=0.1} & \rotatebox{270}{Wi=0.5} & \rotatebox{270}{Wi=1.0} \\
			\midrule
			5 & 105.42 & 100.28 & 121.11 & 107.19 & 101.11 & 114.87 & 97.63 & 91.41 & 95.60 & 107.66 & 99.81 & 105.55 & 99.93 & unsteady & NaN \\
			10 & 125.19 & 120.53 & 168.44 & 126.55 & 123.33 & 204.93 & 122.29 & 118.53 & NaN & 126.54 & 121.90 & NaN & 120.54 & 111.73 & NaN \\
			15 & 128.22 & 123.34 & 170.09 & 129.03 & 125.59 & 157.74 & 126.69 & NaN & NaN & 129.17 & NaN & NaN & 124.45 & 116.53 & 133.87 \\
			20 & 129.42 & 125.17 & 164.26 & 129.96 & 128.00 & 148.30 & 128.43 & NaN & NaN & 130.04 & NaN & NaN & 126.37 & 118.08 & 136.18 \\
			25 & 129.61 & 125.00 & 156.01 & 129.87 & 125.13 & NaN & 128.30 & NaN & NaN & 129.95 & NaN & NaN & NaN & NaN & NaN \\
			30 & 130.36 & 126.31 & 151.31 & 130.63 & 126.65 & 134.39 & 129.67 & NaN & NaN & 130.68 & NaN & NaN & NaN & NaN & NaN \\
			35 & 130.77 & 127.72 & 149.04 & 131.03 & 127.31 & NaN & 130.65 & NaN & NaN & 131.16 & NaN & NaN & NaN & NaN & NaN \\
			40 & 130.79 & 126.79 & unsteady & 130.96 & 125.57 & 132.05 & 130.54 & NaN & NaN & 131.06 & NaN & NaN & NaN & NaN & NaN \\
			\bottomrule
		\end{tabular}
	\end{table}
	
	\begin{table}[htp]
		\caption{Steady-state drag coefficients for viscoelastic flow past a circular cylinder at different grid densities (cylinder radius $R$) and Weissenberg numbers, comparing five different boundary schemes for the conformation tensor at the cylinder surface ($\mathrm{Sc}=10^5$). "Unsteady" indicates no steady state achieved at $t^*=1000$; "NaN" denotes numerical breakdown.}
		\label{Wi_Sc_1.d5_Cd}
		\centering
		\scriptsize
		\setlength{\tabcolsep}{2pt}
		\begin{tabular}{@{}c c c c c c c c c c c c c c c c@{}}
			\toprule
			\multicolumn{1}{c}{\multirow{2}{*}{$R$}} 
			& \multicolumn{3}{c}{$C_D$ (CNEBB)} 
			& \multicolumn{3}{c}{$C_D$ (YLW)}
			& \multicolumn{3}{c}{$C_D$ (HWBB)}
			& \multicolumn{3}{c}{$C_D$ (Malaspinas)}
			& \multicolumn{3}{c}{$C_D$ (NEQE)} \\
			\cmidrule(lr){2-4}  \cmidrule(lr){5-7}  \cmidrule(lr){8-10}  \cmidrule(lr){11-13}  \cmidrule(lr){14-16}
			& \rotatebox{270}{Wi=0.1} & \rotatebox{270}{Wi=0.5} & \rotatebox{270}{Wi=1.0}  
			& \rotatebox{270}{Wi=0.1} & \rotatebox{270}{Wi=0.5} & \rotatebox{270}{Wi=1.0}  
			& \rotatebox{270}{Wi=0.1} & \rotatebox{270}{Wi=0.5} & \rotatebox{270}{Wi=1.0}  
			& \rotatebox{270}{Wi=0.1} & \rotatebox{270}{Wi=0.5} & \rotatebox{270}{Wi=1.0}  
			& \rotatebox{270}{Wi=0.1} & \rotatebox{270}{Wi=0.5} & \rotatebox{270}{Wi=1.0} \\
			\midrule
			5 & 106.04 & 101.73 & 128.69 & 107.45 & 101.66 & 144.18 & 97.79 & 92.44 & NaN & 107.82 & 101.11 & NaN & 100.74 & NaN & NaN \\
			10 & 125.45 & 120.38 & 165.21 & 126.62 & 123.50 & 188.91 & 122.31 & 118.65 & NaN & 126.57 & unsteady & NaN & 120.96 & 111.71 & NaN \\
			15 & 128.32 & 123.27 & 168.92 & 129.05 & 125.67 & 159.53 & 126.69 & NaN & NaN & 129.14 & NaN & NaN & 124.64 & unsteady & NaN \\
			20 & 129.48 & 125.02 & 162.91 & 129.98 & 128.01 & 147.85 & 128.43 & NaN & NaN & 130.02 & NaN & NaN & 126.49 & 118.56 & NaN \\
			25 & 129.64 & 124.86 & 154.90 & 129.88 & 124.00 & NaN & 128.30 & NaN & NaN & 129.94 & NaN & NaN & NaN & NaN & NaN \\
			30 & 130.38 & 126.20 & 149.74 & 130.66 & 126.50 & 133.83 & 129.67 & NaN & NaN & 130.67 & NaN & NaN & NaN & NaN & NaN \\
			35 & 130.80 & 127.60 & 147.22 & 131.05 & NaN & NaN & 130.65 & NaN & NaN & 131.15 & NaN & NaN & NaN & NaN & NaN \\
			40 & 130.81 & 126.72 & unsteady & 130.96 & 125.49 & 131.38 & 130.55 & NaN & NaN & 131.05 & NaN & NaN & NaN & NaN & NaN \\
			\bottomrule
		\end{tabular}
	\end{table}
	
	\begin{table}[htp]
		\caption{Steady-state drag coefficients for viscoelastic flow past a circular cylinder at different grid densities (cylinder radius $R$) and Weissenberg numbers, comparing five different boundary schemes for the conformation tensor at the cylinder surface ($\mathrm{Sc}=10^6$). "Unsteady" indicates no steady state achieved at $t^*=1000$; "NaN" denotes numerical breakdown.}
		\label{Wi_Sc_1.d6_Cd}
		\centering
		\scriptsize
		\setlength{\tabcolsep}{2pt}
		\begin{tabular}{@{}c c c c c c c c c c c c c c c c@{}}
			\toprule
			\multicolumn{1}{c}{\multirow{2}{*}{$R$}} 
			& \multicolumn{3}{c}{$C_D$ (CNEBB)} 
			& \multicolumn{3}{c}{$C_D$ (YLW)}
			& \multicolumn{3}{c}{$C_D$ (HWBB)}
			& \multicolumn{3}{c}{$C_D$ (Malaspinas)}
			& \multicolumn{3}{c}{$C_D$ (NEQE)} \\
			\cmidrule(lr){2-4}  \cmidrule(lr){5-7}  \cmidrule(lr){8-10}  \cmidrule(lr){11-13}  \cmidrule(lr){14-16}
			& \rotatebox{270}{Wi=0.1} & \rotatebox{270}{Wi=0.5} & \rotatebox{270}{Wi=1.0}  
			& \rotatebox{270}{Wi=0.1} & \rotatebox{270}{Wi=0.5} & \rotatebox{270}{Wi=1.0}  
			& \rotatebox{270}{Wi=0.1} & \rotatebox{270}{Wi=0.5} & \rotatebox{270}{Wi=1.0}  
			& \rotatebox{270}{Wi=0.1} & \rotatebox{270}{Wi=0.5} & \rotatebox{270}{Wi=1.0}  
			& \rotatebox{270}{Wi=0.1} & \rotatebox{270}{Wi=0.5} & \rotatebox{270}{Wi=1.0} \\
			\midrule
			5 & 106.93 & 103.88 & unsteady & 107.83 & 102.95 & unsteady & 98.01 & 93.36 & NaN & NaN & NaN & NaN & 102.03 & NaN & NaN \\
			10 & 125.90 & 121.26 & 165.30 & 126.87 & 128.68 & unsteady & 122.37 & unsteady & NaN & NaN & NaN & NaN & 122.06 & NaN & NaN \\
			15 & 128.52 & 123.21 & 161.91 & 129.15 & 126.39 & unsteady & 126.71 & NaN & NaN & NaN & NaN & NaN & 125.42 & NaN & NaN \\
			20 & 129.59 & 124.84 & 158.44 & 130.03 & 128.36 & 147.75 & 128.44 & NaN & NaN & 130.00 & NaN & NaN & 127.07 & NaN & NaN \\
			25 & 129.69 & 124.60 & 152.98 & 129.89 & 124.28 & NaN & 128.30 & NaN & NaN & 129.93 & NaN & NaN & NaN & NaN & NaN \\
			30 & 130.42 & 125.88 & 147.14 & 130.74 & 126.23 & 132.69 & 129.66 & NaN & NaN & 130.66 & NaN & NaN & NaN & NaN & NaN \\
			35 & 130.83 & 127.30 & 143.66 & 131.07 & NaN & NaN & 130.65 & NaN & NaN & 131.12 & NaN & NaN & NaN & NaN & NaN \\
			40 & 130.83 & 126.39 & 140.32 & 130.97 & 125.24 & 130.88 & 130.54 & NaN & NaN & 131.03 & NaN & NaN & NaN & NaN & NaN \\		
			\bottomrule
		\end{tabular}
	\end{table}
	
	\subsubsection{Maximum Achievable Weissenberg Number and Optimal $\Lambda_p$ Parameter Selection}
	
	Having demonstrated the stability of the CNEBB method in the previous analysis, we now focus on determining the maximum Weissenberg numbers that can be achieved with this approach and identifying the optimal selection of the magic parameter $\Lambda_p$. The maximum achievable Weissenberg number serves as a performance metric that reflects the computational robustness of the method for highly elastic flows. To systematically explore the computational limits of our approach, we employ minimal artificial viscosity ($\mathrm{Sc} = 10^6$), representing challenging conditions under which many existing algorithms fail prematurely. 
	
\begin{figure}[htp]
		\centering
		\begin{subfigure}[b]{0.45\textwidth}
			\centering
			\includegraphics[width=\textwidth]{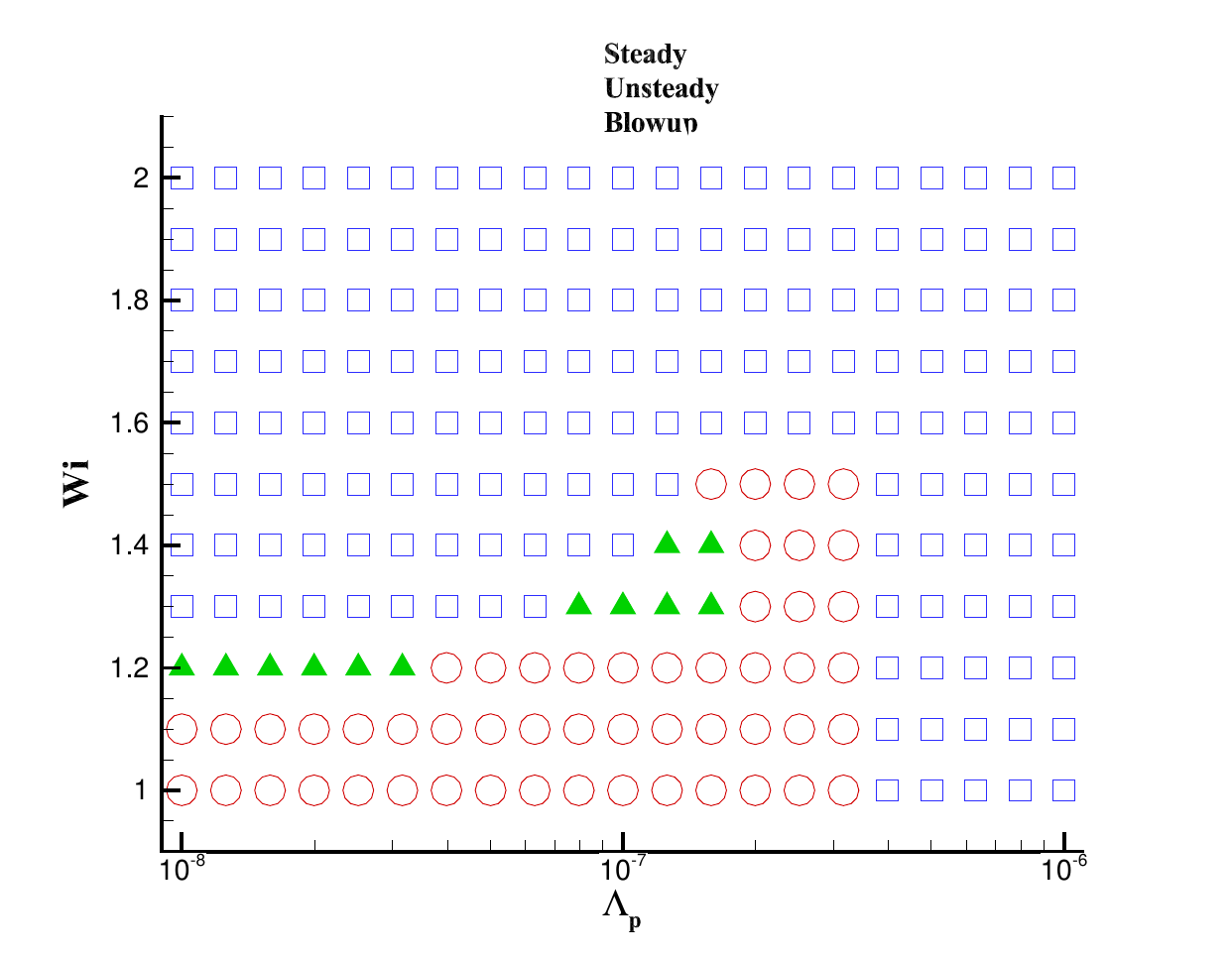}
			\caption{Phase diagram for $R = 20$}
			\label{fig:max_wi_r20}
		\end{subfigure}
		\hfill
		\begin{subfigure}[b]{0.45\textwidth}
			\centering
			\includegraphics[width=\textwidth]{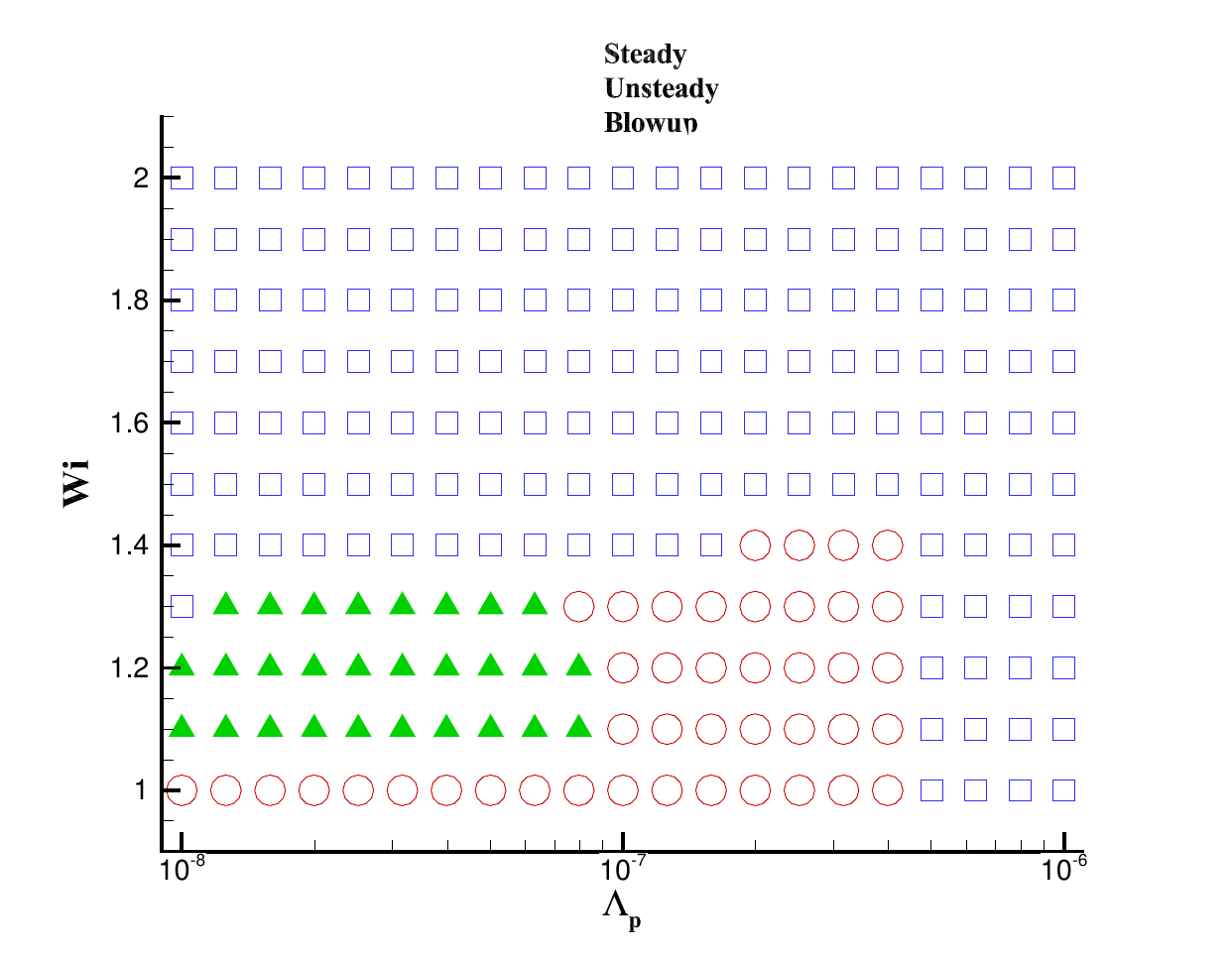}
			\caption{Phase diagram for $R = 30$}
			\label{fig:max_wi_r30}
		\end{subfigure}
		\caption{Phase diagrams showing computational stability regions in the $\Lambda_p$-Wi parameter space at $\mathrm{Sc} = 10^6$, a stringent condition that causes premature failure in many algorithms. Red hollow circles: steady-state solutions achieved; green solid triangles: unsteady solutions obtained; blue hollow squares: numerical breakdown occurred.}
		\label{fig:max_wi_comparison}
	\end{figure}

	The magic parameter $\Lambda_p$ is systematically varied by taking uniformly distributed sample points in the logarithmic space within the range $10^{-8}$ to $10^{-6}$ to identify optimal values that maximize computational stability while maintaining physical accuracy. This parameter plays a critical role in the TRT regularization scheme and requires careful tuning to balance numerical stability and solution quality. Two representative cylinder radii ($R = 20$ and $30$) are considered with Weissenberg numbers ranging from $1.0$ to $2.0$ in increments of $0.1$. The results are presented in Figure~\ref{fig:max_wi_comparison}.
	
	The systematic analysis reveals distinct parameter regimes with varying stability characteristics. For $\Lambda_p$ values in the range $10^{-8}$ to $3.162 \times 10^{-7}$, the maximum achievable Weissenberg number without numerical breakdown is $\mathrm{Wi} = 1.2$. However, when $\Lambda_p$ is optimally tuned within the narrower range from $1.995 \times 10^{-7}$ to $3.162 \times 10^{-7}$, stable computations can be extended to $\mathrm{Wi} = 1.5$. This improvement in the stability limit approaches the benchmark result of $\mathrm{Wi} = 1.7$ achieved by Ma et al.~\cite{ma2023high} using computationally expensive high-order spectral element discontinuous Galerkin (SRCR-DG) methods.
	
	The identification of this optimal parameter window demonstrates the importance of proper magic parameter selection in viscoelastic LBM simulations. Too small values of $\Lambda_p$ lead to insufficient numerical stabilization, while excessively large values introduce spurious diffusion that can destabilize the solution. The discovered optimal range provides practical guidance for future applications of the method to challenging viscoelastic flow problems.
	
	This validation demonstrates the effectiveness of the proposed CNEBB method in handling complex curved geometries while maintaining numerical stability at elevated Weissenberg numbers. The method shows improved performance compared to existing boundary schemes for viscoelastic flow simulations using the lattice Boltzmann framework.
	
	\section{Conclusion}\label{sec5}	
	In this paper, we have proposed an improved lattice Boltzmann method for solving Oldroyd-B fluid flow problems. The hydrodynamic field evolution is solved using an improved two-relaxation-time regularized lattice Boltzmann model, in which we employ a first-order Hermite expansion form for the force term to enhance the accuracy of introducing body forces in the NSE equations. Additionally, a novel discrete velocity source term for direct stress tensor incorporation is proposed, eliminating the need for spatial derivative calculations of the stress tensor required by traditional methods. It should be noted that while this model utilizes the two-relaxation-time regularization operator we previously proposed, this is not mandatory. Any other high-stability collision operator, such as central moment models or cumulant moment models, can be substituted without affecting the core innovation of the present model.
	
	For solving the evolution equation of the conformation tensor, we employ a set of improved convection-diffusion LB solvers. In this model, a novel auxiliary source term (Eq.~\ref{eq_p_G_1}) coupled with the NSE through force terms is proposed to avoid spatial or temporal derivative calculations required in traditional methods, achieving tighter NSE-CDE coupling. The two-relaxation-time collision operator we previously developed is also adopted to enhance model stability, where the magic parameter $\Lambda_p$ controlling the free relaxation parameter significantly influences program stability.
	
	Comprehensive numerical studies of the simplified four-roll mill problem demonstrate that our newly proposed method achieves super-second-order convergence for this doubly periodic boundary problem containing stress singularities in the bulk region without solid boundaries. The numerical results not only agree well with reference solutions in the commonly reported parameter range of $Wi \leq 5$, but also exhibit superior performance compared to existing LBM methods at high Weissenberg numbers. Under extremely low artificial viscosity conditions ($Sc = 10^6$), unlike other LBM approaches that fail prematurely, our method can obtain steady-state solutions at $Wi = 27$ and unsteady solutions at even higher Weissenberg numbers. It should be noted that when artificial viscosity is sufficient (e.g., $Sc = 10^4$ and $10^5$), the selection of the magic parameter $\Lambda_p$ has minimal impact on algorithm stability. However, when artificial viscosity is very low ($Sc = 10^6$), the magic parameter $\Lambda_p$ becomes highly sensitive to algorithm stability. Our numerical studies suggest that for problems without solid boundaries, optimal numerical stability is achieved by selecting $\Lambda_p$ in the range of [specific range to be filled]. Lower values of $\Lambda_p$ can also provide non-divergent solutions at moderate Weissenberg numbers ($Wi \leq 20$), representing an alternative viable range.
	
	To the authors' knowledge, neither traditional methods nor LBM approaches have previously addressed the conservation properties of solid wall boundary schemes for conformation tensor or stress tensor fields. Based on the LBM framework, this paper proposes a novel conservative non-equilibrium bounce-back (CNEBB) scheme for solving conformation tensor fields. Numerical results for Poiseuille flow demonstrate that this method can obtain accurate steady-state velocity and conformation tensor field distributions up to $Wi = 10000$ within our tested range. In our actual computations, higher Weissenberg numbers are also achievable, requiring only extended computational time. Numerical results further show that the algorithm agrees well with transient analytical solutions at arbitrary viscosity ratios up to $Wi = 10000$. Additionally, we observed that convergence order gradually decreases from second-order with grid refinement, with global relative error even stabilizing at a constant value (approximately $O(10^{-4})$). Through extreme reduction of artificial viscosity, we found that the convergence order degradation phenomenon is not significantly alleviated, unlike in the four-roll mill problem where reducing artificial viscosity effectively reduces error. Therefore, we believe this represents a novel systematic error related to boundary schemes.
	
	We also investigated another representative benchmark problem: viscoelastic flow past a circular cylinder. Numerical results demonstrate that three schemes—non-equilibrium extrapolation (NEQE), halfway bounce-back (HWBB), and Malaspinas—exhibit poor stability when handling cylinder flow problems, even with artificial viscosity as low as $Sc = 10^4$. In contrast, the YLW algorithm introduced from multiphase flow and our newly proposed CNEBB scheme inspired by the YLW algorithm show significantly better stability. Among all tested algorithms, our CNEBB algorithm achieves optimal numerical stability performance. Under the most stringent artificial viscosity conditions ($Sc = 10^6$), we can compute maximum Weissenberg numbers up to approximately 1.5, which closely approaches the $Wi = 1.7$ achievable by SRCR-DG methods at lower Reynolds numbers ($Re = 0.01$). Furthermore, we conducted systematic parameter sweeps for the magic parameter $\Lambda_p$, recommending a selection range of $2.0\times10^{-6}$ to $3.0\times10^{-6}$ for problems involving solid boundaries.
	
	In summary, we have developed an improved lattice Boltzmann method and proposed a conservative non-equilibrium bounce-back boundary scheme for computing viscoelastic fluid flow problems. The algorithm demonstrates good numerical stability and accuracy, providing a valuable numerical tool for related research fields. In the future, we plan to improve the computational efficiency of this algorithm at low Reynolds numbers while maintaining its stability, and extend it to three-dimensional applications.
	\section*{Acknowledgments}
	This work was financially supported by the National Natural Science Foundation of China (Grant Nos. 12101527, 12271464, and 12371373), the Natural Science Foundation for Distinguished Young Scholars of Hunan Province (Grant No. 2023JJ10038), the Science and Technology Innovation Program of Hunan Province (Program No. 2021RC2096), and the Scientific Research Fund of Hunan Provincial Science and Technology Department (Grant No. 21B0159). Computational resources were provided by the High Performance Computing Platform of Xiangtan University. 
	
	\bibliographystyle{elsarticle-num} 
	\bibliography{mybibfile}
	
	%
	%
	%
	%
\end{document}